\documentclass[10pt, preprint, a4paper]{aastex}
\usepackage{graphicx}
\newcommand{\Fyris}{\emph{Fyris Alpha}}

\title{A New Computational Fluid Dynamics Code I: Fyris Alpha}
\author{Ralph S. Sutherland}
\affil{Research School of Astronomy \& Astrophysics, Australian National University, Cotter Rd., Weston ACT, Australia 2611}

\begin{document}

\begin{abstract}
A new hydrodynamics code aimed at astrophysical applications has been developed.  The new code and algorithms are presented along with a comprehensive suite of test problems in one, two, and three dimensions.  

The new code is shown to be robust and accurate, equalling or improving upon a set of comparison codes. \Fyris\ will be made freely available to the scientific community.

\end{abstract}

\maketitle

\keywords{hydrodynamics, computational fluid dynamics, algorithms}

\section{Introduction}

Multi-dimensional computational fluid dynamics is a mature field, with roots in the earliest days of computation.  In the astrophysical setting, with compressible, trans--, and super--sonic flows, the hyperbolic Eulerian system of ideal gas equations are typically used, allowing for the formation of shock discontinuities.  The three dimensional non-viscous Euler equations may be written in terms of the conserved variables (density, $\rho$, momentum $m$, and total energy $E$) at time $t$; $U_t$, and flux vectors, $F(U)$, $G(U)$, $H(U)$:
\begin{equation}
U_t + F(U)_x + G(U)_y + H(U)_z = 0,
\end{equation}
with
\begin{eqnarray}
U_t     &=&  \left[ \begin{array}{c} \rho \\ m_x \\ m_y \\ m_z \\ E \end{array} \right]\, , \\
F(U)_x =  \left[ \begin{array}{c} v_x  m_x + P  \\ v_x  m_y       \\ v_x  m_z       \\ v_x ( E + P ) \end{array} \right] \, , \,
G(U)_y &=&  \left[ \begin{array}{c} v_y  m_x   \\ v_y  m_y  + P \\ v_y  m_z       \\ v_y ( E + P ) \end{array} \right] \, , \,
H(U)_z =  \left[ \begin{array}{c} v_z  m_x      \\ v_z  m_y        \\ v_z  m_z + P \\ v_z ( E + P ) \end{array} \right] \, , 
\end{eqnarray}
where the momentum terms are; 
$m_x = \rho v_x$,  $m_y = \rho v_y$, $m_z = \rho v_z$.   The total energy per unit volume is $E = \rho ( \frac {1}{2} {\bf v}^2 + e)$, where ${\bf v}$ is the total velocity and $e$ is the specific internal energy.  $v_x$, $v_y$ and $v_z$ are the velocity components.  $P$ is the gas pressure.

With a baric equation of state $P = EOS_{\rm B}( \rho, e, \gamma)$, relating density, $\rho$, pressure, $P$, and internal energy $e$ ( using the polytropic index $\gamma$), the system can be closed.  When the temperature is required, such as when cooling is used, a caloric equation of state, $T = EOS_{\rm C}( \rho, P, \mu )$ is invoked, using, $\mu$, the mean molecular weight.   
 
 This system has been solved on a grid of cells by considering the cell average values and the flow, or fluxes, between cells.   The methods used have often been founded on the solution of the Riemann problem, a simple non-linear interaction between two adjacent regions of ideal gasses, as first outlined by Godunov in the 1950s, \citep{godunov59}, and \cite{glimm65} in the 1960s.

With the formation of shockwaves in a computed flow, a fundamental problem arises where high order methods inevitably lead to non-physical oscillations of the flow properties in cells near a discontinuity, and the pioneering work of Godunov showed, in the Godunov Theorem, that no monotonic shock capturing method could be higher than first order accuracy.  Since then, the development of hybrid high order methods that give rapid convergence in smooth flows, while also computing locally first order treatments of shock discontinuities, have been the goal of many.    

The first practical methods to localise and manage Godunov's theorem were developed in a series of papers by van Leer in the 1970s ( \cite{vanleerII, vanleerIII, vanleerIV, vanleerV} ).  Godunov's method of solving the Riemann gas problem was combined with the development of adaptive piecewise grid reconstructions, were the order of interpolation was reduced in the vicinity of shock discontinuities, but allowed to be higher than first order in areas of smooth flow.   Conservative fluxes were computed from the solution of inter--cell Riemann problems and used to update the cell quantities.  A key development in 1984 as the use of a conservative 4th order integral polynomial intercell boundary reconstruction, which resulted in the widely used PPM, piecewise parabolic method of Collella \& Woodward ( \citet{cw84, cw84b}).  The PPM method is extremely efficient for the order of accuracy achieved, up to 3rd order in spatial reconstruction in smooth flows on a regular grid.

Apart from higher order inter--cell reconstruction, work progressed on solving the non-linear Riemann problem more quickly, which was especially slow before hardware support for transcendental functions became widespread.   One approach was to linearise the Riemann problem, where an approximate Riemann problem is solved exactly --  resulting in the Roe--Pike approximate solvers in the mid 1980s (see for example \cite{roe1981, roe1986}), which became very widely used.  Other less common approaches included simplifying the iterations when solving the full non-linear problem, giving approximate solutions to the exact Riemann problem \citep{toro99}.  Full, exact solvers of the full non-linear Riemann problem were generally abandoned.

In the 1980s the issues surrounding the control of `post-shock' ringing, or generally non-monotonic behaviour of any method higher than first order were explored and refined with the development of TVD (Total Variation Diminishing) \citep{harten83, harten84} and later ENO/WENO (Essentially Non-Oscillating and Weighted Essentially Non-Oscillating) methods \citep{liu1994,jiang1996}.   The PPM method continues to be effective, despite its somewhat ad-hoc monotonicity constraints compared to the more formal TVD flux limiting methods, due to it's great speed and high order interpolation which -- has a particularly simple form on a regular grid due to exploited symmetries in the polynomial construction.

Since the 1990s, interest in the Astrophysical regime has been to extend the microphysics ( Cooling, dust, ionisation,  magnetic fields etc) and there are a large number of CFD codes targeting different aspects of astrophysical fluids.  With this diversity of applications, the key issues have become similarly diverse, and applying a single code to all problems is not practical.
Here a new code has been developed, to provide a means of computing astrophysical flow problems with more sophisticated microphysics than previous codes, aimed at providing more reliable observational quantities from the simulations, such as H$\alpha$ surface brightness, for example.  

In this work, we present the astrophysical hydrodynamics code, \Fyris, with a focus on those aspects which have been developed specifically for the new code, along with a comprehensive suite of existing and new test problems to verify the new code.  Verification by such test suites is essential, as has been well described in the Lenska \& Wendroff review \citep{liska03}, hereafter LW03, where they showed that even well established codes can have particular weaknesses that are not apparent from a simple description of their algorithms.    LW03 presented a set of 1D and 2D test problems, many with known analytical solutions, and derived robust L1-norm measures of the deviations from the known solutions, providing an objective measure of the complete code performance.   Here we apply the LW03 tests to \Fyris, and present additional 2D and 3D tests to cover aspects either not fully covered in LW03, or not covered at all.  In particular there is a need for analytical 3D test problems for testing code, as large scale 3D simulations are becoming widely used.

Paper II will cover the non-fluid dynamics aspects of the code, such the time-dependent ionisation, along with aspects of the magneto-hydrodynamic treatment, in the code {\em Fyris Beta}.

\section{Fyris Alpha}

\Fyris\ uses a new iterative Riemann solver, that is significantly faster than previously used solvers, and even out performs many non-iterative and linearised solvers.  It allows for a variable equation of state through a generalised polytropic index, $\gamma$.  The grid and cell reconstruction uses the PPM method (\citet{cw84}).  Finally, extensive use of methods to deal with and control floating point representations and `grid--noise' enable optimisations that make the code up 4 -- 10 times faster than a similar PPM  code without the active floating point noise control (AFPNC) and older Riemann solvers.  This additional performance allows for the incorporation of more detailed microphysics, and the computation of larger simulations.    The current implementation uses single and nested refined meshes, and although the capability  for creating and destroying zones and subzones is present, fully adaptive mesh operation (AMR) is not yet implemented and the criterium functions for determining their creation are not developed.  AMR is planned for a future version.

 Extensions including time-dependent ionisation, variable equation of state and molecular cooling, magnetic fields and relativistic treatments are in development and will be presented in later works.

\subsection{Origins}

\Fyris\ is a successor to previous high resolution shock capturing hydrodynamics codes which arose from the original works by \cite{vanleerIII, vanleerIV, vanleerV} and  \cite{cw84, cw84b} on the PPM algorithm, and more recently on codes such as VH1, \citep{christian97} and \emph{ppmlr}, \citep{suth03a},\cite{ suth03b}.   The \Fyris\ code was written to address some of the shortcomings of the older FORTRAN \emph{ppmlr} code, particularly in the areas of: a redesign to allow for more general refined meshes \citep{berger84}, plus significant code improvement and  performance increases to allow more microphysics to be incorporated, and to a lesser extent allow for more ease in modifying and extending the code in the future.

To address these aims, the lower level `C' language was chosen.  It offers comparable high performance (albeit with more work from the writer), a great deal more flexibility than FORTRAN, portability across many platforms, and access to a range of existing utility code and Open Source standards that made the simulation support (I/O, timing etc.) easier to build.

The code is conceptually made up of an overarching framework, called \emph{ZLS} (Zone--Level System), which provides all the standard services needed for any simulation, plus provision for modules of specific algorithm code.  \Fyris\ is the first of these modules, and it encapsulates the split sweep semi--Lagrangian remap ppm method used by \emph{ppmlr} with a number of improvements that have allowed the addition of new micro--physics needed in multi-phase simulations.  Future modules are in development.  A website for the code, including validation suites and user information, will be available soon.

\section{Fyris Alpha Algorithms}

\subsection* {Split Sweep Semi-Lagrangian with Remap}

\Fyris\ uses an iterative Riemann solver in a high order Lagrangian Godunov scheme to integrate the ideal Eulerian hydrodynamic system.  Following earlier shock capturing methods, a Riemann solver is used to compute the states at the boundaries between cells on a grid and the solutions are used to update the grid.

The cells are defined in a standard conservative way, using the coordinates of the left and right boundaries cell $i$and the average variable value $<q_i>$ in the cell.  With a sequence of cells, the PPM algorithm uses the integral of the cells to interpolate and fit parabolae to each cell that preserve the integrals, and hence the average cell values.  With the parabolae, the cell boundary left and right states are estimated by averaging the parabola over a domain determined by the strength of the cell--cell interaction.  \Fyris\ uses the maximum of the flow speed and the sound speeds in the cells to determine these domains.

The Riemann solver computes the speed of the contact discontinuity ($u_*$) and the pressure in that region ($P_*$).  The cell boundaries are updated  to follow this velocity over the time step, an in addition $P_*$ and $u_*$ are used to compute the energy and momentum fluxes from the boundary due to the Riemann waves.  After the modified cells are updated, the boundaries are remapped back to their original positions.  

The grid is swept in 2D in a cyclic order order,and over two time steps is, $x,y-y,x- $, akin to the second order Strang
splitting, (\citep{strang68}),  of xDt y2Dt xDt. 
 In 3D the sweep order is permuted: $x,y,z-y,z,x-z,x,y-$ in order to cycle through orthogonal starting sweeps in each sub--cycle.

Finally, the split semi--Lagrangian method, as opposed to an unsplit fixed grid Eulerian method, has some benefits and drawbacks.  Firstly, it is very memory efficient, not requiring the large number of flux arrays of the unsplit methods.  Secondly, owing to additional numerical dissipation in the remapping step, the order of accuracy is diminished a little.  In 2D advection tests, Fyris achieves a resolution convergence order of 2.4, (like VH1 and \emph{ppmlr}), rather than the full 3rd order seen in an Eulerian PPM method.  The greatly reduced memory requirements however tend to balance out this shortfall in 3D models, where larger grids are feasible with the split semi--Lagrangian method.  The algorithm simplicity of the split semi--Lagrangian method also has performance benefits, and overall the \emph{Fyris} code performs very well in the Liska \& Wendroff tests \citep{liska03} as shown in later sections.   The intrinsic asymmetry of the split sweep method can be overcome with more complex symmetric forms of splitting that are used when absolute symmetry preservation are needed.  Note that the \emph{ZLS} framework is not tied to a Lagrangian method, and is used in the development of Eulerian and unsplit MHD algorithms as well.  

\subsection {Riemann Solvers}

PPM, VH1 and \emph{ppmlr} all used an approximate iterative Riemann solver (hereafter called \emph{CW84}).  From inspection of the code it appears to be based on the exact solver of van Leer, \citep{vanleerV}.  However it had been simplified in logic to assume that both branches of the Riemann problem were shocks, which is exact in the most critical cases, and conservative in general.  

Note that despite the approximate solver form, the resulting multi--cell Riemann grid calculations produce results that converge on the exact global result.  The benefit is that by removing the Boolean branch logic code from the iteration loop, significant speed improvements were obtained.
 
\Fyris\ has been designed to accept a range of Riemann solvers (each identified by a short \emph{codename}), and all the solvers 
of a given system of equations (hydrodynamics here) use a standard interface, making interchange a simple matter.   In addition to \emph{CW84}, we have the solver from \cite{toro99}: \emph{TORO99}, which is exact, albeit relatively slow.  We also have: the Toro hybrid solver, \emph{AIRS}, a two shock only version of the  \emph{TORO99} iterator, \emph{TSS99}, and a non-iterative primitive variable approximate solver, \emph{PVRS}, also from \cite{toro99}.
Apart from the very fast  \emph{PVRS} solver, the solvers from \cite{toro99} are generally slower than others, but are retained for cross-checking and teaching purposes (they are well documented and clear).

The main Riemann solver used by \Fyris\ is based on the solver of \cite{gg88}.  This solver was published there as part of a review of exact solvers after development of approximate solvers of the Roe--Pike variety began to dominate, and as such is probably the last fast exact solver developed.  It differs from the van Leer type,  in that it iterates on the velocity of the star region, rather than the pressure.  This simplifies some of the transcendental functions required, increasing the speed, with a minor degree of complexity added due to the possibility 
of zero and differing signs in the velocity compared to the positive definite nature of the pressure.  \Fyris\ can use the exact solver of Gottlieb \& Groth, \emph{GG88}, or more commonly, a two-shock version, where we followed the Colella \& Woodward example in simplifying the van Leer solver, and applied that to the  \emph{GG88} solver.  The resulting solver, \emph{RSS06}, is an iterative two shock solver that iterates on $u_*$, and is on average about twice as fast as the \emph{CW84} solver, but returns essentially identical results, making it an ideal substitute. 

It has been shown, for example in \cite{cw84b}, that two shock iterative solvers are very stable in the presence of extremely strong shocks, and in astrophysical applications this is an essential property, and as seen in the strong shock tests in subsequent sections, \emph{RSS06} has proven to be robust and accurate as well as being very fast.

With all iterative solvers, reducing the number of iterations is critical to performance. To this end
 \Fyris\ allows for a range of initial guess options for the solver.  In many cases where most of the grid is moving uniformly supersonically, an initial guess for $u_*$ of a simple average of the left and right values works very well.  In cases where initial velocities are mostly zero initially a more complex guess based on a single call to the\emph{PVRS} solver works better. If there are concerns (such as when very strong rarefaction waves are present) a simulation can be run with two different solvers as a test, usually using \emph{RSS06} or \emph{CW84}.  When \emph{GG88} or \emph{TORO99} are used, dissipation settings generally need some adjustment otherwise stronger post-shock ringing effects increase.
 
\subsubsection*{The \emph{RSS06} Riemann Solver Description}

The solution of the Riemann problem reduces to determining the conditions in the so-called star region, between a set of outwardly propagating waves.  These waves are usually either shock waves or rarefaction waves.  Less commonly a vacuum solution can also arise.

With a full non-linear Riemann solver there are different solution branches, depending on the wave type that is expected to form from the interaction of interest.  The two important wave types are a shock or a rarefaction fan.  An exact solver requires internal logic to select the branch of interest and the iterations to solve the problem differ in these two cases. An approximate two shock solver presumes that the two waves on either side of the star region are always shockwaves, enabling some complex logic calculations to be omitted.

The solution in the shock case is iterative, while there is an analytical answer to the rarefaction case.  Use of the non-iterative rarefaction case has been used \citep{toro99} to develop fast non-iterative solvers.  However, the shock case, with iterations,  is used here because of the numerical robustness of the two shock solver in the presence of very strong shocks.

\subsubsection*{Riemann Shock Case}

The  \emph{GG88} solver, of  \citep{gg88} and the \emph{RSS06} solver treat the shock case as follows.

For all cases, the velocity in the star region, $u^*$, a Newton iteration
\begin{equation}
u^*_{i+1} = u^*_i - \frac{P^* (u^*_i) - P^* (u^*_i)}{P^{*\prime} (u^*_i) - P^{*\prime} (u^*_i)}\, ,
\end{equation}
is performed until $ | 1 - P^*_l/P^*_r| < \varepsilon$, where $\varepsilon $ is small, usually $<<10^{-8}$, converging on a uniform pressure across the discontinuity using the following.

When the wave in the direction considered, to the left, (subscript $_l$),  is a shock, the solver evaluates the following pressure functions and derivatives:
\begin{eqnarray}
Z_l & = & \beta_l (u^* - u_l) - \{ 1 + \left[ \beta_l (u^* - u_l)\right]^2 \}^{1/2} \, , \\
P^*_l & = & p_l + c _l( u^* - u_l) Z_l\, .\\
P^{*\prime}_l& = &\frac{ 2 c_l Z^3_l}{1 + Z^2_l}\, .
\end{eqnarray}
where, 
\begin{equation}
a_l  = (\gamma_l P_l/\rho_l)^{1/2}\, , \; c_l  =  a_l \rho_l \, ,  \; \beta_l  =  \frac{\gamma_l +1}{4 a_l} \, .
\end{equation}

When the wave direction is to the right, (subscript $_r$), the $Z_r$ expression has one sign change:
\begin{eqnarray}
Z_r & = & \beta_r (u^* - u_r) + \{ 1 + \left[ \beta_r (u^* - u_r)\right]^2 \}^{1/2} \, , \\
P^*_r& = & p_l + c _l( u^* - u_r) Z_r\, .\\
P^{*\prime}_r& = &\frac{ 2 c_r Z^3_r}{1 + Z^2_r}\, ,
\end{eqnarray}
where, 
\begin{equation}
a_r  = (\gamma_r P_r/\rho_r)^{1/2}\, , \; c_r  =  a_r \rho_r \, ,  \; \beta_r  =  \frac{\gamma_r +1}{4 a_r} \, .
\end{equation}

The sound speed in the star region is not used in the iteration process, and can be derived afterwards if required by,
\begin{equation}
a^*_l  =  a_l \left[ \frac{(\gamma+1)+(\gamma-1) P^*/P_l}{(\gamma+1)+(\gamma-1) P_l/ P^*}\right]^{1/2} \, ,
\end{equation}
and
\begin{equation}
a^*_r  =  a_r \left[ \frac{(\gamma+1)+(\gamma-1) P^*/P_r}{(\gamma+1)+(\gamma-1) P_r/ P^*}\right]^{1/2} \, .
\end{equation}

\subsubsection*{Riemann Rarefaction Case}

Although not used in \emph{RSS06}, the rarefaction case is given here for completeness.  It is used in the full \emph{GG88} solver.

When the wave to the left is a rarefaction wave, the pressures and derivatives are as follows:
\begin{eqnarray}
a^*_l & = & a_l - \eta_l(u^* - u_l)\, ,\\
P^*_l & = & P_l \left[ \frac{a^*_l}{a_l}\right] ^{\gamma_l/\eta_l} \, ,\\
P^{*\prime}& = & -\gamma_l P^*_l/a^*_l\, .
\end{eqnarray}
where, 
\begin{equation}
\eta_l  = \frac{\gamma_l  - 1}{2}\, .
\end{equation}
In the right direction, minor sign differences apply:
\begin{eqnarray}
a^*_r & = & a_r + \eta_r(u^* - u_r)\, ,\\
P^*_r & = & P_r \left[ \frac{a^*_r}{a_r}\right] ^{\gamma_r/\eta_r} \, ,\\
P^{*\prime}& = & \gamma_r P^*_r/a^*_r\, .
\end{eqnarray}
where, 
\begin{equation}
\eta_r  = \frac{\gamma_r  - 1}{2}\, .
\end{equation}

The sound speed in the star region is derived during the iterations in this case.

\subsubsection*{Performance}

The performance of a range of non-linear Riemann solvers, exact and approximate was tested in a two--dimensional test problem (Test 4 from LW03).  The only difference in each test was the substitution of the varying Reimann solvers.  Timing was measured with on-chip hardware support and was able to isolate time spent in specific routines without impacting on code performance.  A comparison run with the FORTRAN \emph{ppmlr} (derived from VH1 \cite{christian97} and \emph{ppmlr}, \cite{suth03a}, Sutherland et al. 2003b) code is shown as well, although the code architecture and structures are different.  The \emph{ppmlr} code uses the \emph{CW84} two shock solver of \cite{cw84, cw84b}.   The results are shown in table \ref{t:performance}.

Overall the \emph{RSS06} solver outperforms all but the simplest non-iterative \emph{PVRS} solver, which is not designed to handle shockwaves.  Subsequent tests will also show that the  \emph{RSS06} solver produces accurate results as good or better than existing codes.

\begin{table}[htdp]
\label{tab:speed}
{\small 
\caption{Performance, 2D Riemann Test 04, to $t = 0.05$, single cpu, optimised PPC 970 (G5) \label{t:performance} }
\begin{tabular}{l r r r r r r l}
                                      & \multicolumn{1}{c}{Total}    & \multicolumn{1}{c}{Riemann}  &\multicolumn{1}{c}{Fraction of}   &                                      & \multicolumn{1}{c}{Per}         &                                       &\\
 \multicolumn{1}{c}{Solver}& \multicolumn{1}{c}{Time (s)}& \multicolumn{1}{c}{Solver (s)}&\multicolumn{1}{c}{Runtime  (\%)} & \multicolumn{1}{c}{Steps}& \multicolumn{1}{c}{Step (ms)}& \multicolumn{1}{c}{Cells/s}&Notes\\
\hline
\hline
PVRS&10.91&0.46&4.2&55&8.33&8.06E+05&non-iterative, approximate\\
RSS06&11.25&0.78&6.9&55&14.1&7.82E+05&2-Shock iterative\\
CW84&13.52&3.08&22.8&55&56.0&6.50E+05&2-Shock iterative\\
GG88&13.65&3.22&23.6&55&58.6&6.45E+05&Exact, iterative\\
AIRS&15.24&4.72&31.0&55&85.9&5.77E+05&Adaptive: PVRS/TORO\\
TORO&19.88&9.24&46.5&55&168.1&4.43E+05&Exact, iterative\\
\hline
ppmlr&18.88&2.76&14.6&46&59.9&3.94E+05&uses CW84, based on VH1\\
\hline
\end{tabular}
} 
\end{table}

\subsection {Dissipation and the Carbuncle Instability}

The principle method of introducing dissipation, beyond that introduced by the approximate two-shock solvers, is through reducing the order of the interpolation scheme on a cell--by--cell basis.   This takes the form of flattening the parabolae, as described in \cite{cw84}, with the slightly more dissipative constants here :$\omega_1 = 0.5, \omega_2 = 10.0$, and $\nu = 0.3$,  reducing to a first order Godunov method in the vicinity of strong shocks, typically the mean interpolation slopes are flattened to zero in three cells accross a shock, and the flattening is reduced to nothing within another cell, depending on the exact form of the shock.   This degenerates the method to a first order Godunov at shock fronts.  A global mininum flattening can be set to introduce a grid wide dissipation for special cases, for example see the Rayleigh Taylor Instability section \ref{s:RTI}.

To combat both `ringing' phenomena as well as the multi-dimensional striping or Carbuncle de--coupling  we use a multi-dimensional criteria for the parabola flattening, which is then fed into the split sweeps.  This introduces dissipation orthogonal to strong shocks where needed.  Following \cite{suth03b}, we use a spatial pattern detection of the onset of the striping in flattened zones and provide additional dissipation in only those cells, essentially eliminating the instability.   In cells where the fitting parabolae have been flattened in an earlier orthogonal sweep, we detect the onset of the striping in the pressure variable by looking for a sequence of three or more consecutive vertices in the pressure.  If found, the remapping process is used to perform a conservative smoothing of the physical variables in the region of the striping, orthogonally to the stripes.   Improving on  \cite{suth03b}, we can limit the smoothing to a subset of the pressure , velocity and/or density in order to perform the minimum of smoothing.    We find that smoothing just the velocity variable is sufficient in most cases to dampen the instability.

This requires adjustment of the smoothing parameters for a given simulation, but generally the required smoothing coefficients are not a sensitive function of shock mach number, and general settings work for a wide range of circumstances.  In some extreme cases, such as a very thin wall double shock structure extending only a few cells, propagating across the grid, we have needed to increase the dissipation across the shocks on a case by case basis.  We test the striping control in sections \ref{s:striping} and \ref{s:stripingii}, and give values for the smoothing coefficients there.

As the code operates with an approximate solver, followed by a Lagrangian step with a subsequent remap, the effect of one single Riemann problem can extend over up to three cells in a single cycle.  We have found that a TVD criterion for flattening the interpolation parabolae and limiting the fluxes in an Eulerian method do not translate to the numerics of the semi-Lagrangian case,  so we are left with a less satisfactory {\em ad-hoc} dissipation scheme.  However we feel that the extremely good shock capturing and shear-layer resolving power of the semi-Lagrangian method outweighs this consideration.

\subsection {Equation of State}

All else being equal,  \Fyris\ is at least twice as (and up to 4--8 times in some cases) fast as the earlier \emph{ppmlr} code in 2D.  This has allowed the incorporation of more calculations allowing for a more general equation of state.  An advantage of the \emph{GG88} and \emph{RSS06} solvers is that they are formulated in a way that allows the adiabatic index to be different on each side of the  boundary.  \emph{ppmlr} allowed for a single global mean molecular weight $\mu$ and adiabatic index $\gamma$.  So a given simulation was restricted to a single state: ionised ($\mu =0.6224$,  $\gamma = 5/3$), neutral  ($\mu =1.30$,  $\gamma = 5/3$), or even molecular ($\mu =2.4$,  $\gamma = 7/5$).   By allowing $\mu$ to vary and using a caloric equation to give temperature 
$T = EOS_{\rm C}( \rho, P, \mu)$, and allowing $\gamma$ to vary and using a baric equations to relate pressure $P$, internal energy $e$ and $\gamma$, i.e., $P = EOS_{\rm B}( \rho, e, \gamma)$, we pave the way to multi-phase models.

 \Fyris\ has a range of options, from a single $\mu$ and $\gamma$ as before, with
 fixed relationships between $(P, \rho, e, T, \mu, \gamma)$. Or to allow $\mu$ and $\gamma$ to be simple functions linked to the cooling function table.  Tabulated as a function of temperature, internal tables are computed to allow $\mu$ and $\gamma$ to be determined as function of $T$, $P/\rho$ or $e$, and so can be varied cell by cell.  In the setup for each Riemann problem the effective $\gamma$ is determined for the averaged left and right states.
 
The Riemann solvers assume constant left and right $\gamma$ values, and when needed the $\gamma$ values for the resulting left and right Star regions are computed.  Variable $\mu$ primarily matters for temperature determination, and is needed for cooling and ionisation calculations when in use.

Appendix A describes the calculation of and effective $\gamma$ for a cell comprising a mixture of ideal gasses, so that the total internal energy is conserved.

\subsection {Microphysics - Cooling and Ionisation}

\Fyris\ generally computes cooling from a cooling function pre-generated for the 
problem as a function of temperature.  When variable $\mu$ and $\gamma$ are in use then the 
table also includes ionisation fractions and the consistent values for $\mu$ and $\gamma$ at each temperature with the corresponding cooling.  Cooling is computed with either an explicit 4th order stable Runge--Kutta integrator, or as an implicit multi--step eulerian integration, both of which give essentially the same results, with the default Runge--Kutta integrator being marginally faster.  Cooling functions and equations of state variables $\gamma$ and $\mu$ are precomputed using the MAPPINGS~III code on a case by case basis.

For the purposes of verification, simple power--law cooling with a single $\gamma$ and $\mu$ are tested in section \ref{s:wallshock}.  Examples with more general astrophysical cooling functions will be presented in paper II.  When a single function of temperature is inadequate, we have implemented time--dependent ionisation implicit integration following the method of Rosen \& Smith \cite{rosen04}, which will also be presented in paper II.

\subsection {Gravity}
Static gravity fields, generalised to an arbitrary grid potential, are used in a similar way to \emph{ppmlr}, as a source term in the Riemann problem.  
Unlike  \emph{ppmlr} and others where potential functions are evaluated continuously to provide grid terms on the fly, \Fyris\
pre-computes a grid potential.  In addition it pre-constructs a set of PPM (3rd order on a regular grid) conservative interpolation coefficients, so the potential differences between any two points in the grid space can be reconstructed using the same PPM method that the fluid dynamics interpolation scheme uses.  This involves pre--computing the potential field 27 times per cell (corners, cell edge centres, face centres and cell centre) once at the start of a simulation and storing the coefficients, but thereafter gravity interpolation is very fast and has similar accuracy and numerical properties as the rest of the calculations.  See Appendix B for details of the PPM interpolation coefficients.

Since the gravity field grid can be arbitrary, it is planned to add a self--gravity calculation step
in the future, and use the grid structures already in use for fixed--gravity calculations.

\subsection {Tracers}
\Fyris\ is built with the eventual aim of performing more complex chemical enrichment and mixing calculations.  The tracer variables have been made more flexible and general than in the earlier \emph{ppmlr} code, and arbitrary number of tracers can be added to 
 a calculation with a simple compile time option.  Ionisation networks also make use of the tracer variables to track ionisation species.

\section{Optimisations and Performance}

\subsection {Active Floating Point Noise Control}

Additional code in the grid coordinates and in updating physical variables has been added to control the introduction of numerical `grid--noise' into calculations.  When used, a 1D supersonic flow and shock can propagate indefinitely across a multi-dimensional level of multiple zones an remain perfectly 1D.  Without careful control of truncation errors in coordinates, within a grid and between zones, propagation of shocks vary enough to generate non-physical disturbances at the $10^{-12}$ relative levels (in pressure for example).  In many models this is immaterial as the noise doesn't grow or is unimportant.  In others - high mach number thin shocks with strong cooling for example, this is critically important as the noise can grow and even dominate.  This has often been the source of perturbations that seed the well known decoupling Carbuncle instability.

Following techniques from \cite{goldberg91} and the Sun Microsystems Numerical Computation Guide \\
( {\small \tt  http://docs.sun.com/source/806-3568/} ), we define an epsilon 16 times (4 bits) larger than the precision limit ( DBL\_EPSILON$*16.0$ in the 64 bit case), which  is used to round off coordinate calculations in a controlled predictable way (such as when pre-generating cell divisions on a grid). In the Lagrangian case the original axis coordinates of the cells is generated using controlled roundoff at the start in to arrays that are reused throughout the code.

When updating grid variables after a sweep calculations, only changes to variables that exceed this threshold are allowed.  Changes less than this amount are deemed to be non--physically small and probably due to numerical noise and are discarded - preventing them from accumulating.

The calculations still retain a very large dynamic range and good precision, and by applying controlled roundoff and non-physical threshold limits, numerical grid noise is essentially eliminated with little if any performance impact.    In addition, when large parts of the grid contain constant values, they retain those constant values indefinitely, and exact floating point comparisons can be safely made to determine if the velocity in a region is equal zero for example.  This allows some optimisations to be made that fail of cells accumulate random numerical noise over time.

\subsection {Pre-flighting sweeps}

With any numerically intensive task great performance improvements can be made if work can be avoided, and calculations bypassed.
\Fyris\ uses a number of pre-flight tests to see if the full calculations need to be done before they are attempted.  With well formed noise free data these tests can be quick enough to not slow the code much even in a worst case.

First, when preparing the interpolation parabolae, the variable being fitted is tested as follows.
First, all the coefficients are set to zero slope, zero parabolic terms.  Then the cells are compared with the first cell, and if any cell is not equal in floating point value to the first cell then the row calculation is performed in full.  If all the cells are equal (it the test never fails) then the coefficients are ready to go, being the right values for a perfectly flat field.  

If numerical noise were not controlled, then this test will almost always fail immediately and the full fitting performed as usual.  If there are any physically real variations likewise the full interpolation is performed as usual.  With floating point noise control, dormant regions of a flat grid stay exactly flat, and this optimisation can speed up the early phases of a calculation a lot, as the parabolic fitting is a significant fraction of the time spent.

The second main pre-flight test is after the Riemann solver has computed the outcome of a row
of Riemann problems.  If all the Riemann solution velocities are exactly zero then nothing can happen, so the evolution and remapping steps are bypassed, and grid values are left untouched.
Looking at the possible outcomes of the Riemann solver, only three cases give $u_* = 0.0$  The trivial case --where the left and right state are the same, and velocity is zero initially -- and the double shock and double rarefaction cases. In both of these latter cases however velocity is always non-zero in adjacent cells.  Only in the trivial case is velocity zero everywhere before and after the Riemann Solver, and hence can be safely skipped, saving evolution, remapping and memory access time.

Again, like the interpolation pre-flight, as soon as the first cell with a non-zero velocity the preflight is aborted and the full calculation proceeds.

With these two optimisations a very large uniform initial grid can be set up with for example a small bubble growing  in the centre.  The bulk of the grid is skipped by the pre-flights and only becomes active as the simulated physical disturbances propagate.   The early stages of the simulation can be run very quickly, getting to the computationally intensive later stages more efficiently.

Finally, the Riemann solvers themselves can be designed to pre-test for trivial cases (i.e. pressure, density and velocity constant from left to right) and reduce iterations and the generation of numerical noise.

\section{ZLS}
A simulation management framework, the Zone--Level System, ZLS, has been built to provide a uniform and efficient means of running a range of computational fluid dynamics simulations. ZLS manages the spatial and temporal coordinates, provides a range of input and output facilities, coordinates simulation events, and manages distributed processing, both preemptive threads and MPI parallel computation.  

The simulation domain is divided into a series of zones, each of which is a self-contained computational element that contains sufficient local information to be computed independently of 
any other zones.  This independence is the basis for parallel processing of all kinds.  Each zone contains spatial coordinates (a grid), as well as timing information (clocks), and a set of simulation variables plus auxiliary data as required by the type of fluid in question.  A set of zones at a given spatial resolution form a level, and the zones may be linked together to form a larger domain.  All the zones clocks and grids in a level are coordinated by a master level grid and clock.

A simulation can comprise of 1 or more levels, with the lowest resolution level denoted as level 0, 
the next higher resolution level 1 and so on.  Despite the increasing numbering, conceptually a level $(n+1)$ is a sub--level of level $n$.  A number of algorithms for updating sub--levels and transferring the results to the level above are implemented, with a recursive method most common.   If level $(n+1)$ has $m$ times the spatial resolution, then it is updated $m$ times before level $n$ is updated.  This is kept consistent with having a single universal Courant--Friedrich--Lewey (CFL) constant for the simulation.  Typical CLF values are 0.5-0.8, and this limits the time-step so that the fastest wave in any cell--cell interaction can only travel CLF$\times dx$ in a time-step, ($dx$ is the cell width).

Computational load balancing requires that the number of cells in zones of high $n$ be limited to prevent simulations taking excessively long times, especially in 3D.  Generally sub-levels are placed in regions of interest that are much smaller than the main level 0 domain, otherwise a single level high resolution grid is needed.

A sub--level must be completely contained in the domain of its super-level, and the coordinates
are defined ultimately by level 0.  Each zone in a sub-level $(n+1)$is a sub-zone of a zone in the level $n$, and maintains that relationship throughout the simulation.  Zones, and entire levels can be activated and deactivated on the run, but to date this has seldom been used.

ZLS handles 1D, 2D and 3D code, a range of coordinate systems; Cartesian, Spherical and Cylindrical (plus lower dimensional versions of these).  

ZLS also provides a uniform set of math and numerical representations, for example: `'Real' (floating point)
`Integer' (singed integer) `Counter' (unsigned integer), in a way that is independent of the architecture in use (32 bit, 64 bit, IEEE big--endian, intel little-endian etc)
so that the code and any embedded codes are automatically portable.

ZLS code usually normalises the simulation variables internally to optimise the numerical dynamic range. Using three physical quantities (4 for MHD, to include B) with a range of dimensions covering the basic physical units of time, mass, and length, normalisation constants can be derived.  ZLS uses density ($\rho_0$), length ($x_0$), and velocity ($v_0$) as fiducial quantities for typical simulations.

\section{Test Problems and Results}

\Fyris\ is now used in a suite of test problems, many from LW03, plus some additional tests, both new and from the literature.  Where possible results are compared to analytical results, and when that is not possible, with re--binned results from very high resolution tests, such as in the blast-wave interaction test.  By re--binning a high resolution result, correctly weighted cell averages are obtained, as opposed to simple cell centre values, can be generated for comparison with low resolution tests and generation of meaningful L1 norms.  Some of the LW03 tests are analysed in more detail, in particular the Rayleigh-Taylor Instability and the multi-dimensional Noh test.

For consistency, the following tests which are shown as 2D images, the same colour table is used throughout.  It is a fairly standard rainbow sequence, with additional 20 percent darker colours at each of 25 intervals, i.e. 2 dark values every 10 indices.   This gives the effect of contouring the images, and the contour lines cover approximately one percent of the range each ($2/256$). 
With this table, otherwise smooth appearing regions may show contours that reflect subtle structure and reveal defects not otherwise apparent.

The lowest index, 0, is black and the highest index, 255, is white.  When auto scaling, the maximum index used is pure red at a value of 254.  When fixed scales are used, the last index is also used, and white then appears in a way that shows regions of clipping, equal or greater than the maximum.  Numerical values for the colour table are included with the \Fyris\ code.

Throughout, the spatial resolution is referred to as either $n$, the total number of cells in the relevant dimensions over the entire domain, or $h$ the number of cells per unit spatial unit when the domain is not a single unit in extent.

\subsection{The Liska and Wenndroff Tests}

First, we present the suite of problems reviewed by  \cite{liska03}, LW03.  The reader is encouraged to obtain a copy of LW03 and to make comparisons of their results for other codes, with the results here.

\subsubsection{1D Riemann Tests}

The 1D LW03 tests consist of eight single Riemann problems with known solutions, plus an interacting two shock test with no analytical solution, but which may be used to compare low and high resolution results.  Note, the final time for the Blast problem was omitted by LW03, but has been determined from previous published results and is presented here.

The eight one dimensional adiabatic tests  are specified as follows:
{\small 
\begin{itemize}
\item Adiabatic index: $\gamma = 1.4$.
\item Grid Domain: $0.0 < x < 1.0$.
\item The left and right boundaries are free, or natural, boundaries that replicate the adjacent grid values.
\item All cells with central coordinates $< x_0$ are set to the Left State, all cells with central coordinates $> x_0$ are set to the Right State ($x_0$ lies on an exact cell boundary). 
\item The tests are run from time $t = 0.0$ until $t = T$, defined for each problem. 
\end{itemize}
}

These parameters are given for each of the problems in table \ref{t:riemann1d}.
\begin{table}[thdp]
{\small 
\caption{1D Single Riemann Problems. \label{t:riemann1d}}
\begin{tabular}{l l l l l l l l l l}
\hline
&\multicolumn{3}{c}{Left State}&\multicolumn{3}{c}{Right State}&&&\\
Test&Density&Pressure&Velocity&Density&Pressure&Velocity&$x_0$&$T$& $n$ (cells)\\
\hline
\hline
1&1.0000&1.0000&0.7500&0.1250&0.1000&0.0000&0.3000&0.2000&100\\
2&1.0000&0.4000&-2.0000&1.0000&0.4000&2.0000&0.5000&0.1500&100\\
3a&1.0000&1000.0000&-19.59745&1.0000&0.0100&-19.59745&0.8000&0.0120&200\\
4&5.9992&460.8940&19.5975&5.9924&46.0950&-6.1963&0.4000&0.0350&200\\
5&1.4000&1.0000&0.0000&1.0000&1.0000&0.0000&0.5000&2.0000&100\\
6&1.4000&1.0000&0.1000&1.0000&1.0000&0.1000&0.5000&2.0000&100\\
\hline
Noh&1.0000&1.0e-6&1.0000&1.0000&1.0e-6&-1.0000&0.5000&1.000&100\\
Peak&0.1261192&782.92899&8.9047029&6.591493&3.1544874&2.2654207&0.5000&0.0039&800\\
\hline
\end{tabular}
} 
\end{table}

The 1D Noh Test and the 1D Peak test differ slightly from the other six tests in the following ways:
{\small 
\begin{itemize}
\item The Noh test uses Adiabatic index: $\gamma = 5/3$.
\item The Peak test uses a grid domain: $0.1 < x < 0.6$.
\end{itemize}}
The 1D Noh test has an analytical solution given in table \ref{t:noh1dsol}.

\begin{table}[thdp]
{\small 
\caption{1D Noh Solution, $\gamma = 5/3$. \label{t:noh1dsol}}
\begin{tabular}{l l l l }
\hline
 \multicolumn{1}{c}{Inside Shock  ($r < t/3$)} &  \multicolumn{1}{c}{Outside Shock  ($r > t/3$)}  \\
\hline
\hline
Density $\rho_0= 4 $  &  Density $\rho_1 = 1.0$     \\
Pressure $P_0 = 4/3$ &  Pressure $P_1 = 0.0$             \\
Velocity $v_0 = 0$        &  Radial Velocity $v(r) = -1.0$    \\
 \multicolumn{2}{c}{Shock Front Expands at $v_s = 1/3$} \\
\hline
\end{tabular}
} 
\end{table}

The 1D Woodward-Collela blast wave test consists of three initial states, resulting in two interacting shockwaves, that cross and reflect from the grid boundaries.
{\small 
\begin{itemize}
\item Adiabatic index: $\gamma = 1.4$.
\item grid domain: $0.0 < x < 1.0$.
\item The left and right boundaries are reflecting boundaries.
\item All cells with central coordinates $< x_0$ are set to the Left State, all cells with central coordinates $> x_1$ are set to the Right State, the remainder are set to the Middle State ($x_0$ and $x_1$ lie on exact cell boundaries), values given in table \ref{t:peak1d}. 
\item The test is run from time $t = 0.0$ until $t = T$. 
\end{itemize}
}

In the absence of an exact solution, the 400 cell simulation is compared with a 2000 cell simulation, re--binned to 400 cells. 

\begin{table}[thdp]
{\small 
\caption{1D Peak Riemann Problem. \label{t:peak1d}}
\begin{tabular}{l l l l l}
\hline
Test&$x_0$&$x_1$&$T$&Cells\\
\hline
\hline
Blast  &0.1000&0.9000&0.038&400/2000\\
\hline
State&Density&Pressure&Velocity\\
\hline
Left     & 1.0000 & 1000.0 & 0.0000 &\\
Middle & 1.0000 & 0.0100 & 0.0000 &\\
Right   & 1.0000 & 100.00 & 0.0000 &\\
\hline
\end{tabular}
} 
\end{table}

\begin{figure}[thdp] \begin{center}
\includegraphics[width = 5in]{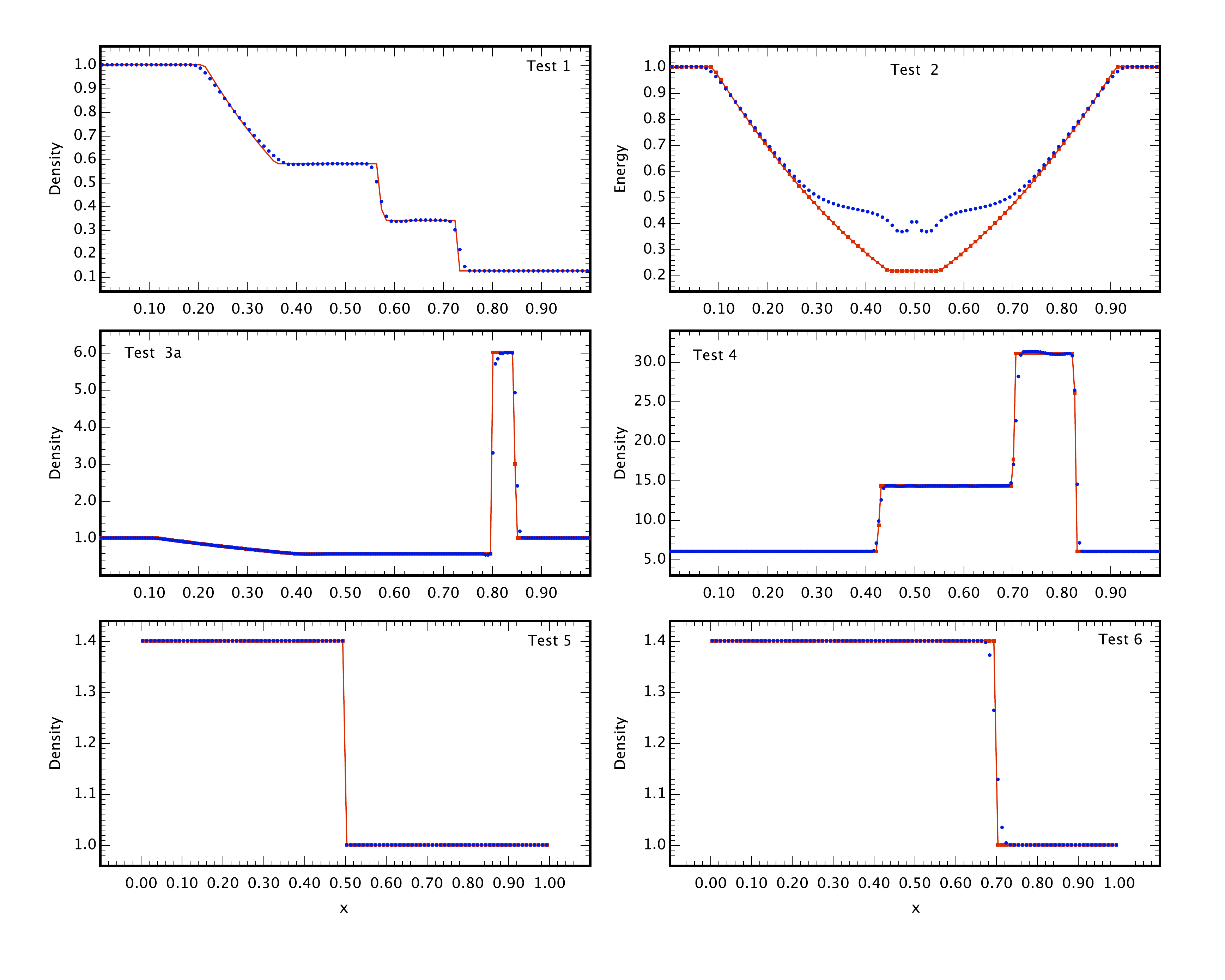}
\caption{\small 1D Riemann problems, tests 1 - 6.}
\label{f:rt1da}
 \end{center} \end{figure}

\begin{figure}[thdp] \begin{center}
\includegraphics[width = 5in]{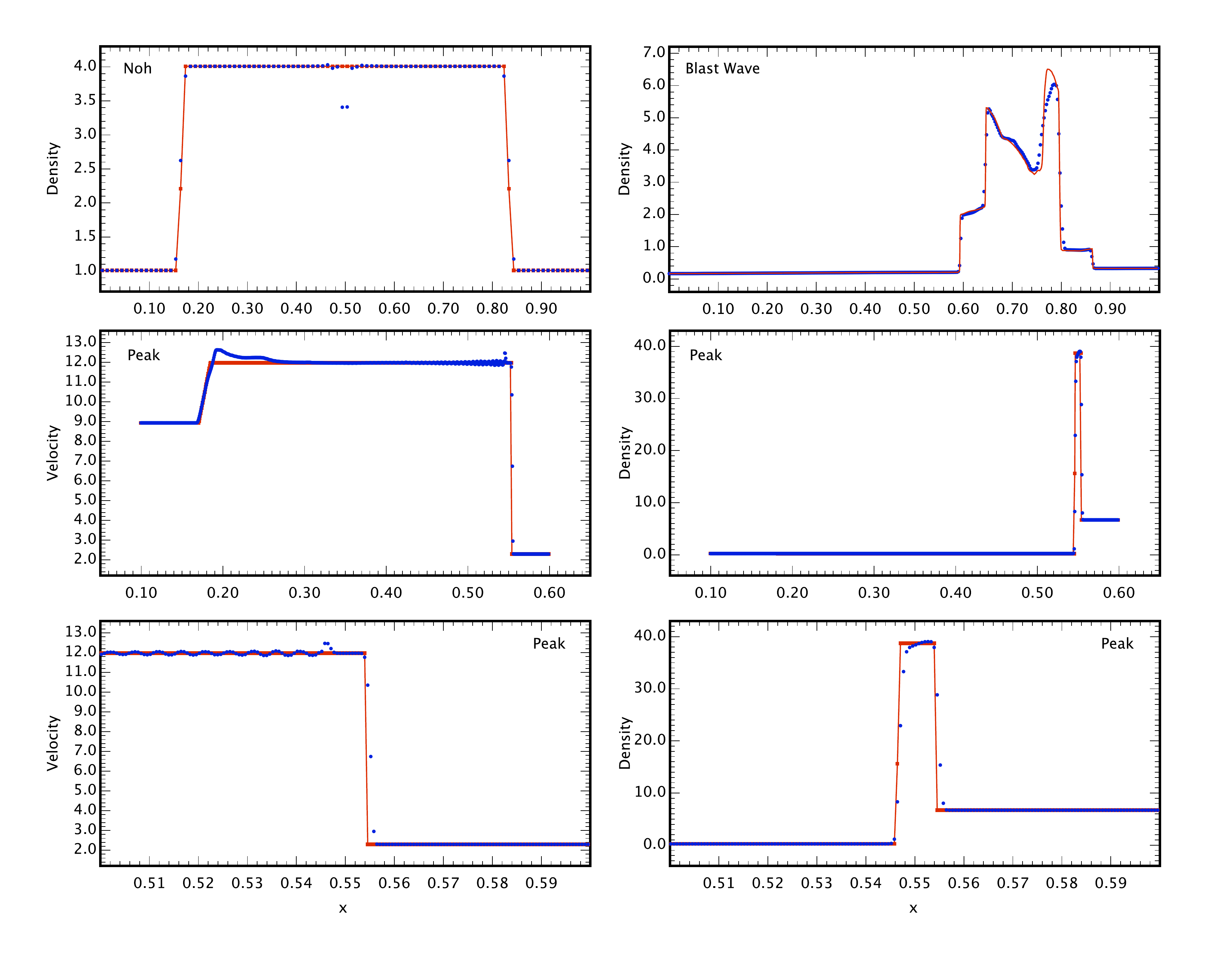}
\caption{\small 1D Riemann problems, Noh, Blast and Peak tests.}
\label{f:rt1db}
 \end{center} \end{figure}

The results are shown as plots (figure \ref{f:rt1da} and \ref{f:rt1db} ) and in the form of $L1$ norms, $\frac{1}{n} \Sigma | \frac{\Delta x}{x}|$, expressed as percentages, showing the deviations from the exact density solution, except for model 2, where the comparison is with the internal energy, $P/(\gamma -1) \rho$.  In computing the norms, the exact solution was integrated and averaged over each cell  to give a comparison at the same resolution as the test, to allow for the case where a discontinuity in the solution lands at a fractional cell coordinate.   Table \ref{t:1dtests} shows the \Fyris\ results compared with the subset of the LW03 codes most similar in method to \Fyris\ ( `--' indicates no result).

The best result for each test is indicated by a bold value.  \Fyris\ performs similarly to the VH1 code, to which it is related, except for the strong shock Noh problem where it performs significantly better than the other codes, which may be due to the new {\em RSS06} Riemann solver.

\begin{table}[thdp]
{\small 
\caption{1D Riemann Test Results, L1 Norm Errors, Percentages. \label{t:1dtests}}
\begin{tabular}{l l l l l l l l l l}
\hline
Code&1&2&3a&4&5&6&Noh&Peak&Blast\\
\hline
\hline
Fyris&1.0&9.8&3.6&1.3&0.0&0.3&{\bf 0.95}&{\bf 0.8}&5.3\\
CLAW&0.8 &--&{\bf3.1}&1.7&0.0&0.4&1.3& -- & --\\
WENO&1.3&23.7&9.2&2.2&0.0&0.4&2.0&2.4& --\\
PPM&{\bf 0.5}&{\bf 6.3}&9.4&{\bf 1.1}&0.0&{\bf 0.1}&4.6&1.3& --\\
VH1&0.9&9.6&3.7&1.3&0.0&0.3&1.5&{\bf 0.8}& --\\
\hline
\end{tabular}
} 
\end{table}

\clearpage 

\subsubsection{2D Advection Convergence}

This test sets up a diagonally travelling sinusoidal density wave, and advects it across the grid, with periodic boundaries, back to the initial position. The L1 density norm is used to measure the accuracy of the final conditions compared to the initial condition. The experiment is repeated by successively doubling the resolution of the grid, with the same domain, and the convergence of the result is estimated. Note: this tests the code's basic advection scheme, not the convergence of the non-linear solution to a Riemann problem.  Table \ref{t:twoconvsetup} gives the problem setup.

Compared to the codes tested by LW03, \Fyris\ performs better or as well, with the exception of the original PPM code, which is a truly 3rd order Eulerian method, and the WENO code which is formally 5th order for smooth flows. \Fyris\ compares well with VH1, and unlike CLAW, and most of the other central codes, it maintains it's convergence degree over all the resolution range.  In table \ref{t:twodconv}, bold entries for the order of convergence show the best of the codes apart from PPM and WENO.

\begin{table}[thdp]

{\small 
\caption{ Setup: 2D Advection Convergence. \label{t:twoconvsetup} }
\begin{tabular}{l l l l }
\hline
 \multicolumn{1}{c}{EOS} &  \multicolumn{1}{c}{Algorithm} &  \multicolumn{1}{c}{Grid} &  \multicolumn{1}{c}{Boundaries}  \\
\hline
\hline
Adiabatic              & CFL, $c = 0.8$       & $n = 400\times400$ cells & All: Periodic \\
 $\gamma = 7/5$ &  flattening: 0 -- 1    &  $-1.0 < x < 1.0 $       &  \\
                            &  $\alpha = 0.001$ &  $-1.0 < y < 1.0 $       &  \\
\hline
 \multicolumn{2}{c}{Initial Conditions}&  \multicolumn{1}{c}{Time Limit} \\
\hline
\hline
{ $\rho = 1.0 +0.2 \sin(\pi (x + y)) $    }  &{ $P = 1.0$    } & \multicolumn{1}{c}{$t = 2.5$} \\
{ $v_x = 1.0$    }& { $v_y = -0.5$    } &   \\
\hline
\end{tabular}
\label{t:advection}

} 
\end{table}

\begin{table}[thdp]
{\small 
\caption{2D Advection Convergence. \label{t:twodconv} }
\begin{tabular}{l l l l l l l l l l}
\hline
Code&L1, $n=25$&order&L1, $n=50$&order&L1, $n=100$&order&L1, $n=200$&&L1, $n=400$\\
\hline
\hline
{\bf Fyris}&4.748E-01& 2.202&1.032E-01&{\bf 2.391}&1.968E-02&{\bf 2.291}&4.022E-03&{\bf 2.348}&7.902E-04\\
CFLFh&2.5&2.293&5.10E-01&2.087&1.20E-01&1.953&3.10E-02&--&--\\
JT&1.1&2.258&2.30E-01&1.963&5.90E-02&1.883&1.60E-02&--&--\\
LL&2&1.322&8.00E-01&1.930&2.10E-01&1.959&5.40E-02&--&--\\
CLAW&4.10E-01&{\bf 2.490}&7.30E-02&2.190&1.60E-02&2.112&3.70E-03&--&--\\
WAFT&5.70E-01&0.074&6.00E-01&1.383&2.30E-01&1.779&6.70E-02&--&--\\
{\em WENO}&3.10E-02&{\em 4.998}&9.70E-04&{\em 4.968}&3.10E-05&{\em 4.576}&1.30E-06&--&--\\
{\em PPM}&2.40E-02&{\em 3.100}&2.80E-03&{\em 3.042}&3.40E-04&{\em 2.983}&4.30E-05&--&--\\
VH1&4.50E-01&2.214&9.70E-02&2.352&1.90E-02&2.284&3.90E-03&--&--\\
\hline
\end{tabular}
} 
\end{table}

\subsubsection{2D Riemann Problems}

A set of six two--dimensional Riemann problems are constructed on a unit square divided into four quadrants.  The spatial resolution for each model is $h = 400$. Each of the four quadrants are set with independent constant gas variables in order to set up boundaries which will generate a variety of shocks, rarefaction waves and, not present in one dimensional models, slip or shear discontinuities.  The problem numbering follows LW03 and are in turn based on the tests of \cite{schulz93a, schulz93b}.   In these tests, shocks, rarefaction waves are produced, and two-dimensional slip layers, which don't occur in one-dimensional problems, are all created.  Overall \Fyris\ performs the problems similarly to the other high-resolution codes, CLAW, PPM, and VH1.  The post-shock stability of \Fyris\ is shown in the smooth density and pressure contours, with only two areas showing ringing behaviour, in the $y-$velocity variable, at the $1\%$ level, in problems 12 and 15.  The LW03 paper does not show the velocity as an image and it is not possible to see if the other codes also produce $y-$velocity stripes in their results.

\paragraph*{ 2D Riemann Problem 3}  Generates a set of four shockwaves which evolve with a complex region where the four shocks meet.  Oblique shocks form between the lower--left region and the upper--left and lower--right regions.  The other two shocks show slight curvature.  The key features of the test are the presence of residual errors or  `glitches' which are formed in the initial steps and are preserved to late times by the high--resolution methods, such as PPM.   \Fyris\ performs similarly to the high-resolution codes, producing sharp shocks and well defined structure in the interaction region.  The colour table used here reveals the low level ($<1$\%) variations, or `glitches',  that evolve from the initial conditions in all the high resolution codes which are dissipated in the more diffusive central codes.

\begin{figure} [thdp] \begin{center}
\includegraphics[width = 5in]{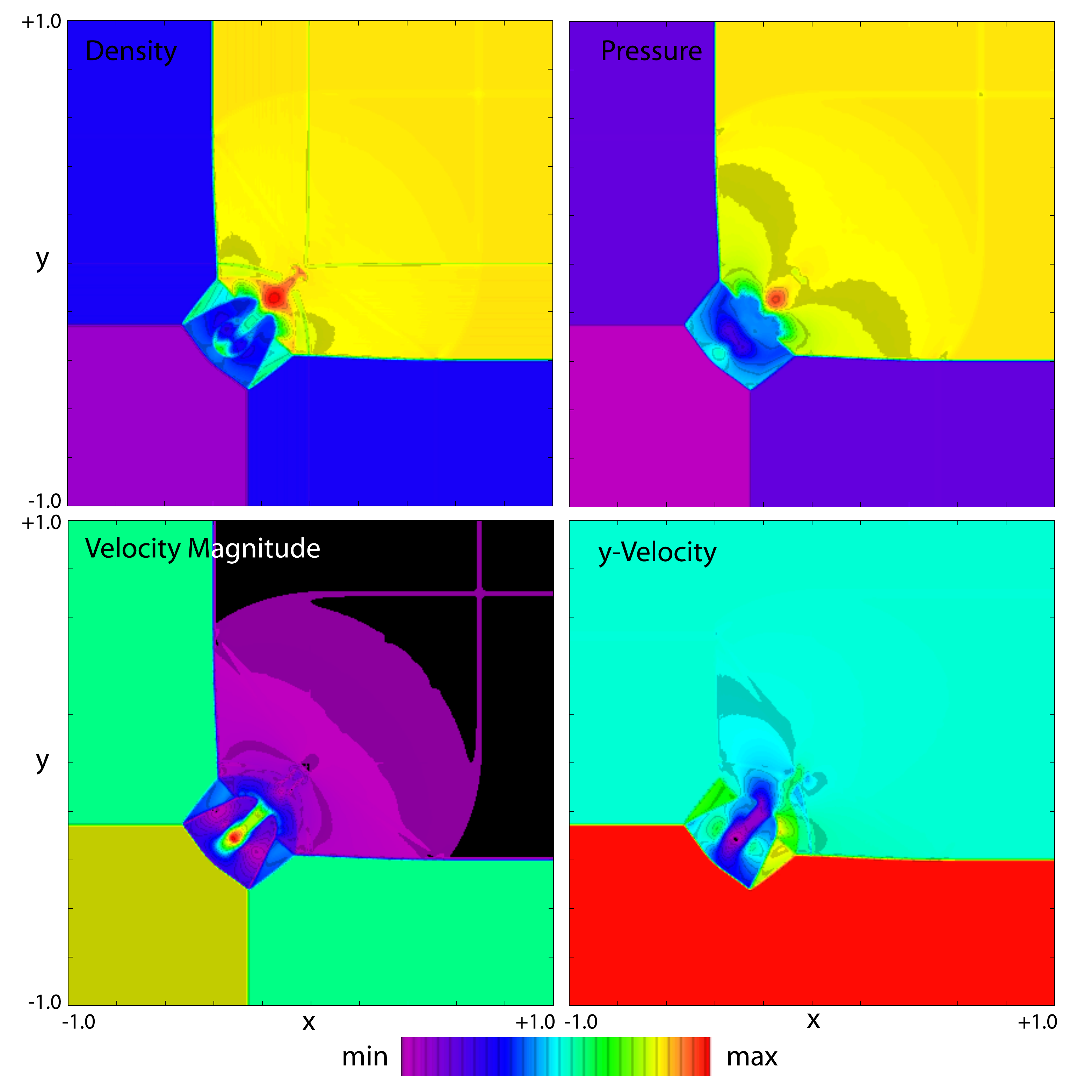}
\caption{\small 2D Riemann Problem 3. Density, Pressure, Speed, $y-$Velocity at $t = 0.3$.  Colour scale black to red, linear. Density 0.0--1.75, Pressure 0.0--1.75, Speed auto-scaled, $y-$Velocity auto-scaled. }
\label{f:rt03}
 \end{center} \end{figure} 

\paragraph*{ 2D Riemann Problem 4}  Generates also generates a set of four shocks, two of which are curved and surround a lens shaped region.
\Fyris\ again reproduces the shock structures well, with little or no sign of post--shock ringing that would affect the smoothness of the colour table contours.
\begin{figure} [thdp] \begin{center}
\includegraphics[width = 5in]{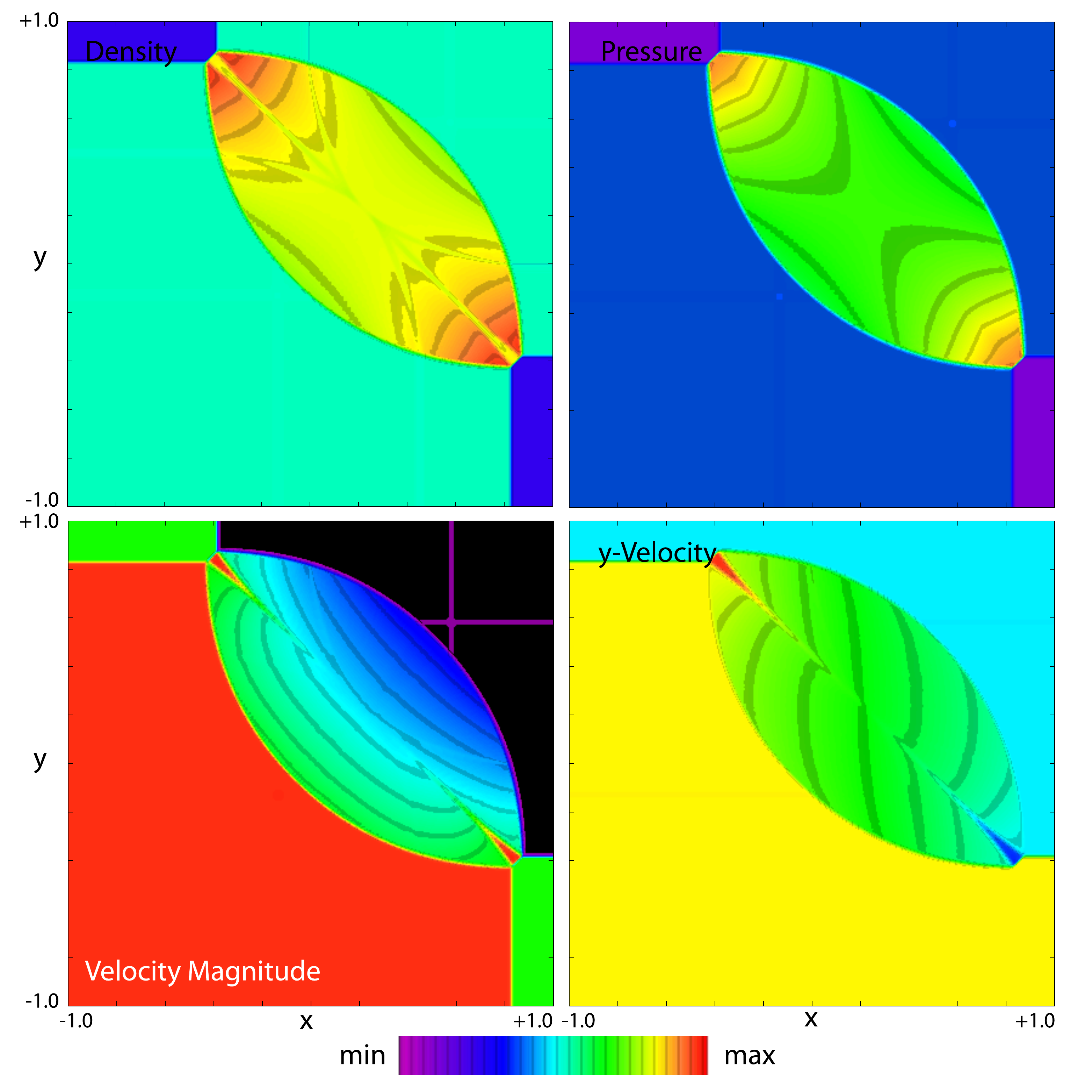}
\caption{\small 2D Riemann Problem 4. Density, Pressure, Speed, $y-$Velocity at $t = 0.25$.   Colour scale black to red, linear. Density 0.0--2.0, Pressure 0.0--2.75, Speed 0.0--1.3, $y-$Velocity -1.3--1.3 }
\label{f:rt04}
 \end{center} \end{figure} 

\paragraph*{ 2D Riemann Problem 6}  Generates a set of four spiralling slip layers.  Codes which are diffusive tend to smear out the spiral structure. \Fyris\ , being a semi-lagrangian method, maintains sharp slip layers right into the centre of the spiral, comparable to PPM and VH1 in the LW03 results.

\begin{figure} [thdp] \begin{center}
\includegraphics[width = 5in]{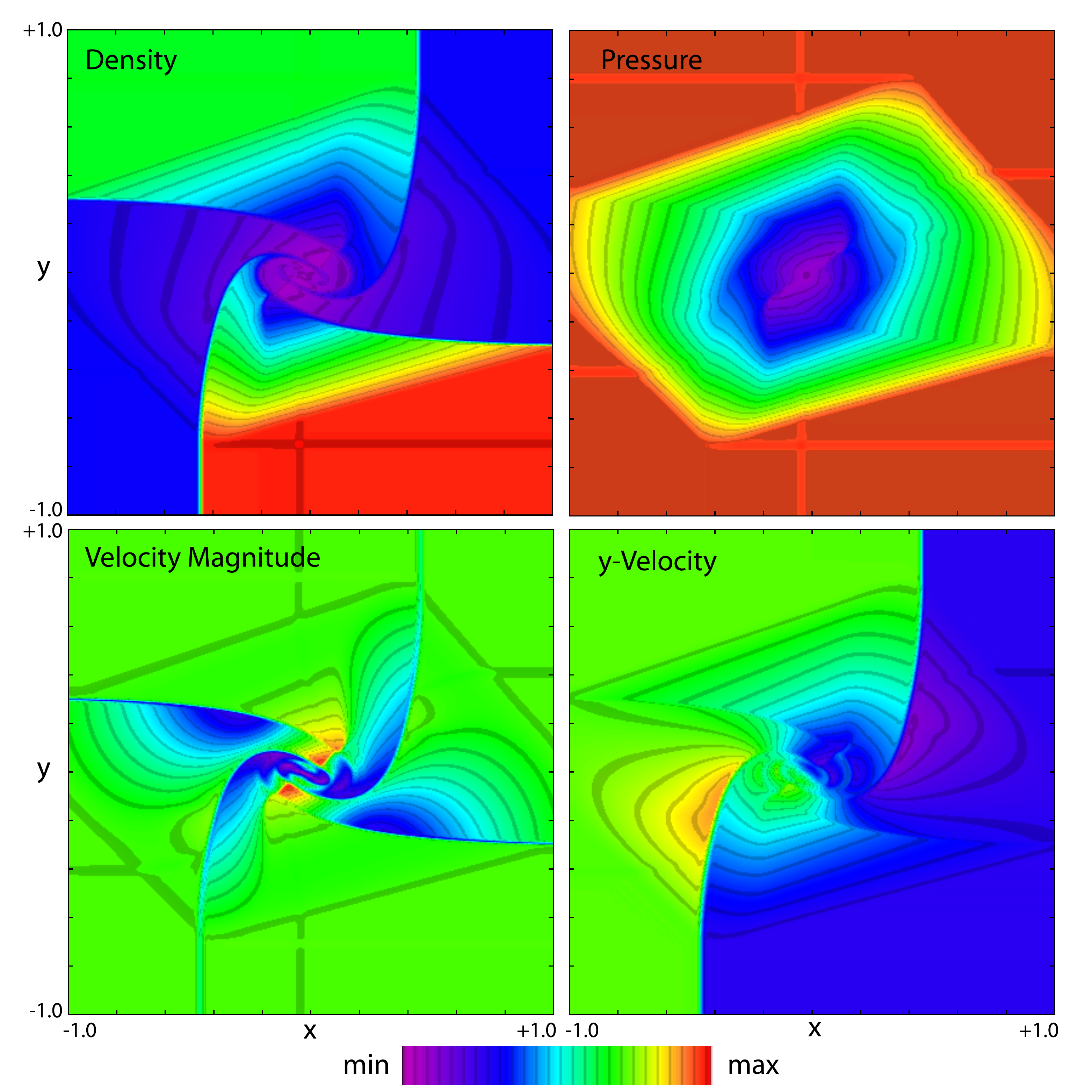}
\caption{\small 2D Riemann Problem 6. Density, Pressure, Speed, $y-$Velocity at $t = 0.3$.   Colour scale black to red, linear. Density 0.0--3.05, Pressure 0.0--1.05, Speed 0.0--1.25, $y-$Velocity -1.1--1.1 }
\label{f:rt06}
 \end{center} \end{figure} 

\paragraph*{ 2D Riemann Problem 12}  Generates a pair of stationary contact layers and a curved pair of shocks, which evolve a complex interaction region.   The stationary contact layers remain perfectly stationary and resolved over a single cell boundary.  The curved shocks are smooth and the post-shock contours  shown only slight initial glitches in the density contours, and some sign of very small pressure waves in the upper--right region in the pressure variable.

\begin{figure} [thdp] \begin{center}
\includegraphics[width = 5in]{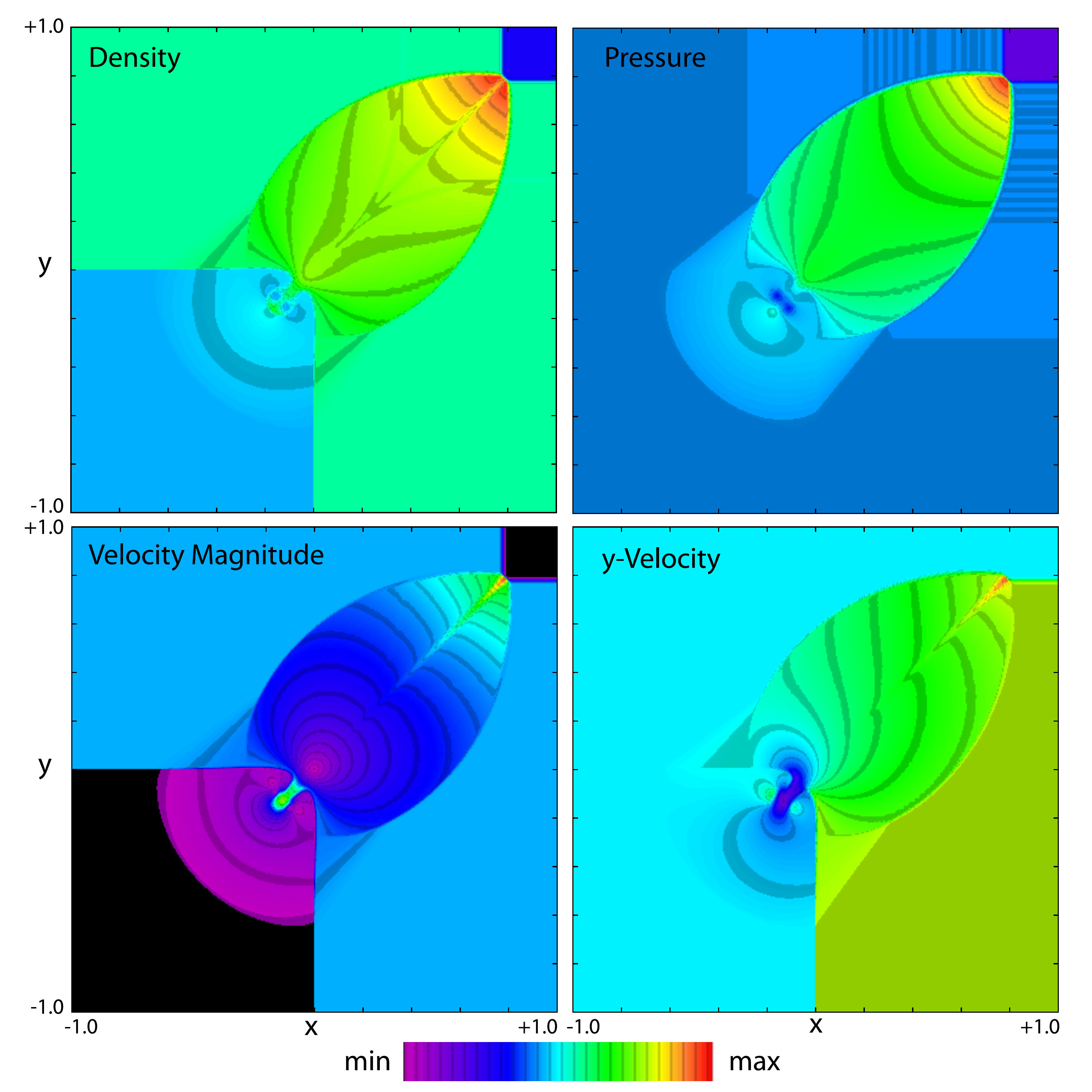}
\caption{\small 2D Riemann Problem 12. Density, Pressure, Speed, $y-$Velocity at $t = 0.25$.   Colour scale black to red, linear. Density 0.0--1.75, Pressure 0.0--2.30, Speed 0.0--1.60, $y-$Velocity -1.25--1.25 }\label{f:rt12}
 \end{center} \end{figure} 

\paragraph*{ 2D Riemann Problem 15}  Generates a pair of slip layers, a shock and a rarefaction wave.  Key features of the test are the resolution of the slowly moving slip layers, and the smoothness of the rarefaction region.  Like test 6, there are small artefacts in the density and pressure variables which propagate from the initial contact locations.  The artefacts are $<1\%$ and are preserved by the very low dissipation in \Fyris, similar to PPM, VH1 and CLAW codes in LW03.  There is a series of striped in the $y-$velocity variable on the right hand side of the upper--right region which may be a sign of post-shock ringing on the slowly downward moving shock. 
 
 \begin{figure} [thdp] \begin{center}
\includegraphics[width = 5in]{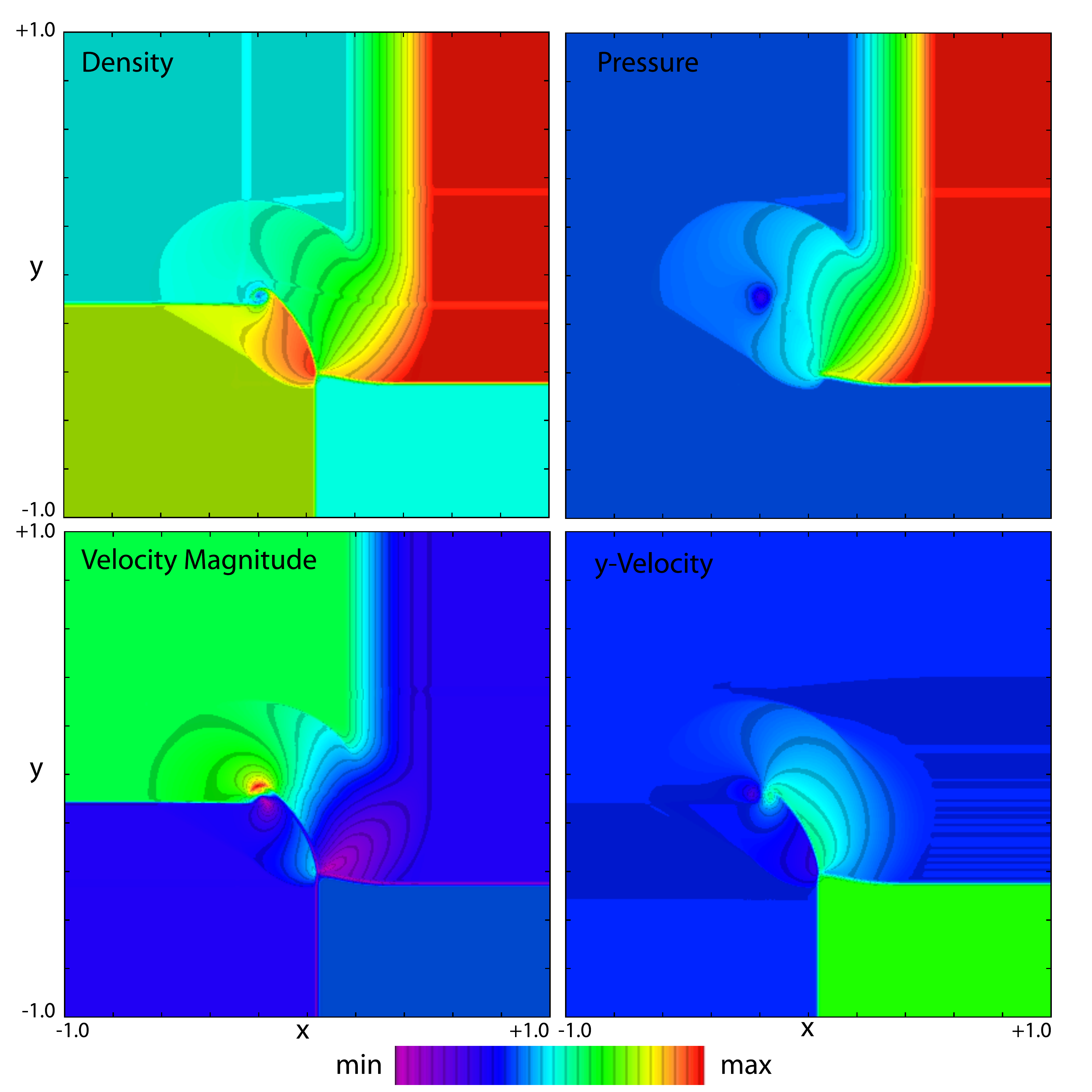}
\caption{\small 2D Riemann Problem 15. Density, Pressure, Speed, $y-$Velocity at $t = 0.2$.   Colour scale black to red, linear. Density 0.0--1.01, Pressure 0.0--1.01, Speed 0.0--1.1, $y-$Velocity -1.1--1.1 }
\label{f:rt15}
 \end{center} \end{figure} 

\paragraph*{ 2D Riemann Problem 17} Also generates two contact layers, a shock and a rarefaction region.  The vertical contact layer at the bottom remains perfectly resolved with no diffusion, the curved shock and rarefactions show no sign of instability, and the only features of note are again the initial contact artefacts that remain $<1\%$ through the simulations, being revealed by the contour colour table used here.
 
 \begin{figure} [thdp] \begin{center}
\includegraphics[width = 5in]{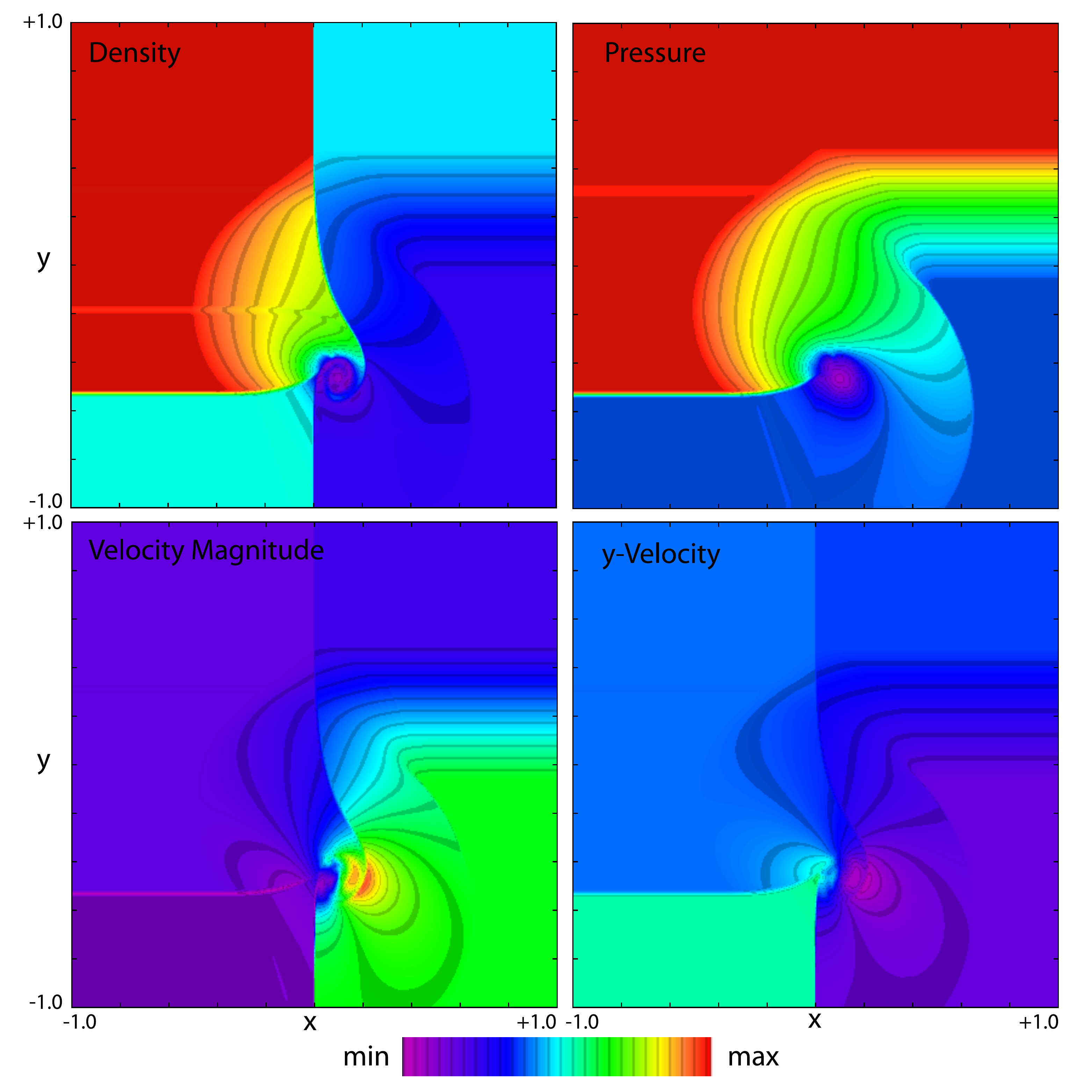}
\caption{\small 2D Riemann Problem 17. Density, Pressure, Speed, $y-$Velocity at $t = 0.3$.   Colour scale black to red, linear. Density 0.0--2.01, Pressure 0.0--1.01, Speed 0.0--1.7, $y-$Velocity -1.7--1.7 }
\label{f:rt17}
 \end{center} \end{figure} 
\clearpage

\subsubsection{2D Symmetric Implosion}

This test probes a number of code/method properties. The initial setup is particularly simple, and on a Cartesian grid can be set up with essentially perfect symmetry. The shocks generated initially have the parameters from the standard Sod test, but here, with reflecting boundaries all around, the simulation is allowed to continue for many crossing times.  The spatial resolution is $h = 400$.

Shocks reflecting from the origin eventually generate a free standing double 2D vortex, that is free to wander over the domain, and is not dissipated by crossing shocks. As the origin of the vortex is a cavity in an unstable contact discontinuity, the path and destiny of the vortex is extremely sensitive to a number of factors.

Firstly, floating point `noise', residual errors from the repeated application of non-linear operations can accumulate and eventually break the perfect symmetry of the initial conditions. Without specific care to control round off errors this noise will be present even in symmetric sweep codes. Secondly, if the standard asymmetric split sweeps are applied, the small asymmetry generated by the initial sweep (x or y) results in an asymmetric grid that will evolve with a small bias from then on. In this simulation that leads to the double vortex drifting in quite different directions over the grid. Only symmetric sweep operations will preserve symmetry and keep the double vortex on the diagonal axis.

\begin{figure} [thdp] \begin{center}
\includegraphics[width = 5in]{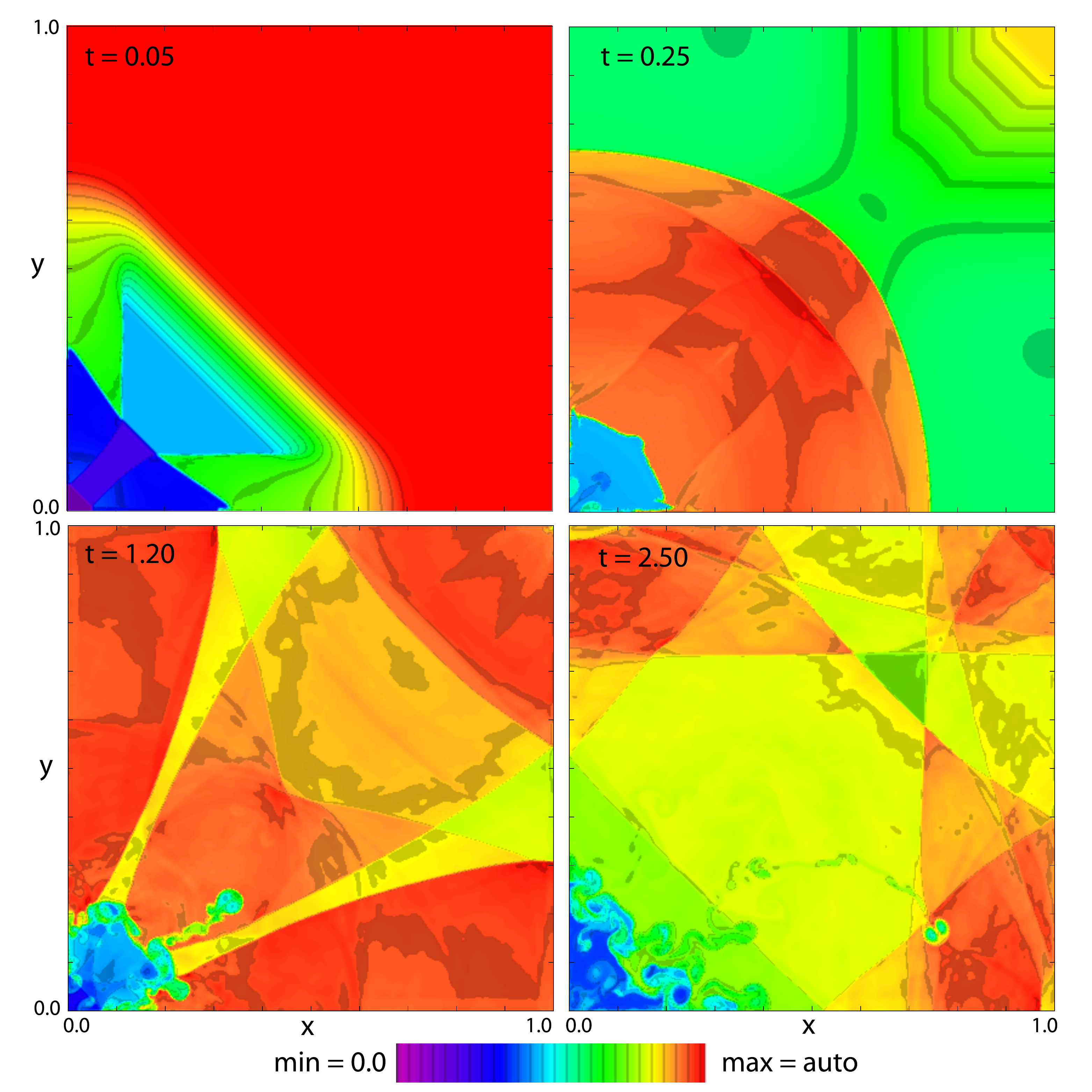}
\caption{\small 2D Implosion, split sweeps. Density at $t = 0.05, 0.25, 1.20$ and $2.5$.  Density colour scale (black) $= 0.0$, red = auto-scaled maximum, linear. }
\label{f:impl400}
 \end{center} \end{figure}

\begin{figure} [thdp] \begin{center}
\includegraphics[width = 5in]{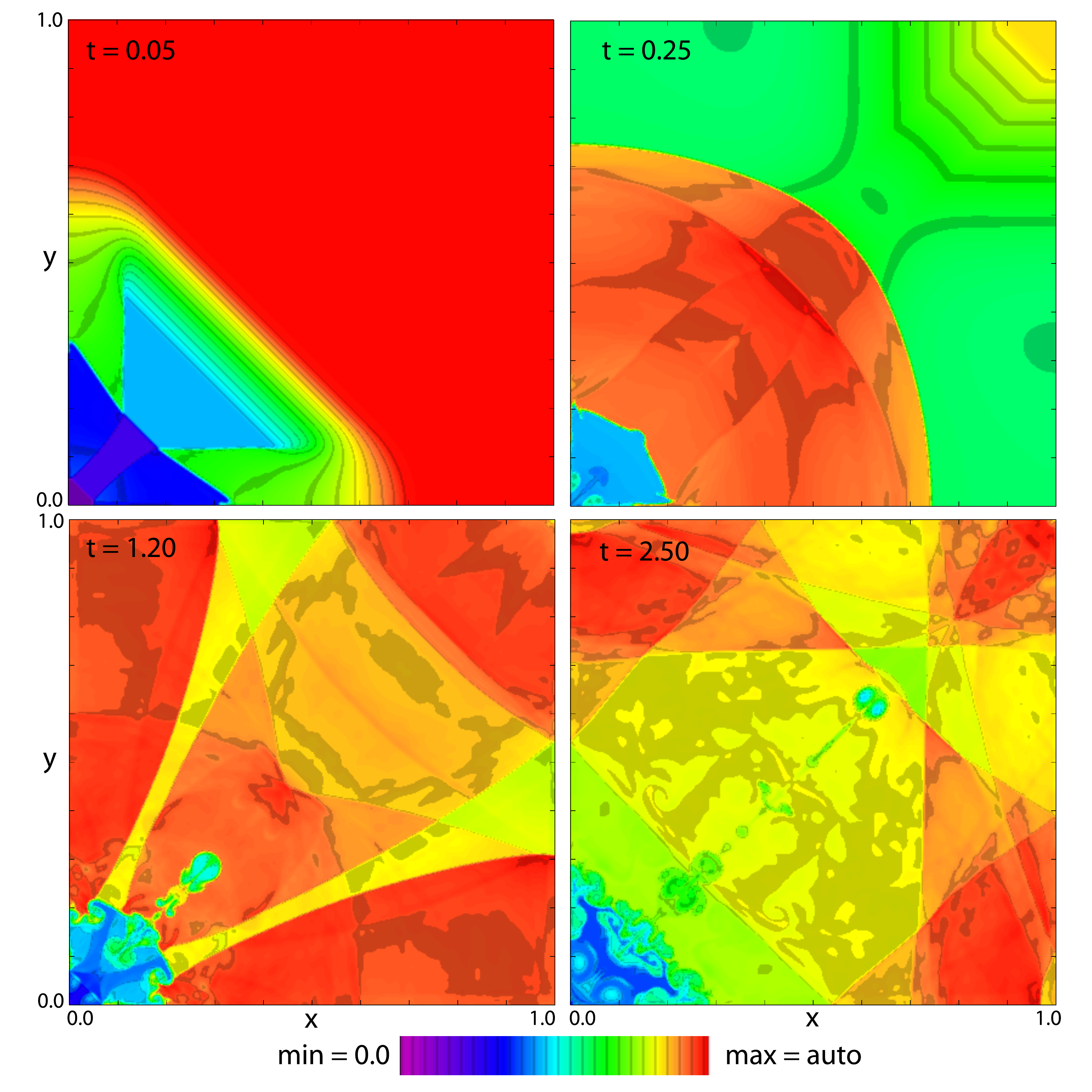}
\caption{\small 2D Implosion, symmetric sweeps. Density at $t = 0.05, 0.25, 1.20$ and $2.5$.  Density colour scale (black) $= 0.0$, red = auto-scaled maximum, linear. }
\label{f:impl400symm}
 \end{center} \end{figure}

Figure \ref{f:impl400} and figure \ref{f:impl400symm} show the density at four points during the simulation.  In the first figure, the normal split sweep algorithm is used, the same one as used in {\em PPM} and {\em ppmlr}.  The initial asymmetry of choosing one sweep direction for the first sweep breaks the symmetry of the initial conditions.  By $t = 0.25$, when the vortex from the origin is forming, the asymmetry has grown so that the vortex drifts to the right of the diagonal.  By $t = 1.2$ the vortex has left the inner region and is crossing the outer denser region, and has been crossed multiple times by shocks.  Overall the simulation remains more or less symmetrical, but the detailed asymmetry is apparent in the vortex and the the boundary between the low and high density regions in particular.  By $t= 2.5$ the vortex is well over to the right and the curved path it took is shown by a low density wake.

Figure \ref{f:impl400symm} shows the case when the split sweeps are symmetrised, by averaging the results of the $x$ and $y$ sweeps before updating the grid values.  This required intermediate storage to save the two sweep results before combining them.  In this case the symmetry of the initial conditions is preserved under the symmetric sweep operation, and so the vortex is constrained to follow the diagonal, in a way similar to the symmetric, un--split, CLAW and WENO codes in the LW03 results.

\begin{figure} [thdp] \begin{center}
\includegraphics[width = 5in]{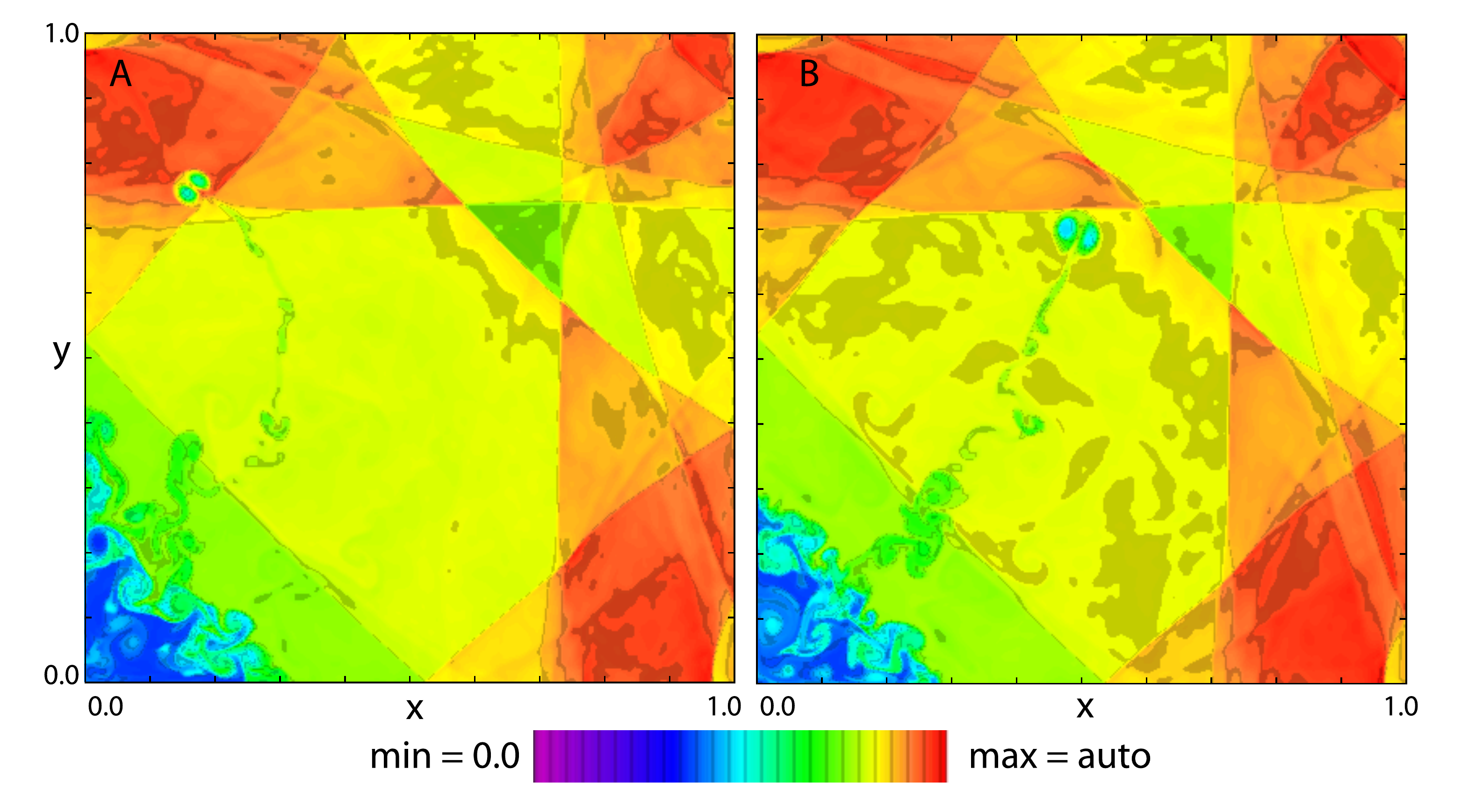}
\caption{\small 2D Implosion.   A, standard split sweeps beginning with the $y$-sweep instead of an $x$-sweep, thereafter proceeding as normal. B, the symmetric sweeps, but the initial diagonal boundary was smoothed in a way that resulted in approximately one percent differences in the initial density field along the interface.  Density at $t = 2.5$.  Density colour scale (black) $= 0.0$, red = auto-scaled maximum, linear. }
\label{f:implsymm}
 \end{center} \end{figure}

This is made clearer in the final figure, figure \ref{f:implsymm}.  This shows two additional  cases, shown at $t = 2.5$, as in the 
last panel of figure \ref{f:impl400} and figure \ref{f:impl400symm}.  The first panel, A,  shows the standard split sweeps, only starting with the $y$-sweep instead of an $x$-sweep, thereafter proceeding as normal. The right panel, B,  used the symmetric sweeps, but the initial diagonal boundary was smoothed in a way that resulted in approximately one percent differences in the initial density field along the interface.  Under the symmetric sweep operation the asymmetry was able to evolve normally, showing that the method does not impose symmetry, only allows it to be preserved if it is present initially.  The instability of the contact discontinuity that forms the vortex is therefore extremely sensitive to the symmetry of the setup and algorithms.

\clearpage 
\subsubsection{2D Explosion}

The 2D explosion test of LW03 centres also around the production of an unstable contact layer, which is crossed by a reflected shock wave. Performing this essentially circular explosion with Cartesian coordinates means that the reflection at the origin of the spherical shock is also reasonably challenging.   The problem setup is given in table \ref{t:expl400}.

\begin{figure} [thdp] \begin{center}
\includegraphics[width = 5in]{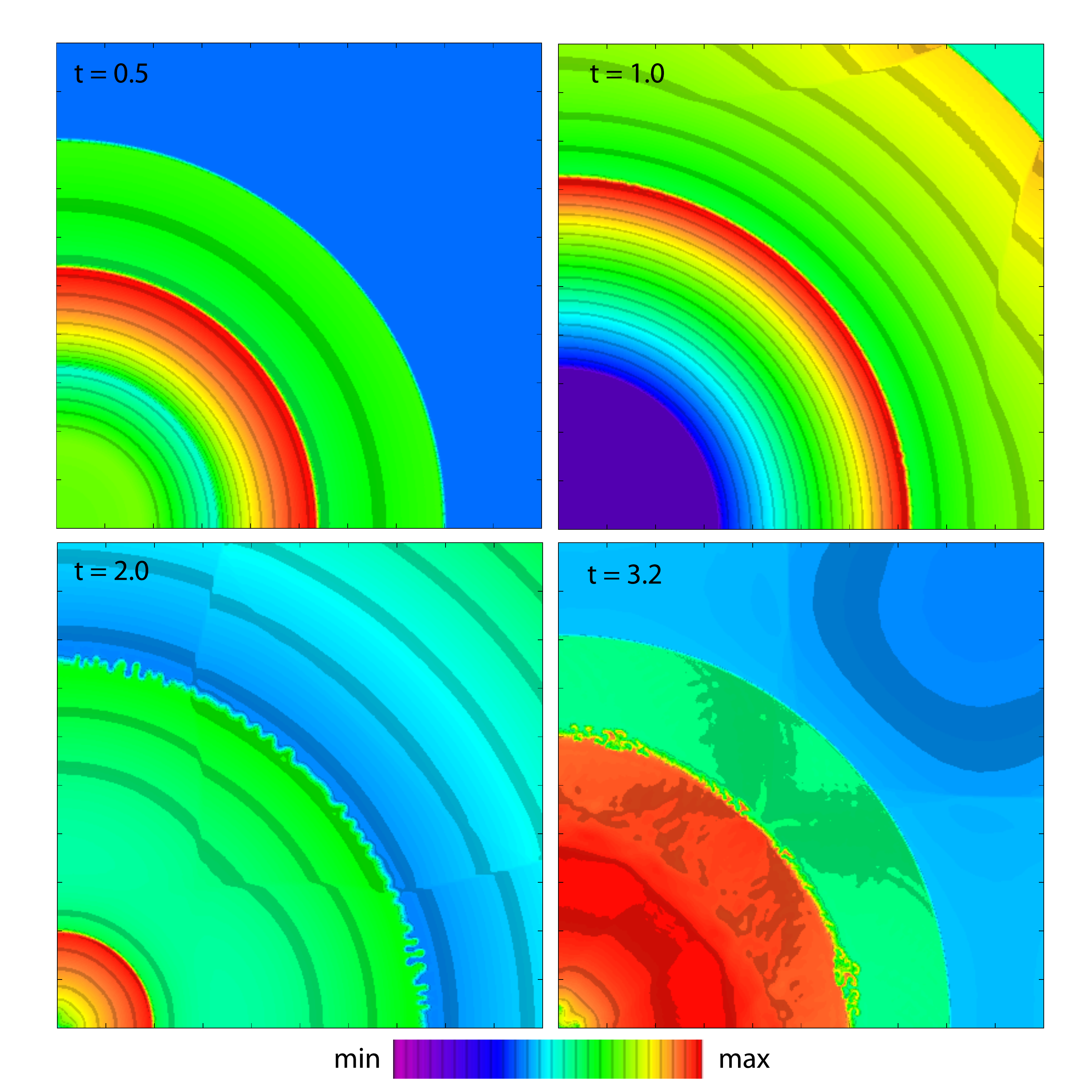}
\caption{\small 2D Noh Explosion at $t = 0.5, 1.0, 2.0$ and $3.2$.  Density colour scale (black) $= 0.0$, red = auto-scaled maximum, linear. }
\label{f:expl400}
 \end{center} \end{figure}

The less dissipative high resolution shock capturing upwind methods, such as PPM, VH1, \Fyris, tend to produce the greatest amount of interface breakup, and the source of the grid noise that seeds the interface breakup appears to lie in the initial boundary definition of the high pressure circular region at t = 0.0. In dissipative schemes, the initial `pixillation' of the circle on the square grid is lost, whereas in the high resolution non-dissipative methods the signature of this pattern survives to late times and the interface breaks up more. 
 See figure \ref {f:expl400}.

Here a second test with the initial circular boundary is smoothed more heavily with a gaussian kernel with a FWHM of 2.0 cells is compared, and the late time interface breakup is less. The standard test is run with a smoothing of FWHM 1.0 cells.

Asymmetry in the late time interface breakup is due to the standard asymmetric split sweep operator used by \Fyris\ (like VH1, PPM). When a symmetric split sweep operator is used, as the expense of extra memory, then the initial grid symmetry is preserved and the late time interface remains symmetric, like the WENO and CLAW codes in the LW03 results.  (See 2D implosion results for description of symmetric sweeps).

\begin{table}[thdp]

{\small 
\caption{Setup: 2D Explosion Problem. \label{t:expl400}}
\begin{tabular}{l l l l }
\hline
 \multicolumn{1}{c}{EOS} &  \multicolumn{1}{c}{Algorithm} &  \multicolumn{1}{c}{Grid} &  \multicolumn{1}{c}{Boundaries}  \\
\hline
\hline
Adiabatic              & CFL, $c = 0.8$       & $400\times400$cells & Left: reflecting \\
 $\gamma = 1.4$ &  flattening: 0 -- 1    &  $0.0 < x < 1.5 $       &  Right: free \\
                            &  $\alpha = 0.001$ &  $0.0 < y < 1.5 $       & Top: free \\
                            &                             &                                  & Bottom: reflecting \\
\hline
 \multicolumn{3}{c}{Initial Conditions}&  \multicolumn{1}{c}{Ending} \\
 \multicolumn{1}{c}{Region 0, $r < 0.4$} &  \multicolumn{1}{c}{Region 1, $r > 0.4$} &  \multicolumn{1}{c}{} &  \multicolumn{1}{c}{Time Limit}  \\
\hline
\hline
{ $\rho_0 = 1.0 $    }  &   $\rho_1 = 0.125$  &  & \multicolumn{1}{c}{$t = 3.2$} \\
{ $P_0      = 1.0$    }  &   $P_1 = 0.1$  &     &   \\
{ $v_0)     =  0.0$    }  &   $v_1 = 0.0$  &   &   \\
\hline
\end{tabular}
} 
\end{table}

\clearpage
\subsubsection{2D Noh Test}

This 2D LW03 test consists of a circular infinite strength shock propagating out from the origin on a square Cartesian grid. There is an analytical solution, and with $\gamma = 5/3$ it is given in table \ref{t:noh2dsol}.  Numerically, pressure is set to $P = 1.0\times10^{-6}$ everywhere, as an approximation to zero pressure.  The radial velocity, $v(r)$, is $-1.0$ (flowing towards the origin). This gives an isothermal Mach number, $M_i = \rho v/ P = 1000.0$.  The simulation setup is given in table \ref{t:noh400}.

\begin{table}[thdp]

{\small 
 \caption{2D Noh Solution, $\gamma = 5/3$.  \label{t:noh2dsol}}
\begin{tabular}{l l l l }
\hline
 \multicolumn{1}{c}{Inside Shock  ($r < t/3$)} &  \multicolumn{1}{c}{Outside Shock  ($r > t/3$)}  \\
\hline
\hline
Density $\rho_0 = 16 $  &  Density $\rho_1 = 1.0 + t/r$     \\
Pressure $P_0 = 16/3$ &  Pressure $P_1 = 0.0$             \\
Velocity $v_0 = 0$        &  Radial Velocity $v(r) = -1.0$    \\
 \multicolumn{2}{c}{Shock Front Expands at $v_s = 1/3$} \\
\hline
\end{tabular}
} 
\end{table}

\begin{table}[thdp]

{\small 
\caption{Setup: 2D Noh Problem, \label{t:noh400} }
\begin{tabular}{l l l l }
\hline
 \multicolumn{1}{c}{EOS} &  \multicolumn{1}{c}{Algorithm} &  \multicolumn{1}{c}{Grid} &  \multicolumn{1}{c}{Boundaries}  \\
\hline
\hline
Adiabatic              & CFL, $c = 0.8$      & $n = 400\times400$cells & Left: reflecting \\
 $\gamma = 5/3$ &  flattening: 0 -- 1   &  $0.0 < x < 1.0 $       &  Right: special \\
                            &  $\alpha = 0.05$  &  $0.0 < y < 1.0 $       & Top: special \\
                            &                            &                                  & Bottom: reflecting \\
\hline
 \multicolumn{3}{c}{Initial Conditions}&  \multicolumn{1}{c}{Ending} \\
 \multicolumn{1}{c}{Whole Grid} &  \multicolumn{1}{c}{Right Boundary} &  \multicolumn{1}{c}{Top Boundary} &  \multicolumn{1}{c}{Time Limit}  \\
\hline
\hline
{ $\rho_0 = 1.0 $    }                  &   $\rho_1 = 1.0 + t/r$  &  $\rho_1 = 1.0 + t/r$ & \multicolumn{1}{c}{$t = 2.0$} \\
{ $P_0      = 1.0\times10^{-6}$ }&   $P_1 = 1.0\times10^{-6}$  &   $P_1 = 1.0\times10^{-6}$    &   \\
{ $v(r)     = -1.0    $      }             &   $v(r) = -1.0$  &   $v(r) = -1.0$  &   \\
\hline
\end{tabular}
} 
\end{table}

\begin{figure} [thdp] \begin{center}
\includegraphics[width = 5in]{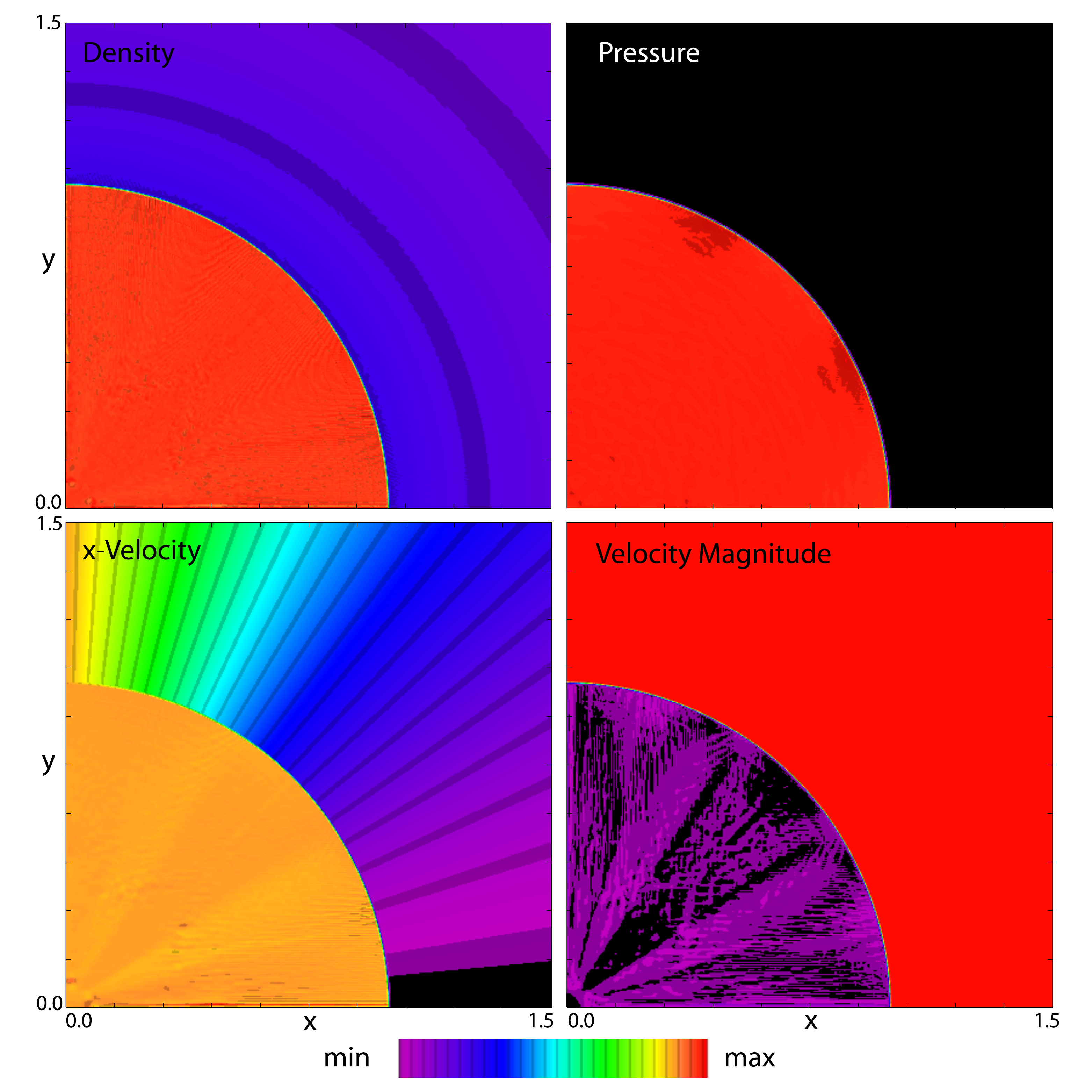}
\caption{\small 2D Noh problem at $t = 2.0$.  Density colour scale (black)$0.0-16.614$(red), linear, auto max. Pressure  (black)$0.0-5.44$(red), linear, auto max. x-Velocity  (black)$-1.0-0.115$(red), linear, auto. Velocity magnitude (black)$0.01.05$(red), linear, fixed scale.}
\label{f:noh400}
 \end{center} \end{figure}

LW03 note that this problem is very difficult to compute, with only limited success achieved using the 8 codes that they tested.  The key problem here is the extreme tendency for the Odd-Even or striping instability to set in in regions where the curved shock becomes aligned with the grid.  Uncontrolled, the striping can extend up to 30 degrees from each axis.  Fyris performs the test well, (see figure \ref{f:noh400}), and we can publish meaningful  L1 norms for the density and pressure over the grid with respect to the analytical solutions.  With the ability to handle very strong shocks, and good suppression of the striping instability, the problem becomes tractable.

$L1$ Norms w.r.t. Analytical model at $t = 2.0$:
{\small 
\begin{itemize}
\item L1 norm Density : 0.74 per cent.
\item L1 norm Pressure : 0.87 per cent.
\end{itemize}
}

\clearpage
\subsubsection{2D Rayleigh--Taylor Instability}
\label{s:RTI}
This 2D LW03 test consists of a domain with a constant gravity field with uniform acceleration.
{\small
\begin{itemize}
\item Adiabatic index: $\gamma = 5/3$
\item Acceleration: $a = -0.1$ (downwards)
\item $0 < t < 8.5$, solution at $t = 8.5$
\end{itemize}
}

\subsubsection*{Hydrostatic Initial Conditions and Compressibility}

The grid is set up in hydrostatic equilibrium. With a constant gravity field this implies an exponential vertical distribution, with a scale height for each component depending on the isothermal sound speed $a^2 = P/\rho$.  

Oddly, the original LW03 test uses different $x-$ and $y-$ resolutions.
The original spatial $x - resolution$ is $n_x = 100$ over a domain of $0.0 < x < 1/6$, for $h_x = 600$, and $n_y = 400$ over a domain $0.0 < y < 1.0$, for $h_y = 400$.  Models with a uniform resolution of $h = 600$ for $x$ and $y$ were also trialled here.

\subsubsection*{Critical Setup of the Initial Interface}

The well known property of the RT instability is the exponential initial growth rates, with the fastest rates corresponding to the highest wave numbers. Consequently, any minute deviations from a smooth initial conditions give rise to high wavenumber  perturbations which will grow rapidly. High resolution methods will model this property best. As the sinusoidal wave is averaged over a finite grid, even with some smoothing of the interface, the high resolution methods will break up into a complex multi-mode interaction between the dense and low density regions.

Here the sinusoidal interface between the high and low density regions is convolved with a Gaussian FWHM of 1.0 cells.

\subsubsection*{Suppression of High Frequency Modes}

With sufficient numerical diffusion, the higher modes can be suppressed, leaving just the principle mode to grow. This is reflected in the results of LW03, where the diffusive central methods produced single RT bubble/stalk structures, while high resolution PPM, WENO, and VH1 tests broke up into complex structures.

\begin{figure} [thdp] \begin{center}
\includegraphics[width = 5in]{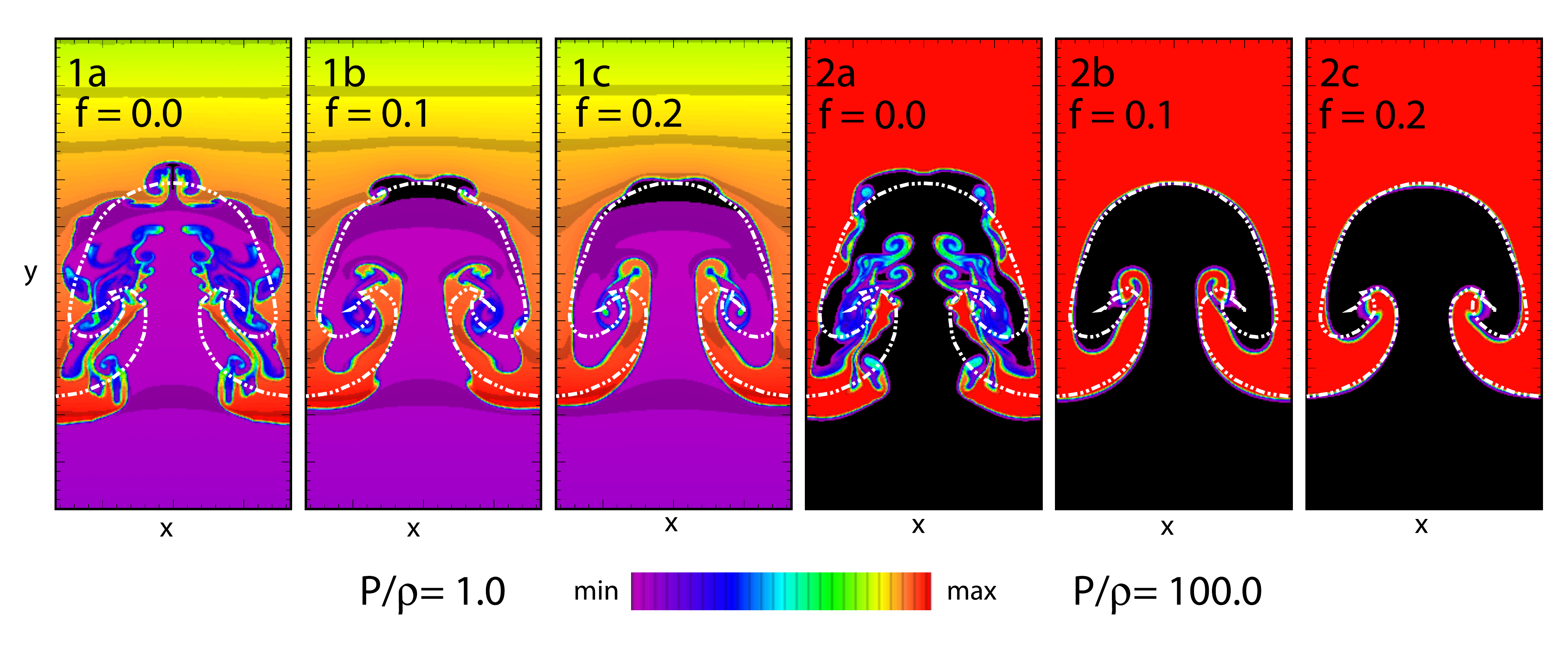}
\caption{\small  2D Rayleigh Taylor Instability test, autoscaled density at $t = 8.5$. Computed on a $100\times 400$ cell domain, as originally defined by LW03. Left row, 1a -- 1c, the isothermal sound speed squared $a^2 = P/\rho =  1.0$ at the interface, defining the temperature and scale heights in the two layers.  Right Row; 2a -- 2c,  $a^2 = 100.0$ In each row, the (a) model used the full PPM range of slopes, $f = 0.0$ resulting in no minimum flattening of the grid parabolae.  The (b) models have mean interpolation slopes reduced by 10\%.  The (c) models have mean slopes reduced by a minimum of 20\%, reducing fractionally the order of the hydrodynamic fluxes.  White Dashed Line: result for the two fluid boundary, using a specialised front-tracking code, see text for details.}
\label{f:rt}
 \end{center} \end{figure}

\Fyris\ performs like the other high resolution methods: small initial high wavenumber deviations from a continuous sin wave interface grow and eventually distort the final state.  Trial 1a in figure \ref{f:rt} may be compared with the LW03 results, with the proviso that the grid is displayed here with square pixels.  Like the other high--resolution codes, complex structure from higher frequency modes are captured and amplified, even with the careful initial boundary smoothing.

However, \Fyris\ also has the option to increase the numerical diffusion incrementally to deliberately suppress the unwanted high modes, while leaving the principle mode to grow. The process used to flatten the interpolation parabola, to reduce the order of the method in the presence of shock waves in order to satisfy the Godunov theorem, is also used to increase numerical diffusion.  

Normally the flattening is assigned dynamically, as described in the PPM method. To add global conservative diffusion in \Fyris, the minimum flattening can be raised from $f=0.0$, normally in the range of $f = 0.0--0.2$ This means that the mean slope from the conservative interpolation of the left and right values is reduced by  up to 20\%, while retaining parabolic interpolation.

When this type of additional diffusion is applied to the RT problem, the higher order modes are suppressed, giving the results in models 1b and 1c.  Notice the simplification to a single stalk-bubble structure, like the output of the diffusive codes in LW03.

\subsubsection*{Compressibility}

The appearance of the LW03 results compared with the front-tracking results they presented from the Los Alamos code (white dashed lines in the figures here) still do not match the more diffusive simulation, 1c, as the extent of the bubble vortices and the $x$-height of the lowest part of the dense stalks differ somewhat.

Although LW03 do not specify the isothermal sound speed, or alternatively the ratio of pressure over density at the interface, it appears by inspection of their figures that the ratio $P/\rho$ was approximately 1.0 at the interface. This results in a dynamically cool system, with some compressibility. The density in the upper, colder, layer is distinctly exponential when viewed with the contoured colour table used here, although it is less obvious in the LW03 figures. This affects the effective Attwood number, which is commonly used to describe the Rayleigh--Taylor, RT, instability, in that the density contrast effectively changes as the bubble/stalk structures grow, whereas in the incompressible limit the upper and lower densities remain constant.

The tests 1a -- 1c use $P/\rho = 1.0$, and give the best comparison to the LW03 codes. Hotter simulations with $P/\rho = 100.0$, 2a -- 2c, giving less compressible fluids, approach the Los Alamos solution.

With the hotter models, the greatly increased scale height in the constant potential results in the vertical density gradients being reduced, giving a nearly uniform mean density above and below the initial interface.  With both diffusion and increased incompressibility, model 2c nearly exactly reproduces the Los Alamos result, suggesting that the Los Alamos test result may indeed be applicable to the incompressible limit, rather than the more compressible cooler models, and hence was not a good comparison for the compressible LW03 tests.

Finally, tests with uniform $x-$ and $y-$ resolution are trialled, in models 3a -- 3c, and 4a -- 4c, corresponding to the first two sets of tests, but on a $100 \times 600$ grid.  Interestingly, the model 4c (corresponding to the best fit model 2c above) shows more deviation from the Los Alamos curve than trial 2c, so some dependence on the grid remains, and more extensive work would be required to properly model the RT instability.

\begin{figure} [thdp] \begin{center}
\includegraphics[width = 5in]{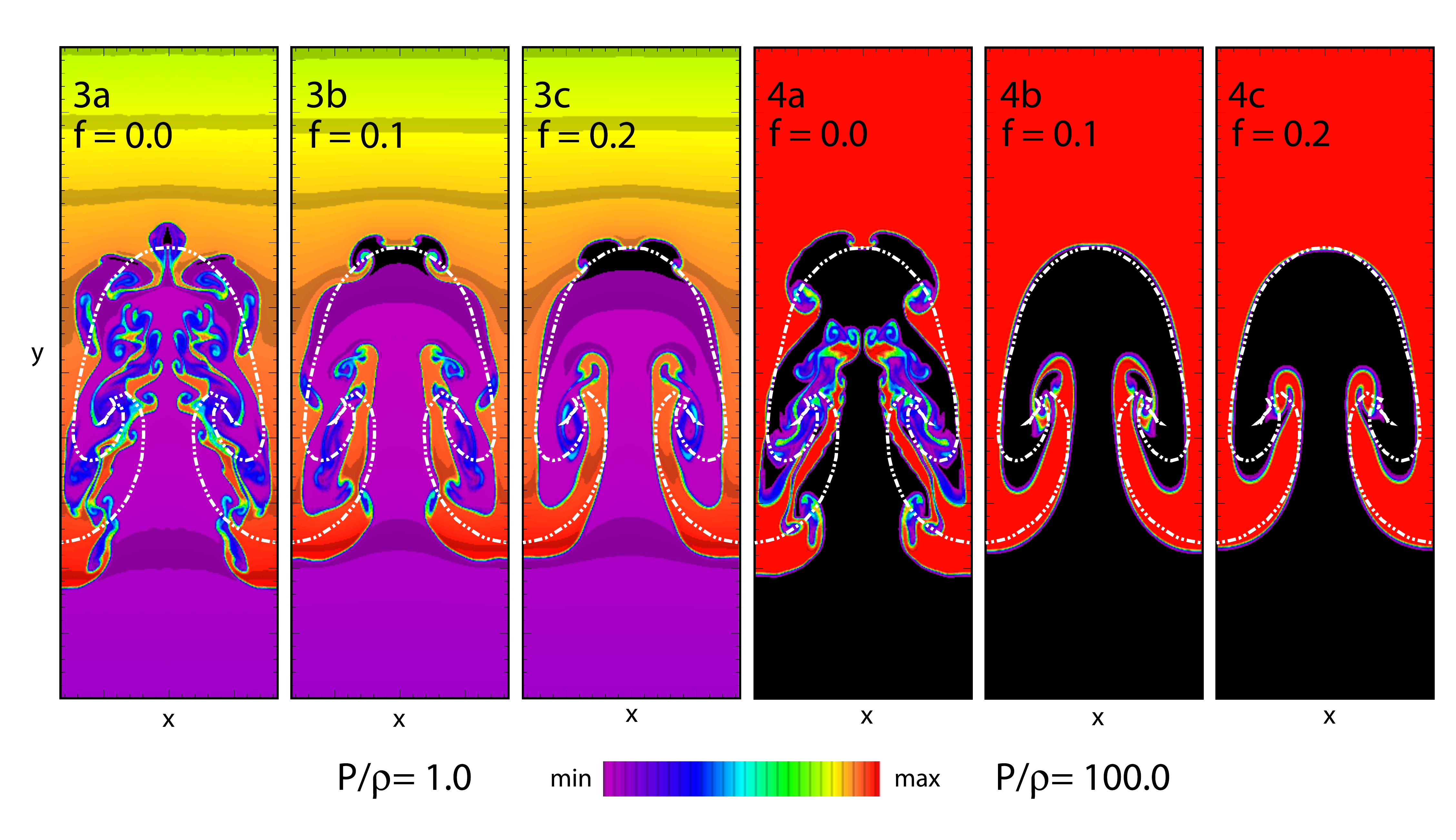}
\caption{\small  2D Rayleigh Taylor Instability test, same as in figure \ref{f:rt} computed on a $100\times 600$ cell domain, with in equal $x$ and $y$ resolution.  Left Row, 3a -- 3c, same parameters as 1a -- 1c, Right Row, 4a -- 4c, same parameters as 2a -- 2c.  White Dashed Line: result for the two fluid boundary, using a specialised front-tracking code. Autoscaled.  See text for details.}
\label{f:rt2}
 \end{center} \end{figure} 

\clearpage 
\subsubsection{2D Gresho Vortex}
The Gresho vortex is a 2D flow pattern where the centrifugal forces are matched by pressure gradients, resulting in a stable time--independent vortex.

The LW03 tests consist of single 2D vortex on a range of grid resolutions, $h = 20 ... 80$. One test models a stationary vortex, and the second test advects the vortex across the grid by three diameters.   In the LW03 results, at the two lower resolutions here, the advected vortex became distorted, particularly with the high resolution methods.  Here, the resolution is increased further to look at whether the stability improves as the vortex is better resolved.

\begin{table}[thdp]

{\small 
 \caption{2D Gresho Vortex Solution, $\gamma = 5/3$. \label{t:greshosol} }

\begin{tabular}{c c c c c}
\hline
 \multicolumn{1}{c}{Radius}& \multicolumn{1}{c}{Density}& \multicolumn{1}{c}{Pressure}& \multicolumn{1}{c}{Azimuthal Velocity}& \multicolumn{1}{c}{Vorticity}\\
 \multicolumn{1}{c}{$r$}& \multicolumn{1}{c}{$\rho(r)$}& \multicolumn{1}{c}{$P(r)$}& \multicolumn{1}{c}{$v_\phi(r)$}& \multicolumn{1}{c}{$V_{yx}(r)$}\\
\hline
\hline
$r \le 0.2$         &1.0&$5.0 +25/2 r^2$                                     &$5r$      &$10.0$\\
$0.2 < r \le 0.4$&1.0&$9-4\ln 0.2 + 25/2 r^{2} -20r + 4 \ln r$&$2-5r$&$2/r - 10.0$\\
$0.4 < r$           &1.0&$3+4\ln 2$                                           &$0.0$    &$0.0$\\
\hline
\end{tabular}

} 
\end{table}

\begin{table}[thdp]

{\small 
\caption{Setup: 2D Stationary Gresho Vortex Problem. \label{t:greshostat} }
\begin{tabular}{c c c }
\hline
\multicolumn{1}{c}{EOS} &  \multicolumn{1}{c}{Algorithm} &  \multicolumn{1}{c}{Grid} \\
\hline
\hline
Adiabatic              & CFL, $c = 0.8$      & $h = 20\times20$ -- $h = 80\times80$cells  \\
 $\gamma = 5/3$ &  flattening: 0 -- 1   &  $0.0 < x < 1.0 $      \\
                            &  $\alpha = 0.001$  &  $0.0 < y < 1.0 $   \\
\hline
 \multicolumn{1}{c}{Initial Conditions}&  \multicolumn{1}{c}{Boundaries}  &  \multicolumn{1}{c}{Ending} \\
\hline
\hline
{ $\rho_0 = 1.0 $    }               & All: free &   \multicolumn{1}{c}{$t = 3.0$} \\
{ $P(r)      = $ Vortex Solution }  \\
{ $v(r)     = $  Vortex Solution  }   \\
\hline
\end{tabular}

} 
\end{table}

\begin{figure} [thdp] \begin{center}
\includegraphics[width = 5in]{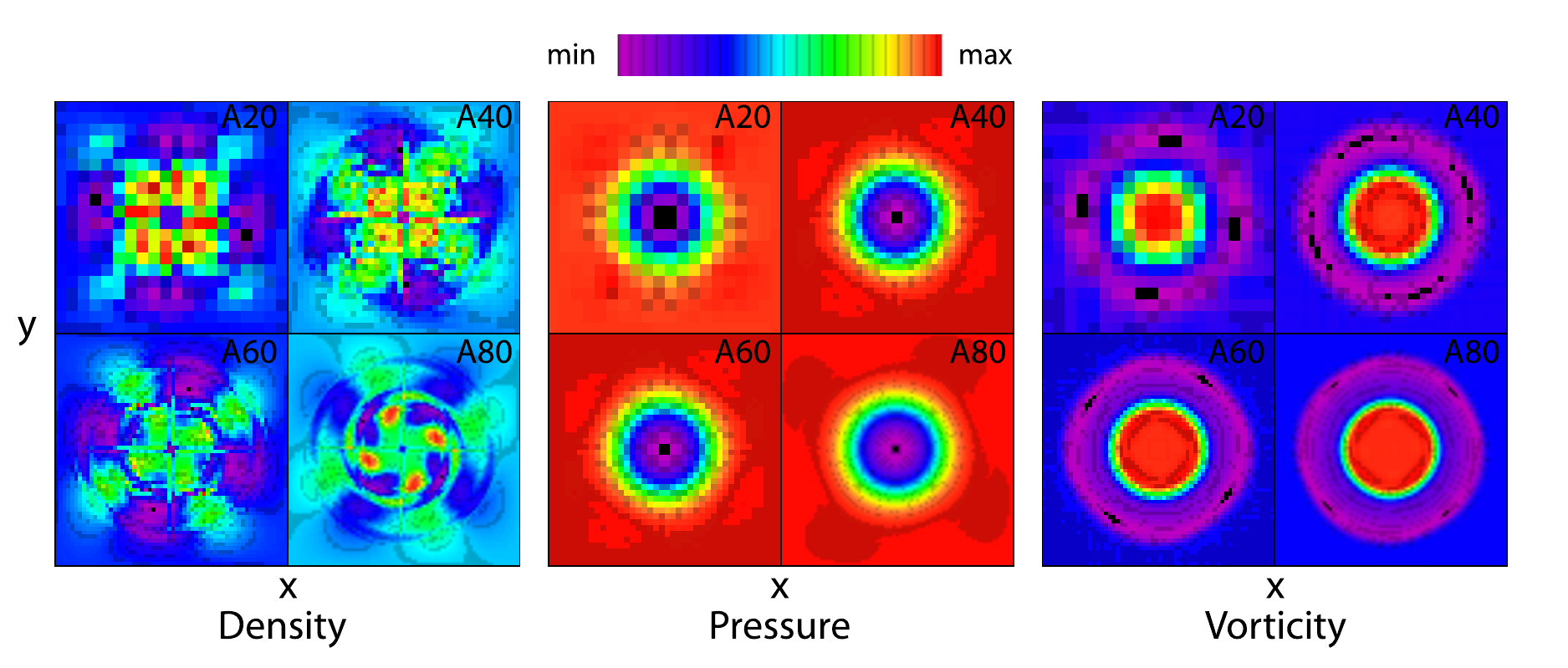}
\caption{\small Stationary 2D Gresho Vortex problem at $t = 3.0$. Left: Density, Middle: Pressure. Right: Vorticity.  Colour scale (black)-(red), linear, auto scaled. }
\label{f:avortex}
 \end{center} \end{figure}

The stationary vortex is defined as a vortex initially centred about $(0.5, 0.5)$, outer radius $r = 0.40$.
Runs are made on a $20\times20$ (run A20), $40\times40$ (run A40), $60\times60$ (run A60) and $80\times80$ (run A80) grid.
Errors and L1 norms are computed over a unit square centred on $(0.5, 0.5)$ at the end of the simulations, for the A models, shown in table \ref{t:g_astat}, along with the results for A20 and A40 for the codes in LW03 in table \ref{t:lw_astat} for comparison.   Figure \ref{f:avortex} shows the stationary vortex at $t = 3.0$.  The autoscaling shows up the small non-uniformities in the final density state, 
which ranges from $\rho = 0.999$ to $\rho = 1.001$ in the A60 case.

\begin{figure} [thdp] \begin{center}
\includegraphics[width = 5in]{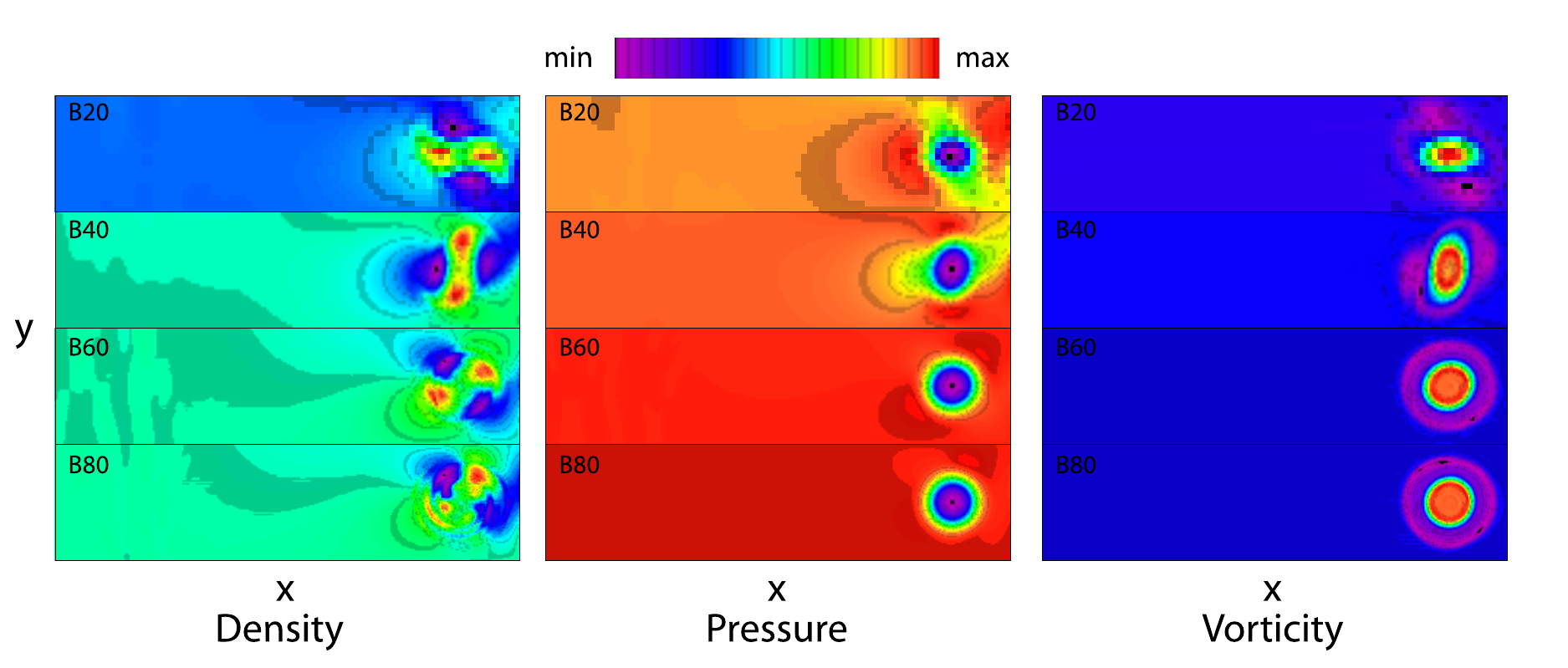}
\caption{\small Moving 2D Gresho Vortex problem at $t = 3.0$. Left: Density, Middle: Pressure. Right: Vorticity.  Colour scale (black)-(red), linear, auto scaled. }
\label{f:bvortex}
 \end{center} \end{figure}

The moving vortex case is the same as the stationary vortex, with a global $v_x = 1.0$ drift velocity added to the whole grid, on top of the initial vortex velocity field. Runs are made on a $20\times80$ (run B20), $40\times160$ (run B40), $60\times240$ (run B60) and $80\times320$ (run B80) grid.
The grid is extended to $0.0< x < 4.0$, and the  L1 norms are computed over a unit square centred on the expected vortex centre at
$(3.5, 0.5)$, at $t = 3.0$, for the B models, in table \ref{t:g_bstat},  along with the results for B20 and B40 for the codes in LW03 in table \ref{t:lw_bstat} for comparison.  Figure \ref{f:bvortex} shows the stationary vortex at $t = 3.0$.  Again, the nearly uniform density state at the end is enhanced with auto--scaling, ranging from $\rho = 0.995$ to $\rho = 1.0039$ for the B60 run.

Note: the vorticity is estimated from the discrete grid by the central finite difference:
$\mbox{vorticity}[i,j] \approx 0.5 \{ (v_y[i+1, j]-v_y[i-1,j])/dx - (v_x[i, j+1]-v_x[i,j-1])/dy  \}$.

The $L1$ errors in density and vorticity in the A20 and A40 models are comparable or better with \Fyris\ than the high--resolution upwind codes in LW03. The density error drops considerably going from the A20 to the A40 model, but thereafter drops only slowly for the A60 and A80 models. For the vorticity and kinetic energy (KE), \Fyris\ is comparable to the high--resolution upwind codes like PPM, and considerably better than the more diffusive central methods.  The KE error continues to drop quickly for the A60 and A80 models, while the vorticity error only falls more modestly.

\begin{table}[thdp]
{\small 
\caption{Fyris Stationary Gresho Vortex L1 Norm Errors, \%. \label{t:g_astat}}
\begin{tabular}{r l l l }
 \hline
 Model& Density & Vorticity & Total KE \\
 \hline
 \hline
A20&{\bf 0.150}& 26.4 &11.64\\
A40&{\bf 0.0276}& 14.5 &1.850\\
A60&0.0181& 11.1 &0.734\\
A80&0.0144& 8.89 &0.383\\
 \hline
\end{tabular}
} 
\end{table}

\begin{table}[thdp]
{\small 
\caption{LW03 Stationary Gresho Vortex. L1 Norm Errors, \%. \label{t:lw_astat}}
\begin{tabular}{l l l l l l l }
 \hline
 & \multicolumn{2}{c}{Density}&\multicolumn{2}{c}{Vorticity}&\multicolumn{2}{c}{Total KE}\\
Code&A20&A40&A20&A40&A20&A40\\
 \hline
 \hline
CFLFh&0.22&0.16&{\bf 22}&20&{\bf 0.2}&{\bf 0.4}\\
JT&0.56&0.22&89&45&55.2&18.3\\
LL&2.27&0.23&71&44&65.6&26.1\\
CLAW&0.33&0.10&50&28&29.9&6.1\\
WAFT&0.24&0.07&47&26&7.7&5.7\\
WENO&0.35&0.06&38&27&30.9&3.7\\
PPM&0.20&0.04&25&{\bf 13}&9.1&0.8\\
VH1&{\bf 0.15}&0.04&26&15&9.6&1.2\\
 \hline
\end{tabular}
} 
\end{table}

 \begin{table}[thdp]
{\small 
\caption{Fyris Moving Gresho Vortex. L1 Norm Errors, \%. \label{t:g_bstat}}
\begin{tabular}{r l l l }
 \hline
 Model& Density & Vorticity & Total KE \\
 \hline
 \hline
B20&{\bf 0.649}& {\bf 62.08}&2.072\\
B40&0.568& 56.47 &{\bf 0.0306} \\
B60&0.104& 18.97 &0.0392\\
B80&0.0418& 14.52 &0.0026\\
 \hline
\end{tabular}
} 
\end{table}

\begin{table}[thdp]

{\small 
\caption{ LW03 Moving Gresho Vortex Errors.  L1 Norm Errors, \%\label{t:lw_bstat}}
\begin{tabular}{l l l l l l l }
 \hline
 & \multicolumn{2}{c}{Density }&\multicolumn{2}{c}{Vorticity}&\multicolumn{2}{c}{Total KE}\\
Code&B20&B40&B20&B40&B20&B40\\
 \hline
 \hline
CFLFh  &1.12   &0.72 &145  &83 &12.8    &0.1\\
JT       &0.81   &{\bf 0.22} &100  &52  &42.8   &22.1\\
LL       &0.65   &0.49 &88    &60  &71.6   &30.9\\
CLAW  &0.72   &0.29 &65    &{\bf 37}&39.9   &8.3\\
WAFT  &0.87    &0.77 &65     &62 &{\bf 1.3}  &12.6\\
WENO &0.37  &0.43 &48     &40 &31.6   &4.0\\
PPM    &1.10  &0.42 &93     &36 &4.9    &1.0\\
VH1   &0.80  &0.66 &65     &55  &11.7  &1.2\\
 \hline
\end{tabular}
} 
\end{table}

\clearpage

\subsubsection{2D Odd-Even Test}
 \label{s:striping}
 
This test is designed to produce a strong numerical instability, known variously as the odd--even decoupling, the striping instability, the carbuncle instability, or more colourfully the red-black instability.

When a strong shock is moving slowly across a grid, such that the shock interface is aligned with the underlying grid (in any coordinate system) a pattern of stripes perpendicular to the shock can develop very quickly. This has been known for many years, with instances seen in the \cite{cw84b} review paper. This numerical instability is very strong, and can be seeded even by errors in the cell coordinates at the limit of floating point precision. In \cite{suth03b} it was seen that by defining the cell coordinates in one part of the code additively, by adding successive deltas, and elsewhere multipicatively by setting a multiples of the grid delta, this resulted in a pattern of cell boundary differences at the level of the least significant bits of precision in use.

Eliminating this grid coordinate `noise' results in a code such that, if an essentially 1D simulation, i.e. a shock aligned with the grid perfectly, is set up on a 2D grid, then the simulation will remain 1D indefinitely, and hence the striping cannot appear, as that would break the 1D symmetry.  This does not mean that the striping instability is not present, it is just a special case where symmetry prevents it being triggered. The Linksa \& Wendroff Odd-Even decoupling test is unfortunately just such an ideal 1D test performed on a 2D grid. As \Fyris\ is designed to allow 1D symmetry to persist indefinitely -- even without any additional dissipation or correction scheme being added.

The initial conditions are from the 1D Blast test, but set on a 2D $n = 800\times10$ cell  grid.  Fyris is completely free of striping in this (special) case. The fact that the other similar codes do produce striping in this test is likely due to their coordinate schemes not being perfectly consistent at all stages of their calculations.

Here the test is run with standard \Fyris\ settings, including the standard striping correction code on (described in the following section), the default, and then with the striping correction code turned off, which is non-standard. The results are identical, with no striping present.

\begin{figure} [thdp] \begin{center}
\includegraphics[width = 5in]{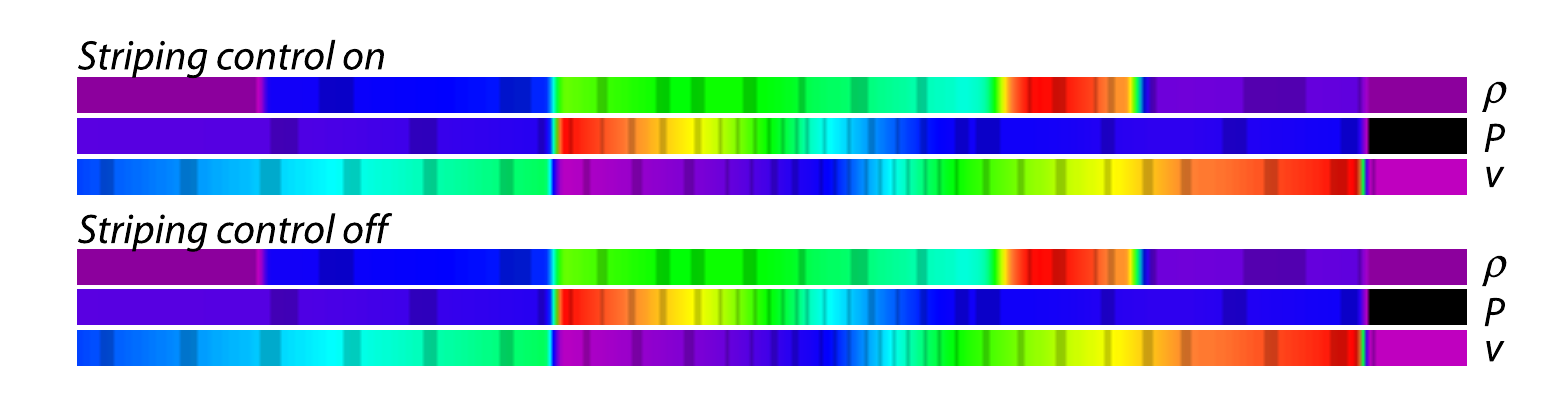}
\caption{\small Odd-Even decoupling problem at $t = 0.038$. Grid shown over $x$-cells 350-750.   Colour scale (black)-(red), linear, auto scaled. }
\label{f:oddeven}
 \end{center} \end{figure} 

\clearpage 

\section{ New 2D and 3D Tests }

\subsection{ 2D Severe Striping Test }
\label{s:stripingii}
 
In order to test the effectiveness of any striping correction in Fyris, a test is needed where perfect 1D symmetry is not initially present.  To this end, a modified adiabatic version of the test used in Sutherland et al (2003) is described, with the aim  to be easier to implement over a wider range of codes, and perform a more rigorous test of any striping corrections.

\Fyris\ uses an enhanced version of striping control to that described in \cite{suth03b}.  Like that earlier method, a viscous dissipation is introduced in cells marked by the flattening algorithm as containing shocks in the preceding orthogonal sweep directions.  The differences here lie in how the post-sweep smoothing is computed, and an additional flux limiting is applied by constraining lateral  the cell boundary motions in the shock fronts if they are orthogonal to the grid.  This proves effective, is 
effectively an additional dampening term, and is somewhat justified as since there should be zero orthogonal fluxes in an ideal shock normal to the grid (as opposed to an oblique shock).  Limiting motion of the cells in sweeps along the face of a shock front is essentially a mass flux limiting process that works against the buildup of the strong alternating density (and pressure) stripes that form perpendicular to the shock front in the carbuncle instability.  If a cell and its neighbours are determined to be in a shock in one sweep, then cell boundary movement is limited in subsequent orthogonal sweeps  -- helping to  suppress the initiation of the striping error.

The second main difference from \cite{suth03b} is that  a more formal velocity divergence is used to scale the smoothing, and only the  velocity is smoothed in the modified remapping process, where \cite{suth03b} applied smoothing to density and pressure as well.

A smoothing parameter, $\eta$, is based on the compressive divergence of the normalized multi-dimensional velocity field.
\begin{equation}
 \eta = \alpha [ -\frac{4}{3} min(\Delta v, 0.0) ]^2 \, \mbox{\rm  for}\,  \Delta v < 0 \, ,
\end{equation}
where $\Delta v$ is the normalized  multi--dimensional velocity gradient per cell, and $\alpha$ is a fractional constant $0 < \alpha < 1$.   The exact value of $\alpha$ for effective striping control is ideally determined on a case by case basis, but in general higher values are needed for higher mach number shocks.  In practise a standard value of $\alpha = 0.001$ is used for a very wide range of simulations, and values of up to 0.05-0.1 are only needed for the most extreme cases (such as the Noh Problem).  

In cells that have flattened parabolae from previous orthogonal sweeps, $eta$ is used to smooth the
velocity variable in the direction of the current sweep during the remapping step.  Pressure and density are not smoothed, unlike \cite{suth03b}.  The remapped velocity in a cell is the weighted average of it and it's neighbouring cells:
\begin{equation}
v_{n+1}[i] = (1 - \eta) v_{n}[i] + 0.5 \eta v_{n}[i-1] + 0.5 \eta v_{n}[i+1]\, ,
\end{equation}
where $\eta << 1$.
 
Finally, again in cells where flattening has occurred in previous sweep directions and are identified as occurring in shock front cells, any lateral velocities of the contact discontinuities that arise from the Riemann solver are suppressed - effectively freezing cell boundary motions orthogonal to shock fronts.  This helps to suppress the initiation of the striping error.

Severe Striping Test Parameters:
{\small 
\begin{itemize}
\item Adiabatic index: $\gamma = 5/3$, Mean Molecular Weight: $\mu = 1.0$. Adiabatic.
\item Grid Domain: $0.0 < x < 2.0$, $0.0 < y < 1.0$.
\item The left and right, $x-$, boundaries are constant, $y-$ boundaries are periodic.
\item Spatial Resolution: $h = 144$, 288 $x-$cells, 144 $y-$cells.
\item All cells with central coordinates $< x_0$ are set to the Left State, all cells with central coordinates $> x_0$ are set to the Right State ($x_0$ lies on an exact cell boundary). 
\item Left State: $P = 1.0$, $\rho = 1.0$, $v = 15.0$.
\item Right State: $P = 300.0$, $\rho = 300.0$, $v = 0.0$.
\item Velocity Perturbation on Left State: $\delta v = 0.1$ added to left and right state velocities on some grid rows as follows: \begin{itemize} 
								\item Rows 0, 72: $v = v + \delta v$, 
								\item Rows 1, 73: $v = v + 0.5*\delta v$, 
								\item Rows 71, 143: $v = v + 0.5*\delta v$.
							\end{itemize}
\item Initial Left-Right Boundary: $x = 1.8$.
\item The tests are run from time $t = 0.0$ until $t = 0.3$. 
\end{itemize}
}

\begin{figure} [thdp] \begin{center}
\includegraphics[width = 5in]{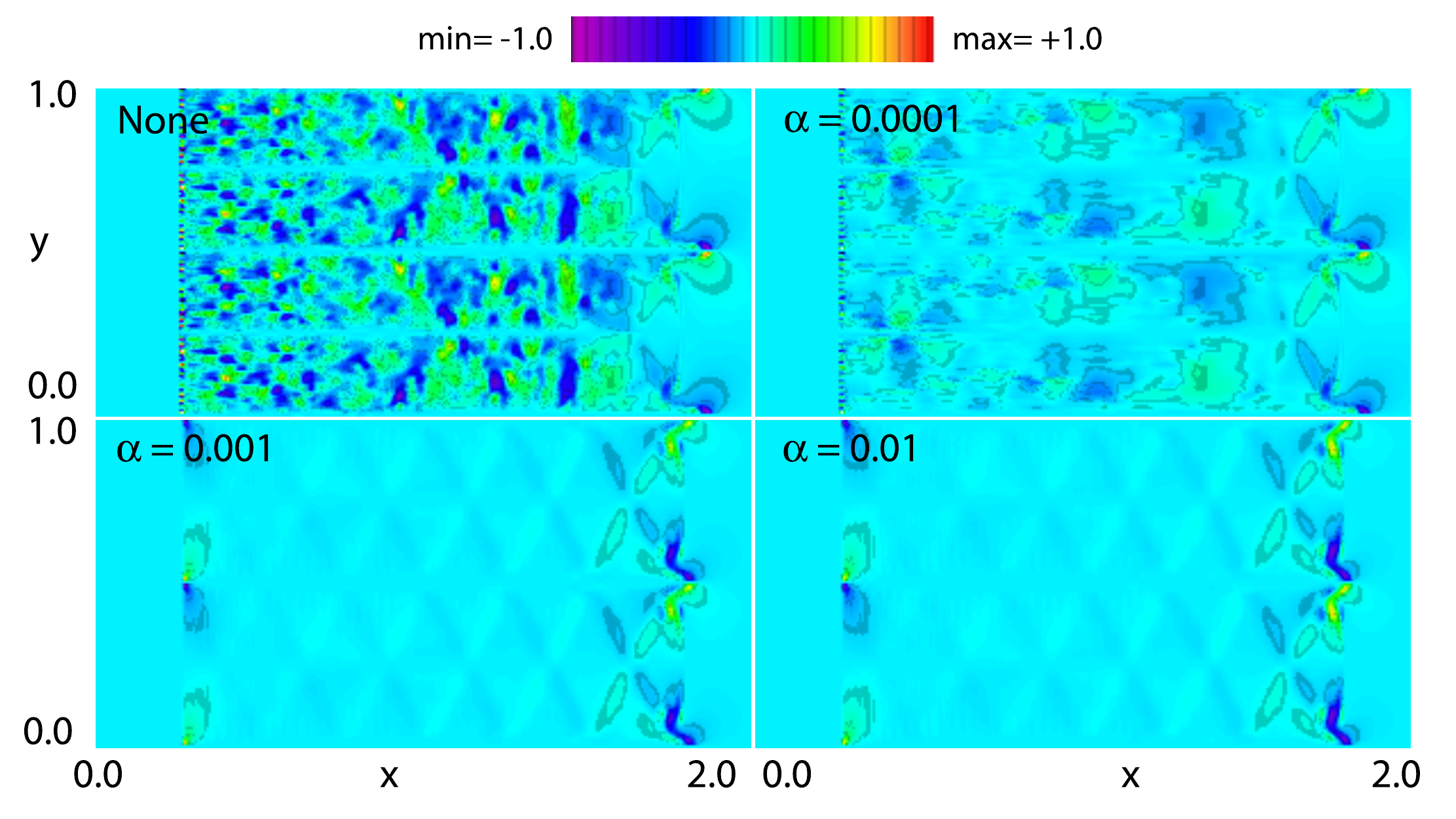}
\caption{\small 2D Striping problem at $t = 0.3$. Density.  Colour scale (black)-(red), linear, 0.0 -- 4.25, (white) > 4.25. }
\label{f:strdens}
 \end{center} \end{figure}

The test is an isothermal mach 15 reflected wall shock in 2D , were some rows of the inflowing gas have velocity variations, $\delta v$.  This results in small sections of the shock front being modified, giving a very slight curvature to the shock front.  This curvature is sufficient to break 1D symmetry, and in the absence of any striping correction, the striping instability sets in very rapidly.  
The test is run without striping correction, and with varying levels of $\alpha$ the diffusion coefficient.  Figure \ref{f:strdens} shows the density variable at $t = 0.3$.  Note that compared to $\alpha = 0.075$ used in \cite{suth03b}, effective suppression is achieved here with $\alpha$ as low as $0.001$. 

\begin{figure} [thdp] \begin{center}
\includegraphics[width = 5in]{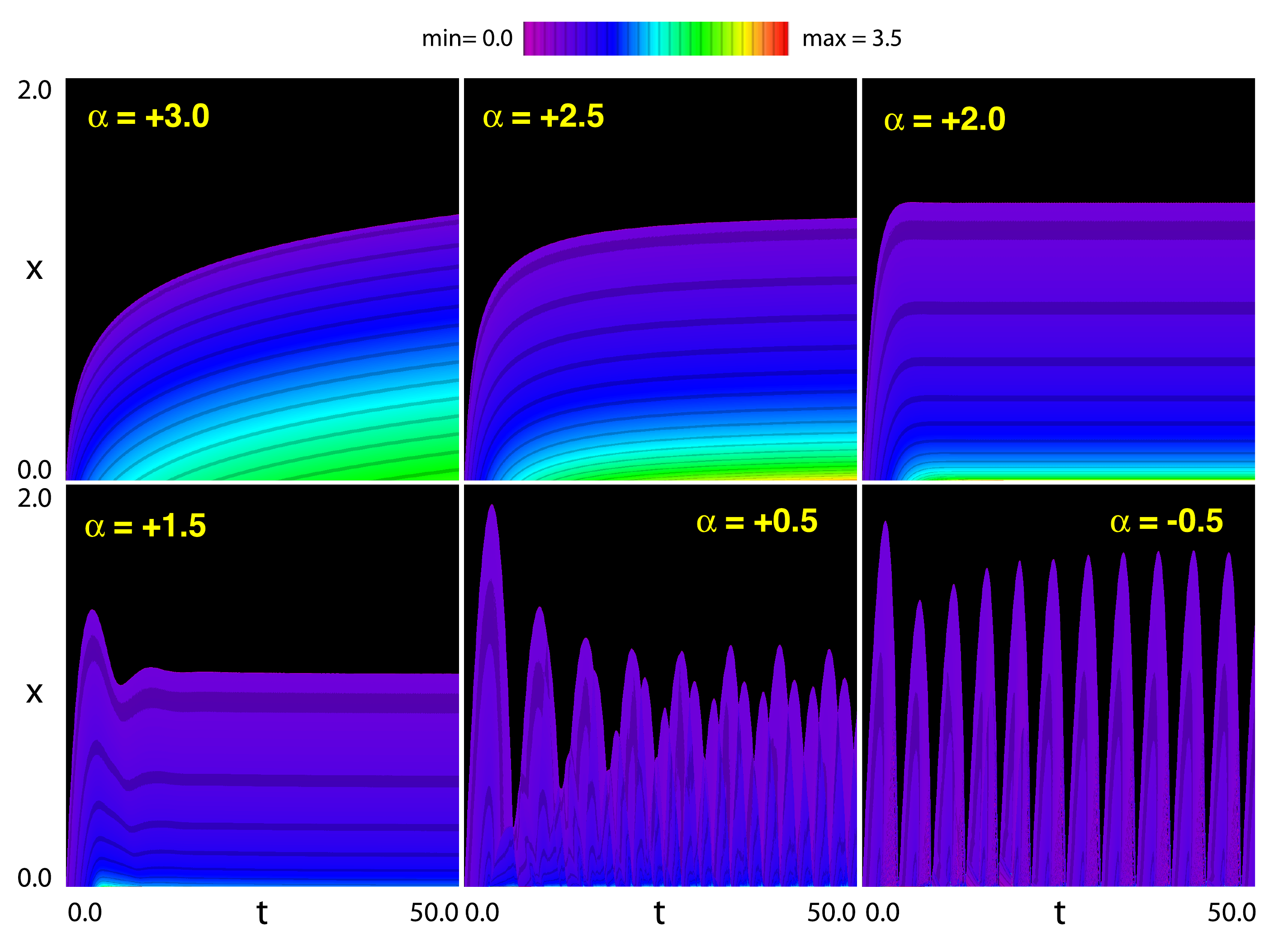}
\caption{\small 2D Striping problem at $t = 0.3$. $y$-velocity.  Colour scale (black)-(red), linear, -1.0 -- 1.0. }
\label{f:strvely}
 \end{center} \end{figure}

Figure \ref{f:strvely} shows how, despite the velocity smoothing in the shock front cells, even with very strong smoothing, the subtle velocity structure in the post--shock region is retained. As the smoothing and flux-limiting is only applied in the spatially localised shock front regions, even when $\alpha$ is set to high values the overall impact of the smoothing on the global simulation outcome is minimal, making the results insensitive to the value of $\alpha$ chosen, once it is sufficient to suppress the striping.  As standard, \Fyris\ uses $\alpha = 0.001$, and has been successfully used up to $\alpha = 0.05$ for the mach 1000 2D Noh problem, yielding results in agreement with the analytical solution even in that severe case.

\clearpage

\subsection{ One Dimensional Power--law Cooling Wall--Shocks }
\label{s:wallshock}

Many astrophysical fluid dynamics simulations have been performed where the cooling of the plasmas
involved is not dynamically negligible (for example \cite{suth03a}).  These non-adiabatic simulations often give
rise to the amplification of density structure. The emission implied by the cooling plasma often constitutes
the most direct observable quantities.  Unfortunately the typical astrophysical cooling function is highly non-linear, 
and dependent of the detailed composition of the plasma, as well as the thermal history of the gas.

Tests to confirm the correct cooling behaviour of such radiative hydrodynamic codes have not be available, in 
the absence of any analytical solutions to compare with.

Here we take a simplifying approach to test the code behaviour under the assumption of pure power--law
cooling, in the expectation that an analytical solution may be obtained for comparison with simulation.

We take a cooling function of temperature (T),  $\Lambda(T) = \Lambda_0 (T/10^6)^{\alpha}$, so that the volume cooling is $n^2 \Lambda(T)$, where
$n$ is the particle density here.  $\Lambda(T)$ has the c.g.s units: ergs~cm$^3$~s$^{-1}$.  \Fyris\ normalizes the cooling internally for numerical accuracy.

A key feature of cooling gas is the expected thermal stability or instability, depending on the cooling behaviour.  \cite{field65} first 
outlined a theory of thermal instability, and in the context of power--law cooling, an index $\alpha \ge 2.0$ should be stable.
If $\alpha < 2.0$, infinitesimal temperature difference may grow as cooling proceeds, making a stable solution unlikely.
\cite{chevalier82} looked at the power--law cooling in the case of the standoff-, or wall---shock, finding self-similar 
time-independent solutions for some values of $\alpha$.  \cite{blondin89} extended this work, and in 2003 Sutherland looked 
at thermal stability in two dimensional shocks, for power--law and astrophysical cooling.

We consider $\alpha$ over a range that are expected to give steady and non-steady solutions, and then focus on steady cases
to confirm that the density profiles agree with the analytical solutions.

\subsubsection{Analytical Density Structures}
Analytical expressions for the self-similar structure of standoff--, or wall--shocks, in the
presence of powerlaw cooling have been determined by \cite{chevalier82}
and \cite{blondin89}.  Here we want to compare the structure, in particular  the
density, obtained by \Fyris\ using a single powerlaw cooling function, with the 
analytical results for time--independent solutions.  

Following \cite{blondin89}, we take the normalised density 
$z = \rho /\rho_0$, with $\rho_0$ the preshock density.
Also, a normalised $x$ coordinate $\chi = x/\lambda_c$, where
$\lambda_c$ is the cooling length, or standoff distance, from 
the shock, $\chi = 0$, to the wall at $\chi = 1$.  
Here we transform to the alternative $x-$coordinate used by \cite{chevalier82},
$\zeta = 1 - \chi$. This makes comparison with the test simulations
easier, since the functions tend  towards very powerlaw like behavior with
plots of $z$ versus the distance from the wall, rather than the shock.

From \cite{blondin89}, the differential relationship between $z$ and $\chi$ is,
\begin{equation}
 \frac{dz }{d\chi} =  2^{-(1+\alpha)} 
 \frac{(\gamma -1)^{3-\alpha}}{(\gamma+1)^{3-2\alpha}}
 \left[ \frac{z^{5-2\alpha}(z-1)^\alpha}{\gamma z - (\gamma + 1)}\right] \, .
\end{equation}
which has been integrated here for values of $\alpha$ not given in the
earlier works, specifically for values that are potentially stable and time
independent.

Integration constants were determined 
to match $z = 4$ at $\chi = 0$ and so that $z$ tends to $\infty$ as 
$\chi$ approaches $0$. For $\gamma = 5/3$, the solutions for 
$\zeta(z) = 1 - \chi(z)$, for $\alpha = 1.5$, $2.0$, and $2.5$ 
for $z = 4 ... \infty$, are given by:
{\small
\begin{itemize}
\item $\alpha = 3/2$
\begin{equation}
\zeta(z) = 1 - \frac{9\sqrt{3}}{7 \pi \sqrt{3}-36}
 \left[
 \frac{14 z( \sqrt{z-1} \arctan(\sqrt{z-1}) + 1 )- 8}{3 z \sqrt{z-1}}
 -\frac{14 \pi \sqrt{3} + 36 }{9\sqrt{3}}
 \right] \; ,
 \label{eq:a32}
\end{equation} 
A powerlaw fit of $z(\zeta)$ over $0.1<\zeta<1.0$ is $z(\zeta) \propto \zeta^{-0.669}$
\item $\alpha = 2$
\begin{equation}
\zeta(z) = 1 - \left(8 \ln\left[4/3\right] - 1 \right)^{-1} 
           \left(
           8 \ln \left[\frac{4(z-1)}{3z}\right]
           +\frac{3}{(z-1)} - 1 
           \right) \; ,
 \label{eq:a2}
\end{equation}
A powerlaw fit of $z(\zeta)$ over $0.1<\zeta<1.0$ is $z(\zeta) \propto \zeta^{-0.985}$
\item $\alpha = 5/2 $
\begin{equation}
\zeta(z) = 1-\frac{3\,\sqrt{3}\, [ 5(z-1) - 1]}{14(z-1)^{3/2}} \; ,
 \label{eq:a52}
\end{equation}
A powerlaw fit of $z(\zeta)$ over $0.1<\zeta<1.0$ is $z(\zeta) \propto \zeta^{-1.946}$
\end{itemize}
}

For each solution, a powerlaw least-squares fit to $z(\zeta)$ between $0.01 < \zeta < 1.0$
 was determined to compare with the test data, and to compute $L1-$norm errors.
 
\subsubsection{Wallshock Test Models}
 
 This test is set up with the values of the key variables normalised from input
 values that are from a regime of interest to astrophysical shocks, in c.g.s. units. 
The c.g.s. values for the problem are given in table \ref{t:wshock}.  The cooling
  is also defined in c.g.s units, to test the normalisation.
 
Using a characteristic length scale of $x_0 = 10^{17}\,$cm, a velocity 
scale, $v_0 = 10^7\,$cm/s, and a density scale of $10^{-24}\,$g/cm$^3$, the 
normalised units are also given in table \ref{t:wshock}.  
The polytropic index, $\gamma = 5/3$ throughout, and the mean molecular weight $\mu = 0.6$ was
held constant as well.
The grid domain is $2.0 > x > 0.0$, or $2\times 10^{17}\,$cm, and the cooling rate
was roughly adjusted to ensure the cooling  length was approximately 1.0 units, $10^{17}\,$cm
for each value of $\alpha$, the power-law index.
Six values of $\alpha$ were used, covering the range $3>\alpha>-1$, ranging from 
slope too steep to cool in a finite time $\alpha = 3$, slopes 
expected to be stable, or nearly so, $\alpha = 2.5, 2.0, 1.5$, and slopes expected to be
unstable or over-stable, $\alpha = 0.5, -0.5$. 

\begin{table}[htdp]
\caption{Wall--shock Cooling Values. \label{t:wshock}}
\begin{tabular}{ r l r l }
$\alpha$ & $\Lambda_0$ (c.g.s) &$\alpha$ & $\Lambda_0$ (c.g.s)\\
\hline
\hline
3.0 & 1.00E-20 & 1.5 & 3.00E-21 \\
2.5 & 5.00E-21 & 0.5 & 3.00E-21 \\
2.0 & 3.00E-21 & -0.5 &  5.00E-21\\
\hline
\end{tabular}
\end{table}%

\begin{figure}[thdp] \begin{center}
\includegraphics[width = 5in]{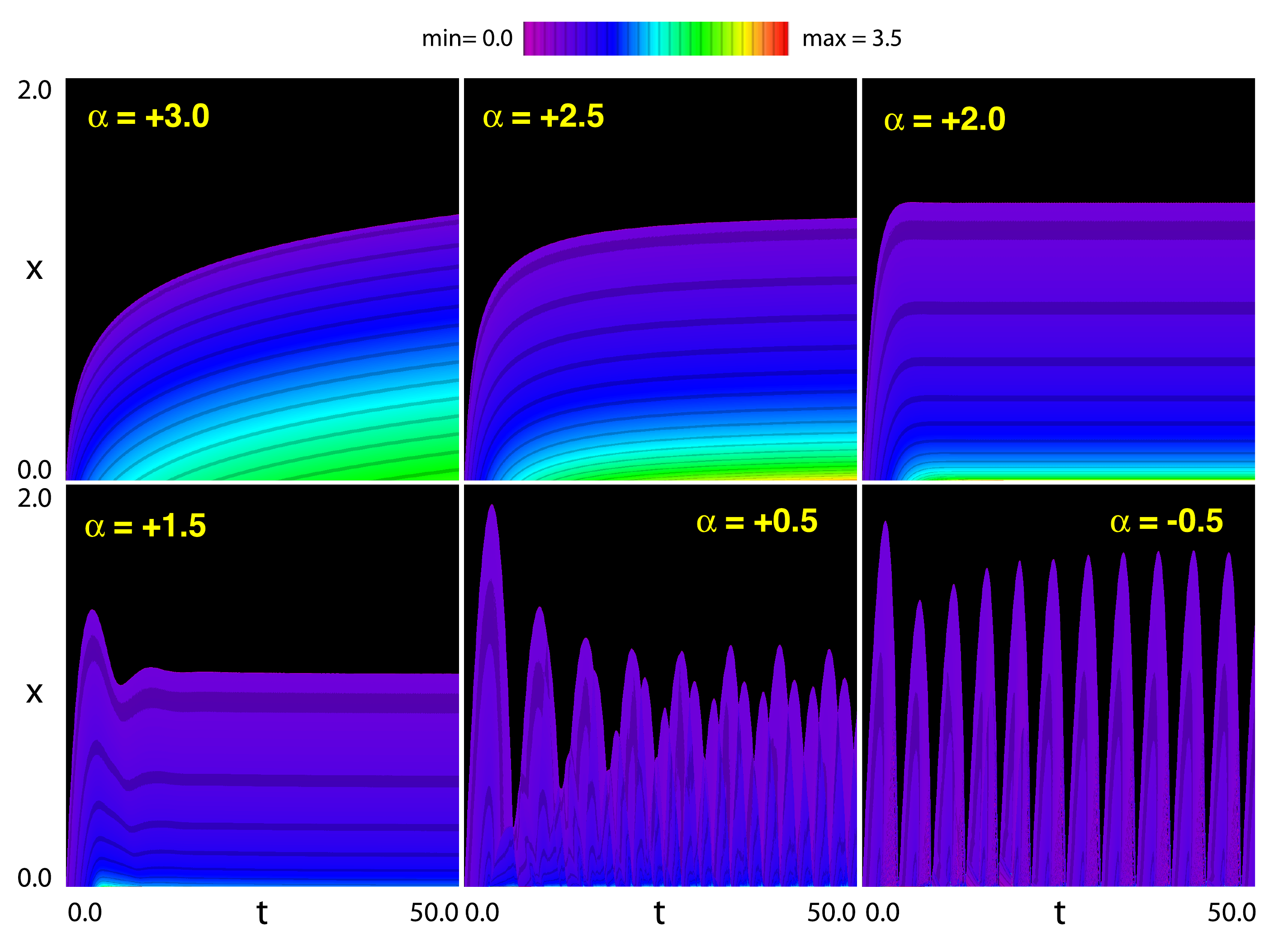}
\caption{\small The logarithm of the density, $\rho$, as $x - t$ diagrams.  The vertical axis represents the spatial dimension, over a domain $0.0 < x < 2.0$ ($2\times 10^{17}$cm).  The horizontal axis represents time, $0.0 < t < 50.0$ ($5\times 10^{11}\,$s).  The standard color table is used with black$ = 0.0$ and red/white$ = 1.0e4$ in density.
Green represents approximately $10^3$, Cyan $~10^2$ and blue $10$ density units.  In each panel the index of the cooling function powerlaw is given. }
\label{f:alphaxt}
 \end{center} \end{figure}

The logarithm of the density is displayed in an $x-t$ diagram  where each pixel column is the density at one output time step from the simulation, one every 0.05 time units,($5\times 10^9\,$s), up to $t = 50.0$ ($5\times 10^11\,$s) .  The resulting $x-t$ diagrams for each of the $\alpha$ values are shown as panels in figure \ref{f:alphaxt}.  Note the well defined, stable, cooling length realised for $\alpha = +2.0$.  Slopes less than $+2.0$ are not steady, and slopes greater than $+2.0$ do not show oscillations.  The $\alpha = 3.0$ case shows linearly spaced contours in the $\log$ density display, indicating a logarithmic density profile.  Formally the cooling time is infinite for $\alpha = 3.0$.

\begin{figure}[thdp] \begin{center}
\includegraphics[width = 5in]{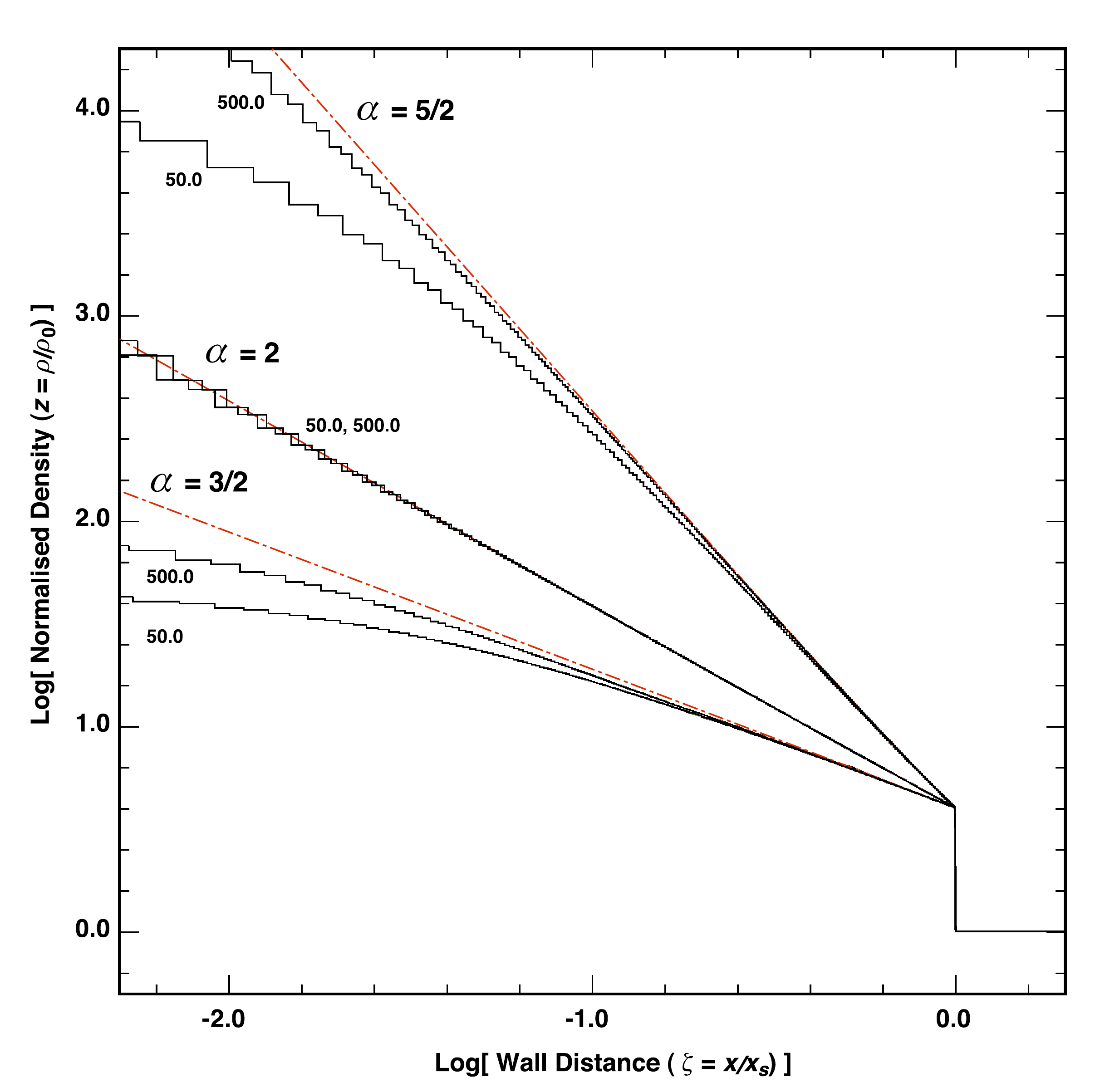}
\caption{\small Comparison of density profiles for $\alpha = +2.5$, $2.0$, and $1.5$.  Red dashed lines represent the analytical solutions, equations \ref{eq:a32} -- \ref{eq:a52}.  The stepped black curves show the density variables, where each cell is represented by a step.  
 The steady $\alpha = 2.0$ case matches the analytical equations very well, reaching equilibrium rapidly.  The two other cases take longer to reach equilibrium.  The density profiles for $t = 50.0$, and $t = 500.0$ tend towards the analytical solutions.}
\label{f:alpha}
 \end{center} \end{figure}

The normalised density, $z = \rho/\rho_0$, vs $\zeta$ the normalised distance from the wall, 
for $\alpha = 2.5, 2.0$ and $1.5$ is shown in figure \ref{f:alpha}, at $t = 50.0$ and $500.0$,  overlaid with the analytical solutions.
Over the range $0.1 < \zeta < 1.0$ the agreement is excellent for all the $\alpha$s, and the deviations from the analytical solutions for $0.01 < \zeta < 0.1$ are decreasing as $t$ is allowed to extend from $t = 50.0$ to $t =500.0$, suggesting that the time required to reach equilibrium is very long for 
$\alpha = 2.5$ and $\alpha = 1.5$.  The $\alpha = 2.0$ simulations agree very well with the analytical solution showing it has reached equilibrium rapidly.  The $\alpha = 2.0$ case is recommended as a test for cooling codes because of it's rapid convergence and stability.

The density profiles at $t = 500.0$ were fit with a least--squares power law over the range $0.1 < \zeta < 1.0$, and the L1-norm error with respect to the semi-analytical power--laws given by fits to equations \ref{eq:a32} - \ref{eq:a52}.  The fits and L1 errors are given in table \ref{t:alphafit}.

\begin{table}[htdp]
\caption{Wall--Shock Density Fits and Errors. \label{t:alphafit}}
\begin{tabular}{r l l }
$\alpha$& powerlaw fit & L1 Error\\
\hline
\hline
1.5	&$4.048 \zeta^{-0.669}$&	2.31\%\\
2.0	&$3.981 \zeta^{-0.985}$&	0.287\%\\
2.5	&$3.660 \zeta^{-1.983}$&	4.76\%\\
\hline
\end{tabular}
\end{table}%

\clearpage

\subsection{ Multi-Dimensional Noh Series }

Here a grid with $h$ = 144 is used to perform a consistent series of Noh shock problems in 1D, 2D and 3D.  The domain extends from $0 - 1.0$ in each dimension with reflecting boundaries on all the inner boundaries, $x = 0.0$, $y = 0.0$, and $z = 0.0$.   The difference here compared the 1D and 2D cases to the previous tests is the uniform use of reflecting boundary conditions, including the 1D case, and a resolution low enough so that a 3D model with the same resolution remains small enough to be easily computed on a wide range of facilities, the other parameters; initial density, pressure and velocities remain the same as before.

This test is designed to confirm that the split sweep method in three dimensions remains capable of solving the extreme shocks present, and that the anti-striping algorithm is working in 3D.   The edges are where the most extreme carbuncle instability occurs, being exactly orthogonal to the other grid axes.  In 3D this is the largest density error, even with the striping control in place, and although it is locally about a 10 \% overshoot the colour table in panel C highlights it.    The origin (left vertical edge in A, lower left corner in the other panels),  still retains the central  density dip seen in the original LW03 Noh tests above, and is a remnant of the early cycles which  persists  in the absence of any internal pressure gradients.

The one and two--dimensional analytical solutions have already been given above, in tables  \ref{t:noh1dsol} and \ref{t:noh2dsol}.  The three--dimensional analytical solution is given below in Table \ref{t:noh3dsol}.

\begin{table}
{\small 
\caption { 3D Noh Solution, $\gamma = 5/3$. \label{t:noh3dsol}}
\begin{tabular}{l l l l }
\hline
 \multicolumn{1}{c}{Inside Shock  ($r < t/3$)} &  \multicolumn{1}{c}{Outside Shock  ($r > t/3$)}  \\
\hline
\hline
Density $\rho_0=  64 $  &  Density $\rho_1 = (1.0 +t/r)^2$     \\
Pressure $P_0 = 64/3$ &  Pressure $P_1 = 0.0 (1.0e-6)$             \\
Velocity $v_0 = 0$        &  Radial Velocity $v(r) = -1.0$    \\
 \multicolumn{2}{c}{Shock Front Expands at $v_s = 1/3$} \\
\hline
\end{tabular}
} 
\end{table}

\begin{figure}[thdp] \begin{center}
\includegraphics[width = 5in]{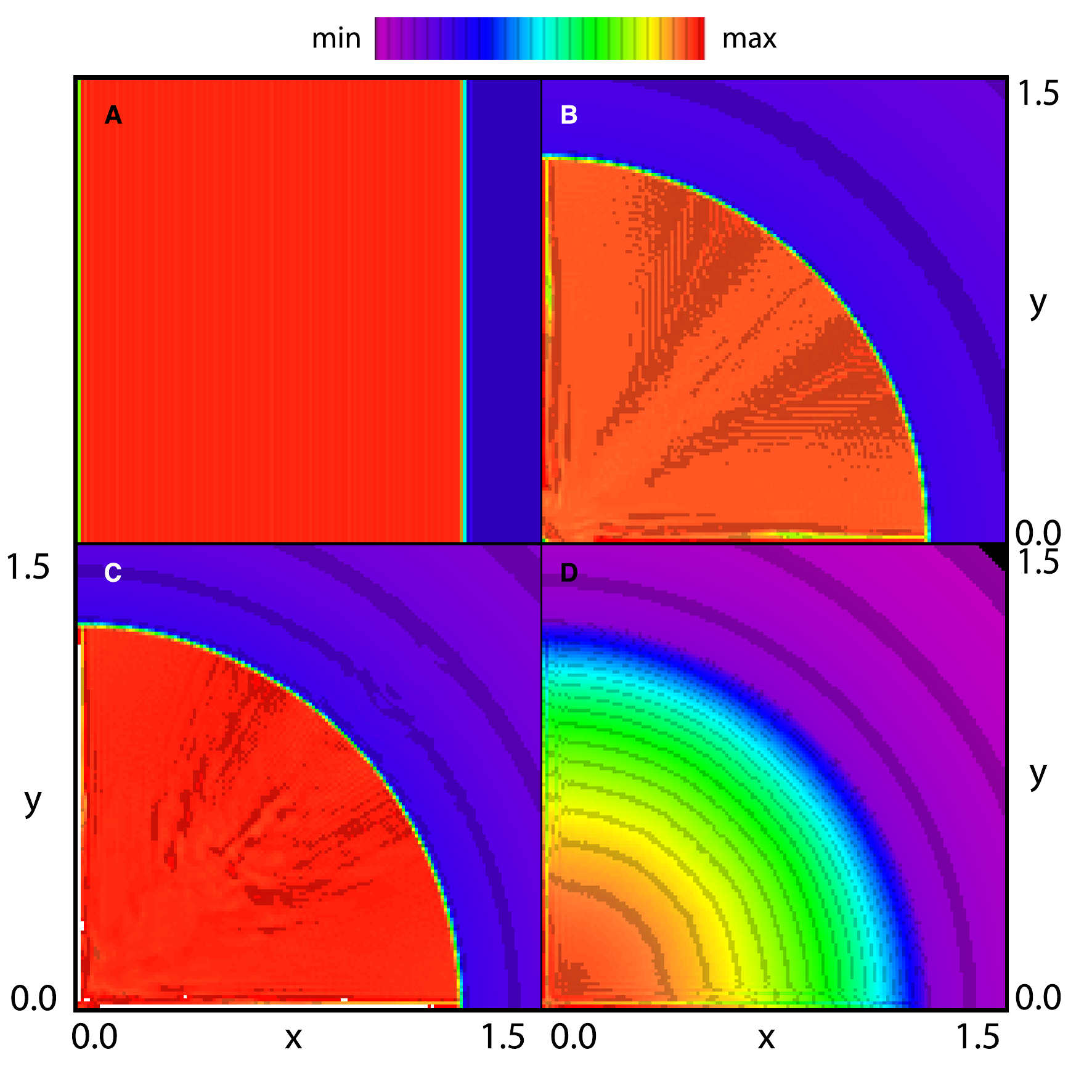}
\caption{\small A: the 1D noh solution at $t = 2.5$, with the $y-$axis stretched to make the 1D structure visible.  The colour maximum white = 4.5. 
B: The 2D solution, colour maximum white = 17.0. C: A 2D slice of the 3D solution on the 10th slice at $k = 9$, colour maximum white = 66.0 copared to the analycal value of 64.0 for the 3D case. D: The summed 3D density, autoscaled, the contours show the overall spherical shape is maintained, albeit with some striping persisting at the axis edges.  }
\label{f:nohseries}
 \end{center} \end{figure} 

As shown, \Fyris\ solves the three versions of the Noh problem, and is stable in the face of the strong striping instability.  The pressure jumps in the shocks exceed a factor of $10^7$, providing a stringent test of the convergence of the code to the correct solution.  The $L1-$norm errors in density and pressure with respect to the analytical solutions at $t=2.5$ are:
\begin{itemize}
\item 1D: $L1-$error Density: 0.76\%, $L1-$error Pressure: 0.45\%.
\item 2D: $L1-$error Density: 1.12\%, $L1-$error Pressure: 1.21\%.
\item 3D: $L1-$error Density: 2.07\%, $L1-$error Pressure: 2.90\%.
\end{itemize}

The errors increase with increasing dimension, however the fact that the pressure error remains small suggests the two-shock solver is working well and the shock velocity is remaining accurate, as any over or under shoot in the shock location introduces very large differences when compared to the analytical solution.

\clearpage

\subsection{ 2D and 3D Constant Energy Bubbles }

The case of a constant  energy driven bubble is a useful multi-dimensional test as it has an analytical solution (\cite{avedisova72,weaver77,koo92a, koo92b}).  
Self-similar structures as a function of radius are known for two and three dimensional cases.  It is to be hoped that a simulation
will tend to  the self-similar structure at late times when $R_{\rm bubble} >> R_{\rm initial}$, and that is tested here.

A circular (2D) or spherical (3D) fast, light wind is driven out from a point source, impacting on a uniform ambient medium.
Taking the wind kinetic energy  $E = \frac{1}{2} \dot m v$, ( $\dot m$ is the mass loss rate, $v$ is the wind velocity), and an
ambient density $\rho$, a combination of these that gives a similarity relationship for radius $r$ and time $t$ can be formed for the two-- and three--dimensional cases, using simple dimensional analysis:
\begin{itemize}
\item 2D The density has the dimensions of mass per unit area, giving the combination:
$$r_{\rm 2D} \lambda = \left( \frac{\dot m v^2}{2 \rho_{\rm 2D}} \right)^{1/4} t^{3/4}$$.
\item 3D The density has the dimensions of mass per unit volume, giving the combination:
$$r_{\rm 3D} \lambda = \left( \frac{\dot m v^2}{2 \rho} \right)^{1/5} t^{3/5}$$.
\end{itemize}
Where, $\lambda$ is a dimensionless scaling factor.

In order to properly set the wind boundary condition with sufficient cells to properly map the divergent flow on a cartesian grid, the simulations must begin with a finite radius, $r_{\rm init}$, whereas the similarity solution can begin with an arbitrarily small radius.  Given a finite initial radius, it may be expected that the simulation may only tend towards a self-similar solution as the radius of the simulated bubble becomes much larger than the initial radius.

In three dimensions in particular, the need to achieve $r >> r_{\rm init}$ places considerable demands on the grid size, requiring a very large grid compared to the initial wind size, and therefore requiring a large amount of computations.

To help deal with this, many codes including \Fyris\ cane make use of grid refinement, to minimise the computation requirements.  Here a very fine mesh may be placed about the initial wind region, and a larger coarse grid can be used to compute the larger scale evolution. 

The constant energy bubble test here is thus designed to test both the energy conservation and bubble behaviour, and to test the ability of the code to compute a solution on a nested grid of varying resolutions. 

\subsubsection{Refined Grid Test Models}

A multi--level simulation with a top level resolution of $n = 144$ is set up with the parameters in table \ref{t:ebubbles}.  For each case, 2D and 3D, three simulations are computed with 1, 2 and 3 levels of grid refinement.  In the refinement, the inner $1/3$ of each level is mapped to a finer mesh with 3 times the resolution.  For example,in the three level case, each level is $144^3$ cells, at 1, 3 and 9 times the resolution of the top level, respectively.
So, the equivalent single level resolution would be $n = 144$, $n = 432$ and $n = 1296$.  In the three level 3D case, instead of $1296^3 = 2.176782336\times10^9$ cells, only $3 * 144^3 = 8.957952\times10^6$ cells are used.

\begin{table}[htdp]
{\small 
\caption{Energy Bubble Parameters. \label{t:ebubbles}}
\begin{tabular}{ l l  l l }
\hline
\multicolumn{2}{c}{EOS}&\multicolumn{2}{c}{Domain}\\
$\gamma$&$\mu$&Grid 2D& Grid 3D\\
\hline
\hline
$5/3$&$1.0$&$-1.0 < x < 1.0$ & $-1.0 < x < 1.0$\\
     &     &$-1.0 < y < 1.0$ & $-1.0 < y < 1.0$\\
     &     &                 & $-1.0 < z < 1.0$\\
\hline
\multicolumn{2}{c}{Wind Boundary}& \multicolumn{2}{c}{Wind Parameters}\\
\hline
2D Origin &$= (0.0,0.0)$      &  Density  &$=  0.1$ \\
3D Origin &$= (0.0,0.0,0.0)$  &  Pressure &$=  0.1$ \\
Radius    &  $= 0.02$         &  Velocity &$= 10.0$ \\
\hline
\multicolumn{2}{c}{Ambient Conditions} &\multicolumn{2}{c}{Ending Time} \\
\hline
Density  &$=  1.0$  & \multicolumn{2}{c}{2D $t = 0.6$} \\
Pressure &$=  0.01$ & \multicolumn{2}{c}{3D $t = 1.6$} \\
Velocity &$=  0.0$  & & \\
\hline
\end{tabular}
}
\end{table}%

Importantly, given the initial radius of the wind, which is the size of the inner wind boundary condition, it is not expected that the single level simulations will correctly sample the wind, having only a few cells in the region to represent a circular or spherical flow.  This may distort the low resolution bubbles, however the energy inflow will still be constant.  So, even if the inflow is under--sampled the bubble growth should still, on average, be a power-law of the same slope as a better resolved model, but most likely with a smaller radius because the poorly sampled wind will generally have a lower effective energy flux.

We measure the bubble radius as a function of time  for each model. In some cases the symmetry of the wind is poorly sampled, and the bubble is not very circular in cross--section, so an effective radius is used.  In 2D, we take the square root of the bubble area, found by counting cells with high pressure, and in 3D we take the cube root of the bubble volume, to derive the effective bubble radius.

\begin{figure}[thdp] \begin{center}
\includegraphics[width = 5in]{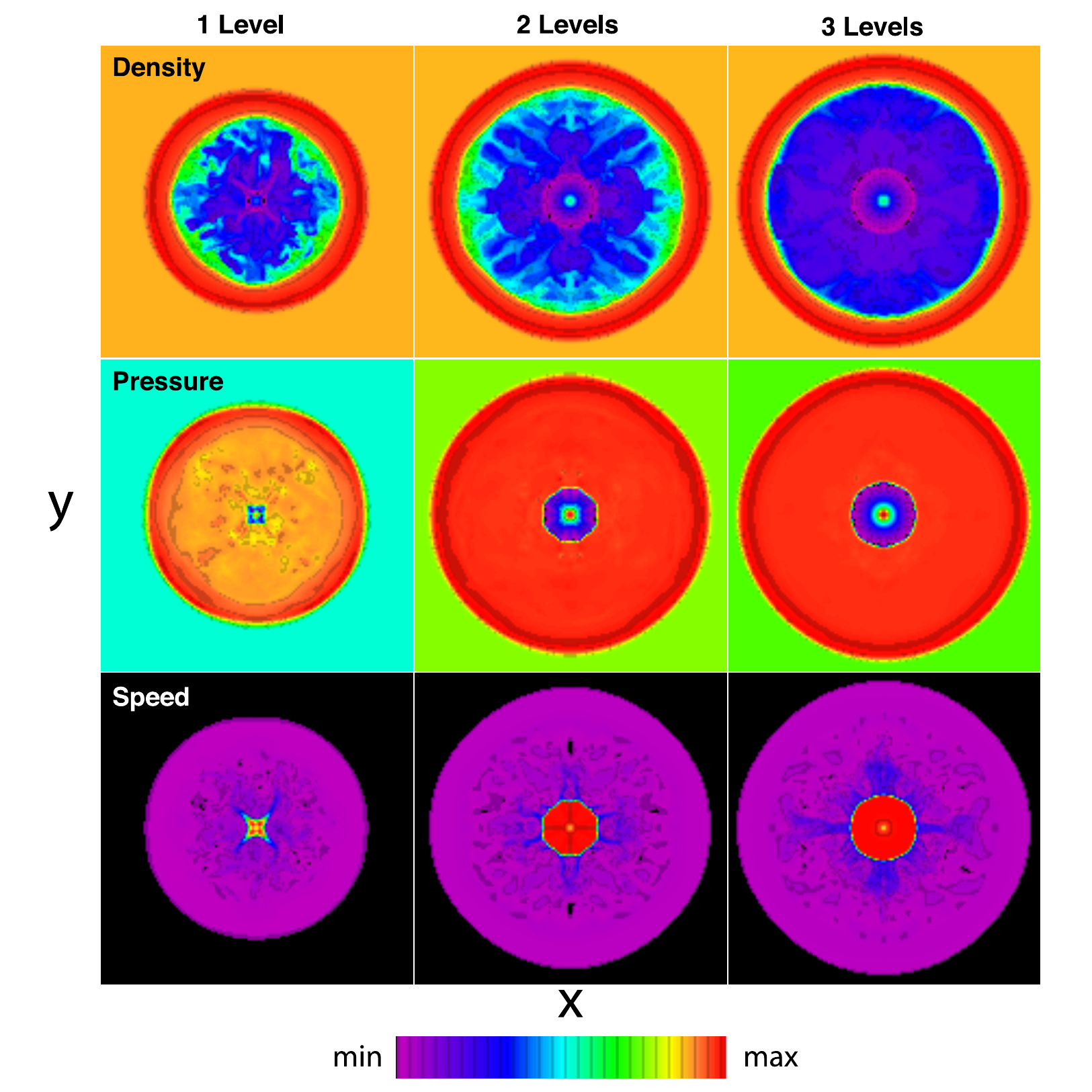}
\caption{\small 2D bubbles.  Autoscaled $\log_{10}$ images of the variables, Density, Pressure and magnitude of the velocity, Speed. In the 1 and 2 level cases there is evidence that the wind is quite non--circular, presumably owing to a poorly sampled wind regions,$r < 0.02$.  Only the 3 level model exhibits truly circular behaviour, however the speed image suggests that the circular wind remains inadequately resolved in that the speed structure is not smooth.  This may be the reason for the less than analytical radius--time power--law slopes exhibited by the 2D tests. }
\label{f:2debubbles}
 \end{center} \end{figure}

\begin{figure}[thdp] \begin{center}
\includegraphics[width = 5in]{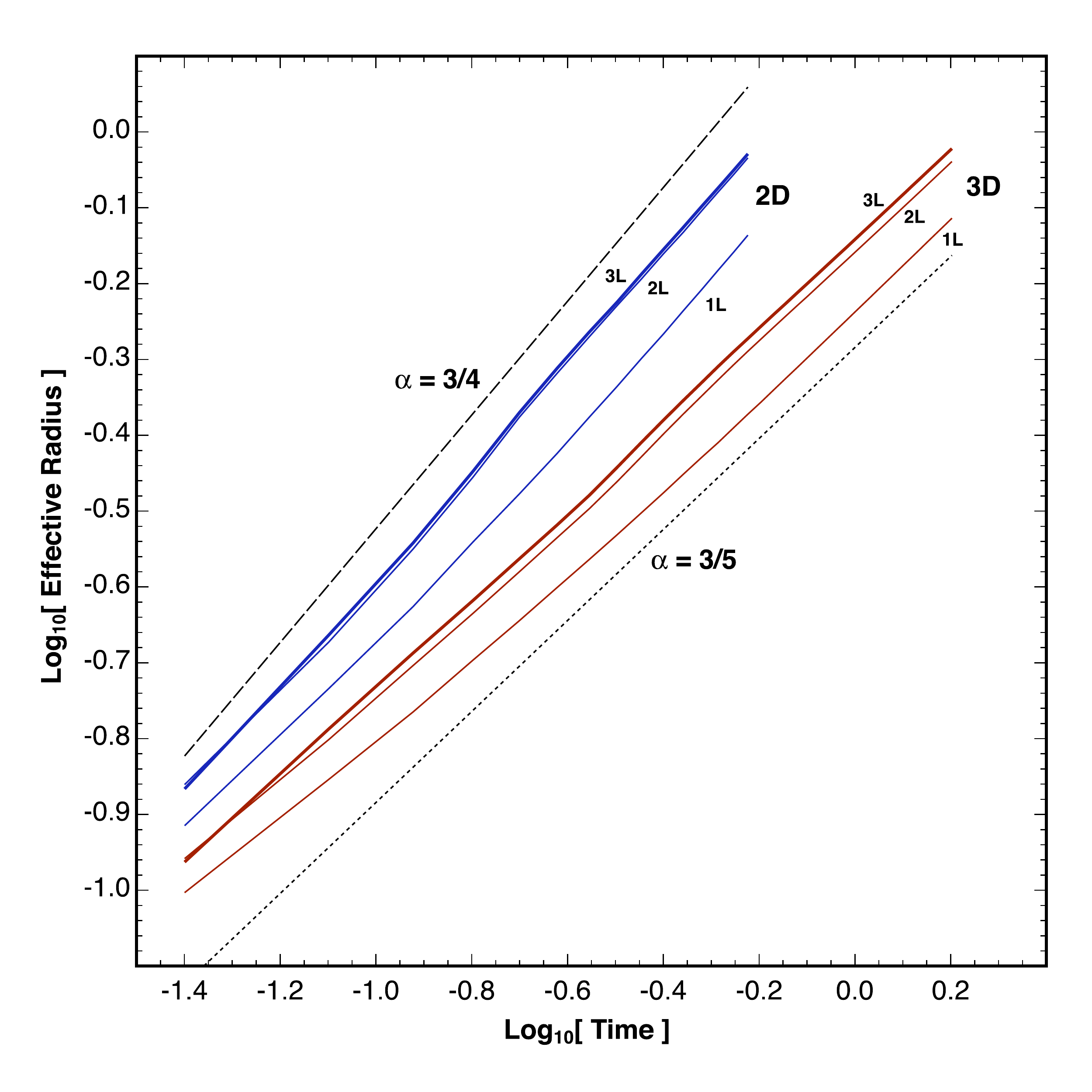}
\caption{\small 3D bubble, central z = 0.0 plane.  Autoscaled $\log_{10}$ images of the variables, Density, Pressure and magnitude of the velocity, Speed.  There is significant diffusion of mass interior to the contact discontinuity seen in the Density plot as cyan and green material in the otherwise blue-purple region.  If the models included cooling this could introduce significant cooling errors in the 1 level and 2 level cases.  In the pressure slices, the interior pressure is very even in the 2 and 3 level runs.  The inner wind shock is best represented in the 3 level model, as the spherical wind is best sampled and gives a good spherical flow. }
\label{f:3debubbles}
 \end{center} \end{figure}

The results for one, two and three level simulations at $t = 0.6$ for the two--dimensional models are shown in figure \ref{f:2debubbles}.
Similarly the three dimensional results at $t = 1.6$ are shown in figure \ref{f:3debubbles}.  The different ending time reflect the difference in the power--law index of the radius--time relationship, terminating each simulation as the bubble fills the top level grid.

\begin{figure}[thdp] \begin{center}
\includegraphics[width = 5in]{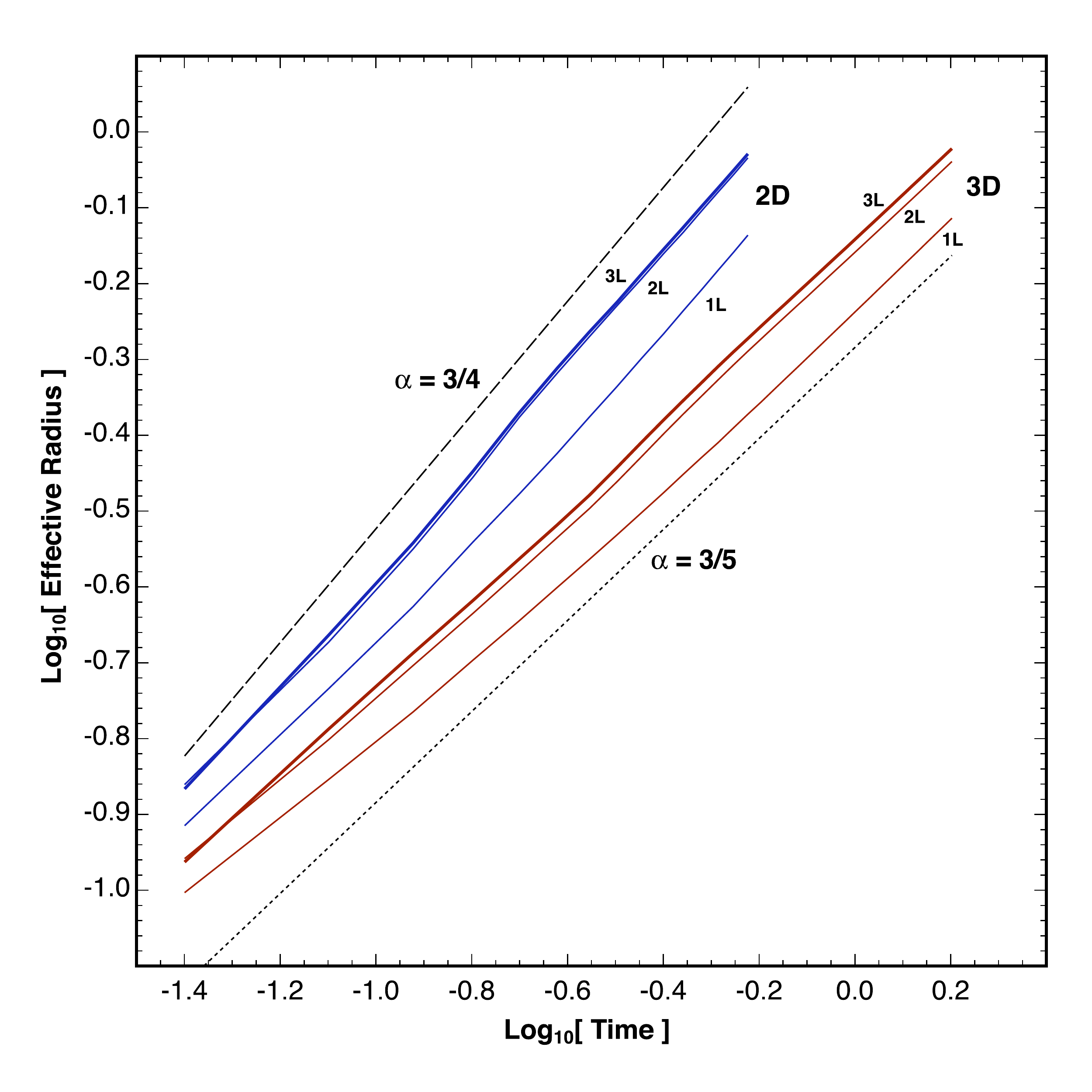}
\caption{\small Constant Energy Bubbles in 2D and 3D, Radius--Time relationships.  Dashed lines give the slopes of the similarity solutions, with their power--law indices marked.  One, two and three level models are marked 1L, 2L and 3L.  The red curves are for 3D models, blue curves are 2D.  The slopes do not stabilise until $\log_{10}[r] > 0.5$. }
\label{f:ebubbles}
 \end{center} \end{figure}

Finally the Radius -- Time relationships for all the tests are shown in figure \ref{f:ebubbles}.  At large radii the curves do tend towards the expected analytical slopes, showing that energy conservation and grid refinement are working, and that even under resolved bubbles are well behaved on average.
These bubbles grow initially a little slower than the similarity solution, when the bubble is comparable to the size of the initial wind boundary regions, $\log_{10}[r] < 0.5$, however, once the bubble is larger than this, the 2 and 3 level models in particular approach the analytical slope. The three--dimensional models finish with a slightly larger radius, suggesting that the wind luminosity has been integrated better with the well sampled wind boundary condition at $r<0.02$.  Fitting for $r>0.5$ gave a power--law of $r_{\rm 2D} = 1.34 t^{0.71}$, and $r_{\rm 3D} = 0.72 t^{0.59}$ for 2D and 3D respectively.  The 3D result is very close to the expected $0.6$ power index, whereas the 2D slope falls somewhat short of the expected $0.75$, and it may be that the 2D case is not well sampled on this small domain.  The accuracy of the 3D result on this small grid is however encouraging for more complex astrophysical simulations where bubbles are an important feature.

\clearpage

\section{ Summary and Conclusions }

\Fyris\ is a new implementation of an established hydrodynamics methodology.  The key new features are;
\begin{itemize}
\item The significant performance improvement, up to a factor of four times faster for the same adiabatic 2D calculations.
\item A new two-shock Riemann solver, based on the \cite{gg88} solver, which is accurate and faster than the equivalent two shock solver of \cite{cw84}.
\item Robust and accurate solutions to test problems demonstrate that \Fyris\ is competitive with other hydrodynamic codes, and in many cases produces better results with smaller errors.
\item Fast and accurate microphysics, including cooling and generalised equation of state allowing for a variable polytropic index, $\gamma$, and variable mean molecular weight $\mu$.
\item Active floating point `noise' control, controlling roundoff errors and floating point representation bit errors that would otherwise affect coordinates and allow non-physical changes to the fluid quantities.
\item Fully parallelized, both in shared memory using pre-emptive threading and in distributed mode using the MPI library ({\small \tt http://www.mpi-forum.org/docs/docs.html}), or a combination of both.
\end{itemize}

\subsection { Code Availability }

Source code, animations and figures for all the problems presented here, along with some additional qualitative problems are available at the website.
{\small \tt  http://www.mso.anu.edu.au/fyris/}

\subsection{ Acknowledgements}
\Fyris\ is approximately 84000 lines of C code, and was begun during a study visit to UAO, supported by the Centre for Complex Dynamics, at the \AA ngstr\"om Labotoriet, Uppsala University, Sweden, and the code is named after the Fyris river that flows through that town. The author would like to thank the Centre for the financial support during the visit.

\clearpage
\section*{ Appendix A: Effective $\gamma$ for Mixtures of Ideal Gasses}

Two co--volume fluids ideally each contribute a partial pressure, $P_0$, and $P_1$, to a total pressure:
\begin{equation}
P_{\rm tot} = P_0 + P_1 \, .
\end{equation}
The total internal energy, by sum of partial pressure contributions, is
\begin{equation}
E_{\rm tot}   =  \frac{1}{(\gamma_{0} -1)} P_{0} + \frac{1}{(\gamma_{1} -1)} P_{1} \, , 
\end{equation}
or with an effective total $\gamma_{\rm eff}$ and the total pressure:
\begin{equation}
E_{\rm tot}  = \frac{1}{(\gamma_{\rm eff} -1)} P_{\rm tot} \, .
\end{equation}

Taking $\beta = P_1/P_0$, $\gamma_{\rm eff}$ can be found, using $\gamma$ for each component,
\begin{equation}
\gamma_{\rm eff} = 1 + \left[ \frac{(1+\beta)(\gamma_0-1)(\gamma_1-1)} {\beta (\gamma_0-1) + (\gamma_1-1)} \right] \, .
\end{equation}

For example, $P_0$ is the thermal pressure for an ideal atomic gas: $\gamma_0 = 5/3$.  If $\beta = 0.5$ (ie $P_{\rm tot} = 1.5 P_0$, and  $\gamma_1 = 7/5$, from a molecular fluid, then $\gamma_{\rm eff} \sim 1.55$.
For $\beta = 8$, largely dominated by $\gamma_1 = 7/5$ gas, $\gamma_{\rm eff} \sim 1.42$, as shown in table \ref{t:gammaeff}.

\begin{table} [thdp]
\caption{ Ideal Gas Mixture $\gamma_{\rm 0} = 5/3 $, $\gamma_{\rm 1} = 7/5 $ \label{t:gammaeff} }
\begin{tabular}{ c c c c }
\hline
$\beta$&$\gamma_{\rm eff} $&$\beta$&$\gamma_{\rm eff} $\\
\hline
\hline
0.000&1.6667&1.000&1.5000\\
0.125&1.6207&2.000&1.4615\\
0.250&1.5882&4.000&1.4348\\
0.500&1.5455&8.000&1.4186\\
\hline
\end{tabular}
\end{table}%

\clearpage
\section*{ Appendix B: Multi-dimensional PPM Interpolation Coefficients on a Regular Grid.}

PPM coefficients, $a_6$, $\Delta$,  are generally defined in terms of cell average, $\langle a \rangle$, and cell boundary values, $a_L$, $a_R$, not cell centre, $a_C$, and cell boundary values, $a_L$, $a_R$, but they can be interchanged. In each case, $\Delta = (a_R - a_L)$.

The standard PPM formulae, for a given $langle a \rangle$, $a_L$, and $a_R$:
\begin{eqnarray}
a_6   & = & 6 [ \langle a \rangle - 0.5 (a_R + a_L)] \, , \mbox{ and }  \\
a_C   & = & \frac{1}{4}[ 6 \langle a \rangle - a_L - a_R].
\end{eqnarray}

The complementary PPM formulae, for a given $a_C$, $a_L$, and $a_R$:
\begin{eqnarray}
a_6   & = & 4 [ a_C - 0.5 (a_R + a_L)] \, , \mbox{ and }  \\
\langle a \rangle   & = & \frac{1}{6}[ 4 a_C + a_L + a_R].
\end{eqnarray}

When evaluating a function of spatial coordinates, $r$, on a regular grid, it is often convenient to evaluate the central value at the cell centre, $a_C = f(r_C)$, as well as the cell boundaries, $a_L = f(r_L)$, $a_R = f(r_R)$.  From these the cell average, $\langle a \rangle$, and the PPM coeficients $\Delta$, and $a_6$ can be derived and stored.  This is straightforward in 1D, evaluating the complementary formulae above.

\begin{figure}[thdp] \begin{center}
\includegraphics[width = 5in]{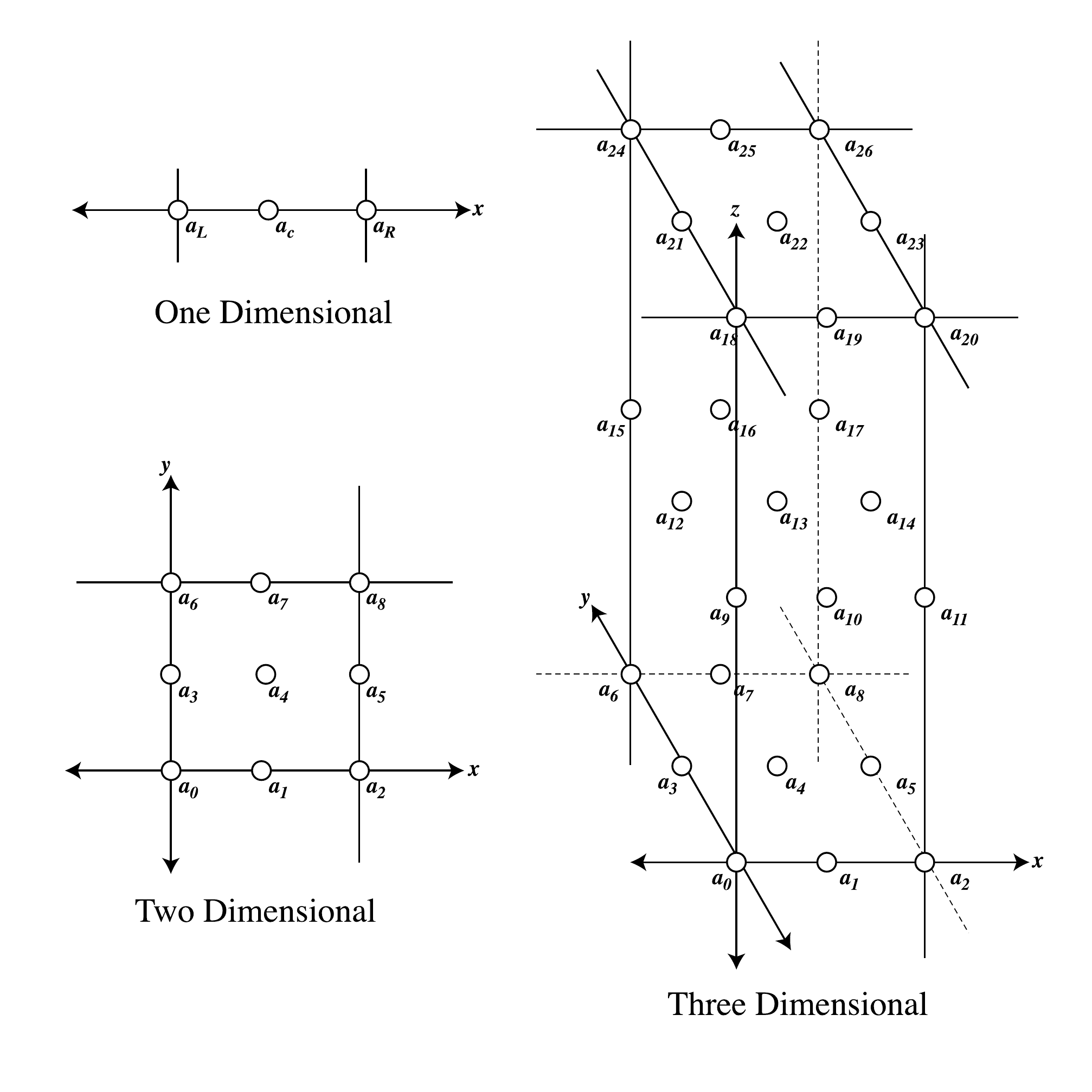}
\caption{\small PPM interpolation notation for one, two and three dimensions.  The circles represent evaluation points on a compuation cell, which are then combined to derive PPM interpolation coefficients.  Note the three dimensional case has the $z-$direction exaggerated. \label{f:figPPM}}
 \end{center} \end{figure}

In 2D the cell corners, as well as the cell boundary centres and cell centre are used, to give the PPM coefficients for each direction. By evaluating the function, $f(r)$ at the cell corners, edge centres and cell centre, the cell average will be accurate even when the cell is large compared to the radius of curvature.  The 9 points are designated in figure \ref{f:figPPM} for two dimensions.  The coefficients in the $x-$direction are:
\begin{eqnarray}
          a_{xL} & = &\frac{1}{6}[ 4 a_3 + a_0 + a_6]\, ,  \label{e:axl}\\
          a_{xC} & = &\frac{1}{6}[ 4 a_4 + a_1 + a_7]\, ,  \label{e:axc} \\
          a_{xR} & = &\frac{1}{6}[ 4 a_5 + a_2 + a_8]\, .  \label{e:axr}
\end{eqnarray} 
The cell average and PPM coeffs for the $x-$direction are: 
\begin{eqnarray}
          \langle a_x \rangle & = & \frac{1}{6}[ 4 a_{xC} + a_{xL} + a_{xR}]\, . \label{e:ax}\\
          a_{x6}  & = & 4 [ a_{xC} - 0.5 ( a_{xL} + a_{xR})]\, . \label{e:ax6}\\
         \Delta_x & = & (a_{xR} - a_{xL})\, . \label{e:deltax}
\end{eqnarray}

In the $y-$direction:
\begin{eqnarray}
          a_{yL} & = &\frac{1}{6}[ 4 a_1 + a_0 + a_2]\, , \label{e:ayl}\\
          a_{yC} & = &\frac{1}{6}[ 4 a_4 + a_3 + a_5]\, , \label{e:ayc}\\
          a_{yR} & = &\frac{1}{6}[ 4 a_7 + a_6 + a_8]\, . \label{e:ayr}
\end{eqnarray} 
The cell average and PPM coeffs for the $y-$direction are: 
\begin{eqnarray}
          \langle a_y \rangle & = & \frac{1}{6}[ 4 a_{yC} + a_{yL} + a_{yR}]\, .  \label{e:ay}\\
          a_{y6}  & = & 4 [ a_{yC} - 0.5 ( a_{yL} + a_{yR})]\, .  \label{e:ay6} \\
         \Delta_y & = & (a_{yR} - a_{yL})\, .  \label{e:deltay}
\end{eqnarray}
Note, $\langle a_x \rangle = \langle a_y \rangle = \langle a \rangle$ as expected.  

In 3D, corners, edge centres, face centres and cell centres are used to give PPM coefficients for each dimension, using the assignments in figure  \ref{f:figPPM} for three dimensions.
The coefficients in the $x-$direction are evaluated from the 27 points as:
\begin{eqnarray}
          a_{xL} & = &\frac{1}{36}[ 16 a_{12} + 4 ( a_3 + a_9 + a_{15} + a_{21} )   +   ( a_0 + a_6 + a_{18} + a_{24}) ]\, , \\
          a_{xC} & = &\frac{1}{36}[ 16 a_{13} + 4 ( a_4 + a_{10} + a_{16} + a_{22} )  +   ( a_1 + a_7 + a_{19} + a_{25}) ]\, , \\
          a_{xR} & = &\frac{1}{36}[ 16 a_{14} + 4 ( a_5 + a_{11} + a_{17} + a_{23} )  +   ( a_2 + a_8 + a_{20} + a_{26}) ]\, .
\end{eqnarray} 
The cell average and PPM coeffs for the $x-$direction are as before, using equations \ref{e:ax} - \ref{e:deltax}.

In the $y-$direction:
\begin{eqnarray}
          a_{yL} & = &\frac{1}{36}[ 16 a_{10} + 4 ( a_1 + a_9 + a_{11} + a_{19} )   +   ( a_0 + a_2 + a_{18} + a_{20}) ]\, , \\
          a_{yC} & = &\frac{1}{36}[ 16 a_{13} + 4 ( a_4 + a_{12} + a_{14} + a_{22} )  +   ( a_3 + a_5 + a_{11} + a_{23}) ]\, , \\
          a_{yR} & = &\frac{1}{36}[ 16 a_{16} + 4 ( a_7 + a_{15} + a_{17} + a_{25} )  +   ( a_6 + a_8 + a_{24} + a_{26}) ]\, .
\end{eqnarray} 
The cell average and PPM coeffs for the $y-$direction are as before, using equations \ref{e:ay} - \ref{e:deltay}.

In the $z-$direction:
\begin{eqnarray}
          a_{zL} & = &\frac{1}{36}[ 16 a_{ 4} + 4 ( a_{ 1} + a_{ 3} + a_{ 5} + a_{ 7} )  +   ( a_0 + a_2 + a_{6} + a_{8}) ]\, , \\
          a_{zC} & = &\frac{1}{36}[ 16 a_{13} + 4 ( a_{10} + a_{12} + a_{14} + a_{16} )  +   ( a_9 + a_{11} + a_{15} + a_{17}) ]\, , \\
          a_{zR} & = &\frac{1}{36}[ 16 a_{22} + 4 ( a_{19} + a_{21} + a_{23} + a_{25} )  +   ( a_{18} + a_{20} + a_{24} + a_{26}) ]\, .
\end{eqnarray} 
The cell average and PPM coeffs for the $z-$direction are as before:
\begin{eqnarray}
          \langle a_z \rangle & = & \frac{1}{6}[ 4 a_{zC} + a_{zL} + a_{zR}]\, .  \label{e:az}\\
          a_{z6}  & = & 4 [ a_{zC} - 0.5 ( a_{zL} + a_{zR})]\, .  \label{e:az6} \\
         \Delta_z & = & (a_{zL} - a_{zR})\, .  \label{e:deltaz}
\end{eqnarray}
Note, $\langle a_x \rangle = \langle a_y \rangle = \langle a_z \rangle = \langle a \rangle$ as expected.

\clearpage
\begin{thebibliography}{}

\bibitem[\protect\citeauthoryear{{Avedisova}}{{Avedisova}}{1972}]{avedisova72}
{Avedisova}, V.~S. 1972, Soviet Astronomy,{ 15}, 708--+
\bibitem[\protect\citeauthoryear{{Berger} \& {Oliger}}{{Berger} \&
  {Oliger}}{1984}]{berger84}
{Berger}, M. \& {Oliger}, J. 1984, Journal of Computational Physics,{ 53},
  484--512
\bibitem[\protect\citeauthoryear{{Blondin} \& {Cioffi}}{{Blondin} \&
  {Cioffi}}{1989}]{blondin89}
{Blondin}, J.~M. \& {Cioffi}, D.~F. 1989, \apj,{ 345}, 853--861
\bibitem[\protect\citeauthoryear{{Chevalier} \& {Imamura}}{{Chevalier} \&
  {Imamura}}{1982}]{chevalier82}
{Chevalier}, R.~A. \& {Imamura}, J.~N. 1982, \apj,{ 261}, 543--549
\bibitem[\protect\citeauthoryear{{Christian} \& {Blondin}}{{Christian} \&
  {Blondin}}{1997}]{christian97}
{Christian}, W. \& {Blondin}, J. 1997, in Bulletin of the American Astronomical
  Society, Volume~29 of Bulletin of the American Astronomical Society American
  Astronomical Society, 787
\bibitem[\protect\citeauthoryear{{Colella} \& {Woodward}}{{Colella} \&
  {Woodward}}{1984}]{cw84}
{Colella}, P. \& {Woodward}, P.~R. 1984, Journal of Computational Physics,{
  54}, 174--201
\bibitem[\protect\citeauthoryear{{Field}}{{Field}}{1965}]{field65}
{Field}, G.~B. 1965, \apj,{ 142}, 531--+
\bibitem[\protect\citeauthoryear{{Glimm}}{{Glimm}}{1965}]{glimm65}
{Glimm}, J. 1965, Comm. Pure. Appl. Math,{ 18}, 697--715
\bibitem[\protect\citeauthoryear{{Godunov}}{{Godunov}}{1959}]{godunov59}
{Godunov}, S.~K. 1959, { Mat. Sb. \& US Joint Publ. Service, JPRS 7226, 1969
  },{ 47}, 271--306
\bibitem[\protect\citeauthoryear{{Goldberg}}{{Goldberg}}{1991}]{goldberg91}
{Goldberg}, D. 1991, Computing Surveys,{ 23}, 5--48
\bibitem[\protect\citeauthoryear{{Gottlieb} \& {Groth}}{{Gottlieb} \&
  {Groth}}{1988}]{gg88}
{Gottlieb}, J.~J. \& {Groth}, C.~P.~T. 1988, Journal of Computational Physics,{
  78}, 437--458
\bibitem[\protect\citeauthoryear{{Harten}}{{Harten}}{1983}]{harten83}
{Harten}, A. 1983, J. Computational Physics,{ 49}, 357--393
\bibitem[\protect\citeauthoryear{{Harten}}{{Harten}}{1984}]{harten84}
{Harten}, A. 1984, SIAM J. Numer. Anal.,{ 21}, 1--23
\bibitem[\protect\citeauthoryear{{Jiang}}{{Jiang}}{1996}]{jiang1996}
{Jiang}, G. 1996, Journal of Computational Physics,{ 126}, 202--228
\bibitem[\protect\citeauthoryear{{Koo} \& {McKee}}{{Koo} \&
  {McKee}}{1992a}]{koo92a}
{Koo}, B.-C. \& {McKee}, C.~F. 1992a, \apj,{ 388}, 93--126
\bibitem[\protect\citeauthoryear{{Koo} \& {McKee}}{{Koo} \&
  {McKee}}{1992b}]{koo92b}
{Koo}, B.-C. \& {McKee}, C.~F. 1992b, \apj,{ 388}, 103--+
\bibitem[\protect\citeauthoryear{Liska \& Wendroff}{Liska \&
  Wendroff}{2003}]{liska03}
Liska, Richard \& Wendroff, Burton 2003, SIAM Journal on Scientific
  Computing,25(3), 995--1017
\bibitem[\protect\citeauthoryear{{Liu}, {Osher}, \& {Chan}}{{Liu}
  et~al.}{1994}]{liu1994}
{Liu}, X.-D., {Osher}, S., \& {Chan}, T. 1994, Journal of Computational
  Physics,{ 115}, 200--212
\bibitem[\protect\citeauthoryear{{Roe}}{{Roe}}{1981}]{roe1981}
{Roe}, P.~L. 1981, Journal of Computational Physics,{ 43}, 357--+
\bibitem[\protect\citeauthoryear{{Roe}}{{Roe}}{1986}]{roe1986}
{Roe}, P.~L. 1986, Annual Review of Fluid Mechanics,{ 18}, 337--365
\bibitem[\protect\citeauthoryear{{Rosen} \& {Smith}}{{Rosen} \&
  {Smith}}{2004}]{rosen04}
{Rosen}, A. \& {Smith}, M.~D. 2004, \mnras,{ 347}, 1097--1112
\bibitem[\protect\citeauthoryear{{Schulz-Rinne}}{{Schulz-Rinne}}{1993}]{schulz%
93a}
{Schulz-Rinne}, C.~W. 1993, SIAM Journal of Mathematical Analysis,{ 24}, 76--88
\bibitem[\protect\citeauthoryear{{Schulz-Rinne}, {Collins}, \&
  {Glaz}}{{Schulz-Rinne} et~al.}{1993}]{schulz93b}
{Schulz-Rinne}, C.~W., {Collins}, J.~P., \& {Glaz}, H.~M. 1993, SIAM J. Sci.
  Comput.,{ 14}, 1394--1414
\bibitem[\protect\citeauthoryear{Strang}{Strang}{1968}]{strang68}
Strang, Gilbert 1968, SIAM Journal on Numerical Analysis,5(3), 506--517
\bibitem[\protect\citeauthoryear{{Sutherland}, {Bicknell}, \&
  {Dopita}}{{Sutherland} et~al.}{2003}]{suth03b}
{Sutherland}, R.~S., {Bicknell}, G.~V., \& {Dopita}, M.~A. 2003, ApJ,{ 591},
  238
\bibitem[\protect\citeauthoryear{{Sutherland}, {Bisset}, \&
  {Bicknell}}{{Sutherland} et~al.}{2003}]{suth03a}
{Sutherland}, R.~S., {Bisset}, D.~K., \& {Bicknell}, G.~V. 2003, ApJS,{ 147},
  187
\bibitem[\protect\citeauthoryear{{Toro}}{{Toro}}{1999}]{toro99}
{Toro}, E.~F. 1999, Riemann Solvers and Numerical Methods for Fluid Dynamics: A
  Practical Introduction(2nd ed.). (Berlin: Springer-Verlag), 624

\bibitem[\protect\citeauthoryear{{van Leer}}{{van Leer}}{1974}]{vanleerII}
{van Leer}, B. 1974, Journal of Computational Physics,{ 14}, 361--370
\bibitem[\protect\citeauthoryear{{van Leer}}{{van Leer}}{1977a}]{vanleerIII}
{van Leer}, B. 1977a, Journal of Computational Physics,{ 23}, 263--299
\bibitem[\protect\citeauthoryear{{van Leer}}{{van Leer}}{1977b}]{vanleerIV}
{van Leer}, B. 1977b, Journal of Computational Physics,{ 23}, 276--299
\bibitem[\protect\citeauthoryear{{van Leer}}{{van Leer}}{1979}]{vanleerV}
{van Leer}, B. 1979, Journal of Computational Physics,{ 32}, 101--136
\bibitem[\protect\citeauthoryear{{Weaver} et~al.}{{Weaver}
  et~al.}{1977}]{weaver77}
{Weaver}, R., {McCray}, R., {Castor}, J., {Shapiro}, P., \& {Moore}, R. 1977,
  \apj,{ 218}, 377--395
\bibitem[\protect\citeauthoryear{{Woodward} \& {Colella}}{{Woodward} \&
  {Colella}}{1984}]{cw84b}
{Woodward}, P. \& {Colella}, P. 1984, Journal of Computational Physics,{ 54},
  115--173
\end{thebibliography}
\end{document}